# Quantifying Charge Noise Sources in Quantum Dot Spin Qubits via Impedance Spectroscopy, DLTS, and C-V Analysis

Tyafur Rahman Pathan,[1] and Daryoosh Vashaee[1,2,*]

[1]Electrical and Computer Engineering Department, North Carolina State University, Raleigh, North Carolina 27616, USA

[2]Materials Science and Engineering Department, North Carolina State University, Raleigh, North Carolina

## Abstract

The coherence and fidelity of quantum dot (QD) spin qubits are fundamentally limited by charge noise originating from electrically active trap states at oxide interfaces, heterostructure boundaries, and within the bulk semiconductor. These traps introduce electrostatic fluctuations that couple to the qubit via spin-orbit interactions or charge-sensitive confinement potentials, leading to dephasing and gate errors. In this work, we present a general framework for identifying and quantifying the spectral signatures of these trap states using AC impedance spectroscopy, deep-level transient spectroscopy (DLTS), and conventional capacitance-voltage (C-V) analysis. While demonstrated using strained Ge/SiGe quantum well heterostructures, the approach is broadly applicable to other qubit platforms. We show that oxide interface traps dominate low-frequency conductance and Nyquist spectra, quantum well (QW) interface traps are best resolved via multi-exponential decays in time-domain response, and bulk traps affect high-frequency admittance and leakage currents. By correlating each trap type to its characteristic time constant, spatial location, and frequency response, we offer a diagnostic toolset for disentangling material-induced noise sources that limit qubit performance. This unified methodology connects traditional semiconductor defect metrology with emerging qubit noise analysis and provides actionable guidance for material and process optimization in scalable quantum information technologies.

Keywords: Quantum dot spin qubits, charge noise, impedance spectroscopy, deep-level transient spectroscopy, Ge/SiGe heterostructures

## I. Introduction

Gate-defined quantum dot (QD) spin qubits in Ge/SiGe heterostructures have emerged as leading candidates for scalable quantum information processing, owing to their high hole mobility, strong and tunable spin-orbit coupling (SOC), and compatibility with CMOS fabrication processes.[1-3] Among these, hole spin qubits in strained Ge quantum wells (QWs) offer the key advantage of all-electrical spin manipulation via electric-dipole spin resonance (EDSR), enabling ultrafast single- and two-qubit gate operations.[4,5] However, the same SOC that facilitates electrical control also enhances susceptibility to charge noise, which perturbs the spin states through electric field fluctuations and ultimately limits coherence times and gate fidelities.[6,7]

Spin qubits in Ge/SiGe and GeSn/Ge heterostructures can be realized in either heavy-hole (HH) or light-hole (LH) configurations, each exhibiting distinct spin dynamics and noise sensitivities. Recent comparative modeling has shown that LH-based qubits can achieve longer coherence times and higher Rabi frequencies due to reduced spin relaxation and favorable g-factor anisotropy, underscoring the importance of heterostructure and strain engineering in optimizing qubit performance.[8]

* Corresponding author: dvashae@ncsu.edu

Charge noise in gate-defined Ge QD spin qubits arises primarily from electrically active defects in the surrounding material stack, including oxide-semiconductor interfaces, QW interfaces, bulk semiconductor layers, and near-interface border traps within the oxide. These defects act as fluctuating charge centers, whose dynamic occupancy modulates the local electrostatic environment, inducing time-dependent field noise that couples to the qubit via spin-orbit interaction. The spatial location, trap density, and emission timescale of each trap type govern the spectral shape of the resulting charge noise, with direct implications for qubit dephasing times ($T_2^*$). In particular, slow traps at the oxide interface and within the oxide bulk (border traps) contribute predominantly to low-frequency ($1/f$-type) noise, while faster traps near or inside the QW generate higher-frequency fluctuations in the kilohertz-to-megahertz range, frequencies that overlap with single- and two-qubit gate operations. Accurate characterization and modeling of these trap dynamics are therefore essential to mitigating decoherence and guiding quantum device design.[9-11]

Despite the central role of charge noise in limiting qubit performance, existing studies have largely focused on phenomenological models or indirect measurements of noise power spectral density. What remains lacking is a comprehensive, physics-based, spatially resolved framework for identifying and quantifying the sources of charge noise in actual device architectures. In particular, there is a need to distinguish among oxide interface traps, quantum well interface traps, and bulk defects, each of which exhibits different capture/emission dynamics, spatial coupling to the QD, and frequency response.

To address this gap, we present a detailed numerical study of impedance spectroscopy and DLTS in Ge/SiGe QW heterostructures tailored to gate-defined QD geometries. These techniques are widely used in conventional semiconductor device analysis to characterize electrically active defect states, but their application to quantum-confined systems for charge noise diagnostics remains largely unexplored. Impedance spectroscopy, through capacitance-voltage (C-V) and conductance-frequency (G/ω) measurements, enables the extraction of trap density, capture cross-sections, and relaxation times across a broad frequency range under varying gate biases[12,13]. DLTS extends this capability by providing time-domain signatures of trap emission, allowing identification of trap location and dynamics that are otherwise buried in steady-state measurements[1-3].

In this study, we consider three distinct categories of electrically active trap states based on their physical location within the Ge/SiGe quantum well heterostructure.

1. Oxide Interface Traps ($N_{t,ox}$) are modeled as surface states located precisely at the semiconductor side of the $Al_2O_3/Si_{0.2}Ge_{0.8}$ interface. These traps are electrostatically coupled to the gate and typically result from bonding defects or dangling bonds at the oxide-semiconductor boundary.
2. Quantum Well Interface Traps ($N_{t,QW}$) reside at the interfaces between the Ge quantum well and the adjacent $Si_{0.2}Ge_{0.8}$ barrier layers. These are also modeled as surface states and arise from interface roughness, strain relaxation defects, or interdiffusion at the epitaxial heterojunctions.

3. Bulk Traps ($N_{t,bulk}$) are modeled as a volumetric distribution of deep-level defects within the semiconductor stack, including the Ge quantum well and surrounding SiGe barriers. These traps can originate from point defects, dislocations, or unintentional doping and are distributed uniformly throughout the active region.

By explicitly distinguishing these spatial categories, the simulation framework enables analysis of how the proximity and density of each trap type influence both frequency-domain admittance and time-domain transient responses.

Using frequency-domain perturbation theory and transient time-domain simulations, we model the complex impedance and current response of MIS and QW devices under varying conditions. The resulting data are interpreted using SRH recombination theory and trap models with Gaussian energy distributions, enabling us to extract key parameters such as emission time constants, spectral density contributions, and their dependence on device geometry, gate voltage, and material choice.

Most importantly, we establish direct connections between these electrical characterization outputs and the dominant sources of charge noise in Ge QD spin qubits. We show how different trap populations manifest in impedance and DLTS signals, and how their temporal response maps to noise components in the qubit operating band. For example, we find that oxide traps dominate the slow response and contribute to low-frequency quasistatic noise, while fast traps near the QW region can be isolated using DLTS and contribute disproportionately to fast decoherence and leakage in spin rotations. These insights are synthesized into guidelines for qubit device optimization, including dielectric choice, epitaxial interface quality, and acceptable trap density thresholds.

This work represents the first comprehensive study to bridge electrical defect spectroscopy with quantum coherence analysis in Ge quantum dot systems. Our methodology provides a scalable and generalizable framework for diagnosing and mitigating charge noise in next-generation quantum devices. Although this study focuses on strained Ge/SiGe QWs as a representative platform, the modeling framework (including AC impedance, DLTS-style transient analysis, and trap spectral coupling) directly applies to Si/SiGe QW systems. Both architectures rely on high-$\kappa$ gate stacks, strained quantum wells, and are limited by interface and bulk trap-induced charge noise. Differences in effective mass, carrier type, and band structure may affect quantitative results such as trap time constants, but the overall methodology and physical interpretation remain applicable to electron-based Si/SiGe quantum dots.

All frequency- and time-domain simulations in this work were performed using COMSOL Multiphysics, incorporating realistic material parameters and device geometries representative of gate-defined Ge/SiGe quantum well structures.

## II. Device Structures and Simulation Framework

We investigate two heterostructure configurations representative of gate-stack geometries commonly employed in Ge/SiGe quantum device research. The first is a metal-insulator-semiconductor (MIS) capacitor composed of a gold top electrode, a 30 nm $Al_2O_3$ high-$\kappa$ gate dielectric, and a 55 nm relaxed $Si_{0.2}Ge_{0.8}$ buffer layer grown on a virtual substrate. This structure serves as a reference for isolating and characterizing traps at the oxide/SiGe interface and within the SiGe itself.

The second configuration introduces a compressively strained Ge QW into the stack: a 16 nm Ge layer is epitaxially embedded between two 55 nm $Si_{0.2}Ge_{0.8}$ barriers, forming a type-I band alignment. This Au-$Al_2O_3$-$Si_{0.2}Ge_{0.8}$-Ge-$Si_{0.2}Ge_{0.8}$ QW device captures the essential physics relevant to gate-defined hole spin qubits, where the Ge layer acts as the active quantum-confined region. The introduction of the Ge QW adds a second heterointerface, allowing for the exploration of additional trap states associated with strain relaxation, interfacial roughness, or misfit dislocations.

Unless otherwise specified, all simulations and analyses assume a gate surface area of 1 $mm^2$, representative of large-area test structures used in DLTS and impedance measurements. Because the gate area is large and the device is laterally homogeneous, the charge dynamics can be effectively modeled as one-dimensional, with all field and carrier variations occurring along the growth direction.

Three categories of electrically active traps are considered: (1) oxide interface traps located at the $Al_2O_3$/SiGe boundary, (2) QW interface traps at the SiGe/Ge heterointerfaces, and (3) bulk traps distributed within the Ge or SiGe layers. These trap types differ in their spatial localization, energy depth, and dynamic response, making them distinguishable under varying electrical stimuli.

Simulations are performed over a broad frequency range to resolve the characteristic response of each trap type. Additionally, we conduct transient simulations in the time domain to demonstrate a more robust methodology for distinguishing between trap classes based on their emission time constants and spatial coupling to the gate.

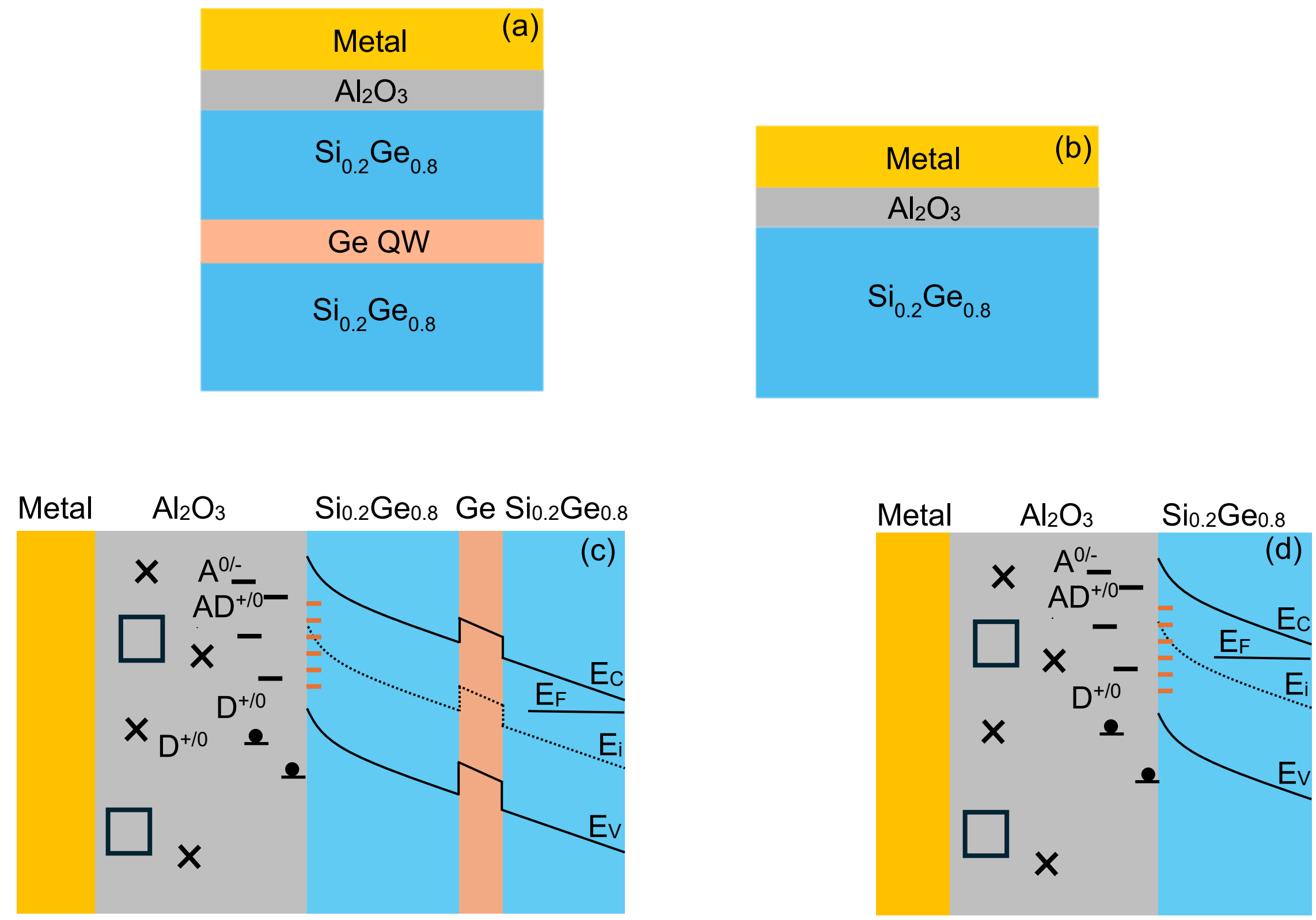

Figure 1: Cross-sectional schematics of the two device structures studied: (a) an Au-$Al_2O_3$-$Si_{0.2}Ge_{0.8}$-Ge-$Si_{0.2}Ge_{0.8}$ QW structure, and (b an Au-$Al_2O_3$-$Si_{0.2}Ge_{0.8}$ metal–

insulator–semiconductor (MIS) capacitor. Panels (c) and (d) show the corresponding energy band diagrams for the MIS and QW devices, respectively, illustrating band bending under equilibrium conditions. The thicknesses of the gate dielectric ($Al_2O_3$), $Si_{0.2}Ge_{0.8}$ barrier layers, and the Ge quantum well are 30 nm, 55 nm, and 16 nm, respectively. These geometries represent the vertical stack used in simulations to analyze trap-induced effects in impedance and transient responses relevant to Ge/SiGe quantum devices.

Unless otherwise noted, all simulations presented in this work were performed at T = 300 K, consistent with standard practice in MOS and DLTS modeling to ensure full trap activity and parameter accuracy.

Various electrically active traps can exist in different regions of Ge/SiGe QW devices, each influencing charge transport, recombination dynamics, and noise behavior. These traps arise from both intrinsic defects and process-induced imperfections, and include bulk defects, interfacial disorder, and fixed charges at dielectric interfaces. Their spatial location, charge state, and energy level within the bandgap determine their impact on the device's AC and transient response. Table 1 summarizes representative trap types observed in Ge/SiGe QW systems, categorized by charge type and energy level. Accurate modeling of these traps is essential for realistic simulation of device performance, particularly for qubit applications where decoherence is highly sensitive to charge fluctuations.[4-14]

Table 1: Representative Trap Types in Ge/SiGe Quantum Well Devices

| **Bulk Traps** | **Acceptor/Donor** | **Shallow/Deep** |
|---|---|---|
| Oxygen Vacancies | Acceptor/Donor | Deep [4,5] |
| Al Interstitials | Donor | Shallow [6] |
| Oxygen Interstitials | Acceptor | Deep [5] |
| **Interface Traps** | **Acceptor/Donor** | **Shallow/Deep** |
| Dangling Bonds | Acceptor/Donor | Deep [7,9] |
| Interfacial Defects & Disorder | Acceptor/Donor | Shallow/Deep [10,11] |
| $SiO_x$ Suboxides | Acceptor/Donor | Deep [12] |
| $GeO_x$ Suboxides | Acceptor | Deep [12] |
| Ge Segregation | Acceptor/Donor | Deep [10] |
| Fixed Charges/Dipoles | Donor/Acceptor | Shallow [13,14] |
| Vacancy Clusters | Acceptor | Deep [6] |
| Interstitial O atoms | Donor | Shallow/Deep [5] |
| Grain Boundaries | Acceptor/Donor | Deep [6] |

### III. Trap Modeling and Spectroscopic Formalism

Understanding how electrically active defects influence device behavior requires a detailed physical description of trap-assisted processes under both steady-state and dynamic conditions. In this section, we present the theoretical framework used to model the effects of trap states on AC impedance and transient response. The formalism includes recombination dynamics via Shockley-Read-Hall (SRH) theory, Gaussian energy distributions for interface and bulk traps, and the derivation of small-signal equations for frequency-domain analysis. Together, these models allow us to simulate and interpret the frequency- and time-dependent conductance behavior of Ge/SiGe quantum well structures under various biasing conditions.

### III.A. Physical Models for Trap States and Recombination Mechanisms

To model recombination via traps, we use SRH formalism. This approach models carrier capture and emission events mediated by deep-level defects located within the bandgap. The total recombination rate is given by:

$$R_n = R_p = \frac{np - \gamma_n \gamma_p n_i^2}{\tau_p(n + n_1) + \tau_n(p + p_1)}$$

Here, $\tau_n$, $\tau_p$ represent the lifetimes associated with electron and hole capture, and $n_1$, $p_1$ are the trap occupancy terms dependent on the trap energy level $E_t$. The SRH model accounts for both forward and reverse recombination pathways and allows fitting to experimentally extracted lifetimes.

$n_1$ ($p_1$) is the equivalent electron (hole) concentration due to the trap energy $E_t$ coinciding with fermi level. $\gamma_n$ and $\gamma_p$ are electron and hole degeneracy factors respectively, which were assumed to be 1.

$$n_1 = \gamma_n n_i \exp\left(\frac{\Delta E_t}{k_B T}\right)$$

$$p_1 = \gamma_p n_i \exp\left(-\frac{\Delta E_t}{k_B T}\right)$$

$$\Delta E_t = E_t - E_i$$

To resolve spatial and energetic variations in traps, we use explicit energy-distributed trap model as implemented in the simulation framework. A Gaussian distribution of trap states is defined within the bandgap:

$$g_t(E) = \frac{N_t}{\sqrt{2\pi}\sigma_E} \exp\left(-\frac{(E - E_t)^2}{2\sigma_E^2}\right)$$

Total trap density is given by:

$$N_t = \int_{E_{t,min}}^{E_{t,max}} g_t(E_t) dE_t$$

This allows modeling of broad defect bands due to interface roughness, dislocations, or alloy disorder. The occupancy function of traps is governed by Fermi-Dirac statistics:

$$f_t(E) = \frac{1}{1 + \frac{1}{g_d} \exp\left(\frac{E - E_f(E_t)}{k_B T}\right)}$$

Net charge contributed by traps considering both electrons in occupied traps and holes unoccupied traps is given by:

$$Q_t = -q \int_{E_0}^{E_{t,max}} f_t(E_t) g_t(E_t) dE_t + q \int_{E_{t,min}}^{E_0} \left(1 - f_t(E_t)\right) g_t(E_t) dE_t$$

Here, $E_0$ is the neutral trap level.

Electron and hole recombination rates via these traps are integrated over energy:

$$R_n = \int r_n dE, \quad R_p = \int r_p dE$$

$$r_n = C_n g_t(E) n \left(1 - f_t(E)\right) \left(1 - e^{\frac{E_f(E_t) - E_{fn}}{k_B T}}\right)$$

$$r_p = C_p g_t(E) p f_t(E) \left(1 - e^{\frac{E_{fp} - E_f(E_t)}{k_B T}}\right)$$

$$C_n = \sigma_{Cn} v_{thn} \; , \; C_p = \sigma_{Cp} v_{thp}$$

The rates $r_n$ and $r_p$ describe how quickly electrons and holes are captured or emitted by traps. The balance between $r_n$ and $r_p$ determines the time evolution of $f_t(E_t)$ and the steady-state response.

$$g_t(E_t) \, \frac{\partial f_t(E_t)}{\partial t} = r_n - r_p$$

As discussed, we consider three spatial configurations of traps relevant to the Ge/SiGe quantum well device: oxide interface traps located at the $Al_2O_3$/SiGe boundary (semiconductor side), QW interface traps situated at the Ge/SiGe heterointerfaces, and bulk traps distributed throughout the SiGe-Ge-SiGe stack. Oxide interface traps are modeled as a two-dimensional sheet located on the semiconductor side of the $Al_2O_3$/SiGe interface, consistent with standard capacitance and conductance analysis approaches where interface states are strongly coupled to the semiconductor band edge. These spatially resolved trap models allow us to isolate and quantify the contribution of each trap type to the overall frequency-dependent impedance and energy dissipation in the device.

### III.B. Frequency-Domain Formalism for AC Impedance Analysis

To analyze the influence of trap states on the AC response of Ge/SiGe quantum well devices, we employ a frequency-domain small-signal perturbation approach. This formalism captures the system's linear response to a sinusoidal voltage excitation over a broad frequency spectrum, enabling detailed extraction of impedance and conductance characteristics associated with trap dynamics. The electrical behavior is governed by a self-consistent solution of the Poisson, drift-diffusion, and continuity equations, which are linearized around a steady-state bias point to compute the complex admittance response.

At the core of the model is the determination of carrier concentrations, which are governed by the quasi-Fermi levels under nonequilibrium conditions:

$$n = N_c e^{\frac{q(E_{fn}-E_c)}{k_B T}}, \qquad p = N_v e^{\frac{q(E_v-E_{fp})}{k_B T}}$$

Here, $E_{fn}$ and $E_{fp}$ represent the electrochemical potentials for electrons and holes, which are perturbed by the applied AC signal. A complete list of symbols and physical constants used is provided in Table 2.

These relations assume nondegenerate carrier statistics, which are valid for the low-doping, low-injection regimes typical in quantum devices.

The electrostatic potential is governed by Poisson's equation:

$$\nabla^2 V = -\frac{\rho}{\epsilon} = -\frac{q(p - n + N_d^+ - N_a^-)}{\epsilon}$$

This equation ensures self-consistency between the potential and the net space charge density, incorporating both free carriers and ionized dopants. The solution determines how band edges bend under applied bias, directly influencing trap energy alignment and occupancy.

Carrier transport is modeled by drift-diffusion current equations that include both field-driven and diffusive contributions:

$$\boldsymbol{J_n} = qn\mu_n \nabla E_{fn} + \left(\frac{\left(E_c - E_{fn}\right)\mu_n + Q_{th,n}}{T}\right)\nabla T$$

$$\boldsymbol{J_p} = qp\mu_p \nabla E_{fp} + \left(\frac{\left(E_v - E_{fp}\right)\mu_p - Q_{th,p}}{T}\right)\nabla T$$

The first terms represent electrical drift under the gradient of the quasi-Fermi levels, while the second terms capture thermoelectric effects (Seebeck contribution), which may be relevant for thermally activated carrier exchange with traps.

Band edges are computed using spatially varying expressions:

$$E_c = -(V + \chi_0),\ E_v = -\left(V + \chi_0 + E_{g,0}\right)$$

Charge conservation is enforced by the continuity equations:

$$q\frac{\partial n}{\partial t} = \nabla.\boldsymbol{J_n}, \qquad q\frac{\partial p}{\partial t} = -\nabla.\boldsymbol{J_p}$$

, which in the frequency domain become:

$$j\omega qn = \nabla \cdot J_n + G_n - R_n, \qquad j\omega qp = -\nabla \cdot J_p + G_p - R_p$$

Here, $\omega$ is the angular frequency of the AC signal. The complex nature of these equations captures both amplitude and phase shifts due to carrier inertia and delayed trap response.

The simulation proceeds in two sequential steps. First, a DC steady-state solution is performed by solving the full set of nonlinear coupled equations, including Poisson's equation and the continuity and drift-diffusion equations. This yields the equilibrium values of the electrostatic potential $V_0$, electron and hole concentrations $n_0$ and $p_0$, and the corresponding steady-state current densities. Once the system reaches this equilibrium, the second step involves an AC small-signal analysis. Here, the equations are linearized around the previously obtained DC operating point and solved in the frequency domain. This step produces frequency-dependent electrical quantities such as the complex impedance $Z(\omega)$, admittance $Y(\omega)$, and conductance $G(\omega)$, enabling detailed analysis of the device's response to small AC perturbations.

To interpret the simulated impedance, we model the device as a series combination of the gate oxide capacitance and a parallel RC network representing the semiconductor response. The oxide contributes a fixed geometric capacitance $C_{ox}$, while the semiconductor is represented by a parallel combination of capacitance $C_s$ and conductance $G$, where $G$ captures dissipative processes such as trap-assisted recombination and hopping conduction. By subtracting the known oxide contribution from the total simulated admittance, we isolate the trap-induced conductance originating from the semiconductor layers. Subtracting the oxide impedance yields:

$$Y_{sem}(\omega) = \frac{1}{Z(\omega) - 1/(j\omega C_{ox})}$$

The real part of this admittance gives the frequency-dependent conductance:

$$G(\omega) = \Re[Y_{sem}(\omega)]$$

The shape and position of $G(\omega)$ peaks provide insight into the characteristic emission times of traps and their coupling to the gate.

### III.C Time-Domain Modeling for DLTS: Method and Interpretation

While impedance spectroscopy provides frequency-resolved information on trap response, DLTS offers complementary insight into trap emission dynamics in the time domain. Both methods rely on the same underlying electrostatic and recombination physics, particularly Poisson's equation, drift-diffusion transport, and SRH recombination, but differ in how the trap occupancy dynamics are probed.

In a DLTS experiment, the system is typically brought into nonequilibrium by applying a voltage pulse that alters the trap occupancy, either by filling traps (forward bias) or emptying them (reverse bias). After the pulse is removed, the device is held at a fixed reverse bias and the transient response of the capacitance (or current) is monitored. The decay reflects the thermal emission of carriers from traps back into the conduction or valence bands, returning the system to equilibrium.

In the context of simulations, this process is modeled by explicitly solving the time-dependent drift-diffusion and continuity equations:

$$q\frac{\partial n}{\partial t} = \nabla \cdot J_n + G_n - R_n, \quad q\frac{\partial p}{\partial t} = -\nabla \cdot J_p + G_p - R_p$$

where $G_n$, $G_p$ are generation terms (negligible under dark, low-bias conditions), and $R_n$, $R_p$ are recombination rates including the SRH contribution. These rates evolve in time as trap occupancy changes.

The key quantity in DLTS is the thermal emission rate of traps, which determines the time constant of the capacitance or current transient. For a trap at energy $E_t$ below the conduction band (for electron emission), the emission rate $e_n$ is given by:

$$e_n = \sigma_n v_{th,n} N_c exp\left(-\frac{E_c - E_t}{k_B T}\right)$$

Likewise, for hole emission:

$$e_p = \sigma_p v_{th,p} N_v exp\left(-\frac{E_t - E_v}{k_B T}\right)$$

These expressions show that emission rates are exponentially sensitive to the trap depth and temperature, which is why DLTS is often performed over a temperature sweep to extract the activation energy and capture cross-section of the traps. The pre-factors include the capture cross-section $\sigma$ and the thermal velocity $v_{th}$, which represent the microscopic probability and rate at which carriers interact with trap sites.

The observed capacitance transient $C(t)$ typically follows an exponential decay governed by these emission rates:

$$C(t) = C_0 + \Delta C \cdot e^{-t/\tau}$$

where $\tau \approx 1/e_n$ or $1/e_p$, depending on the dominant carrier species involved. In practice, real systems often involve multiple traps with overlapping time constants, resulting in more complex decay curves that require fitting with a sum of exponentials or stretched exponential forms, as we will see for the device under study.

From a modeling perspective, this time-domain transient is computed by applying a voltage pulse to the gate (in accumulation or depletion mode), then holding the structure at a fixed DC bias and tracking the time evolution of the terminal capacitance or charge. This is implemented using a time-dependent solver, with trap models defined as in the frequency-domain case. Interface and bulk trap distributions are included using the same Gaussian energy profile $g_t(E)$ and recombination integrals $R_e$, $R_h$ discussed previously.

In our simulations, we apply this approach to distinguish between oxide interface traps, QW interface traps, and bulk traps. The spatial location of the trap affects how strongly the emission current couples to the gate electrode, altering the amplitude and time constant of the observed signal. Interface traps typically exhibit faster response times and more pronounced transients due to their proximity to the gate, whereas bulk traps, especially those deeper in the Ge channel, can exhibit slower transients with lower signal amplitude.

In contrast to impedance spectroscopy, which resolves trap behavior in the frequency domain, DLTS simulations provide direct access to the trap emission kinetics and allow correlation of observed time constants with trap energy levels and spatial localization. When combined, these two techniques offer a comprehensive picture of the trap landscape in Ge/SiGe QW devices, critical for understanding decoherence and charge noise in spin qubit applications.

This formalism also offers a pathway to interpret experimental DLTS data using calibrated numerical models, enabling quantitative extraction of trap parameters such as energy level, spatial depth, and capture cross-section. This capability is particularly important for quantum devices, where even small numbers of interface or bulk traps can dramatically affect qubit coherence.

Table 2: Definition of Variables Used in Frequency-Domain and Trap Models

| **Symbol** | **Description** | **Units** |
|---|---|---|
| $n, p$ | Electron and hole concentrations | $cm^{-3}$ |
| $N_c, N_v$ | Effective density of states in conduction/valence band | $cm^{-3}$ |
| $E_c, E_v$ | Conduction and valence band edge energies | eV |
| $E_{fn}, E_{fp}$ | Electron and hole quasi-Fermi levels | eV |
| $V$ | Electrostatic potential | V |
| $\varepsilon$ | Dielectric permittivity | F/cm |
| $\rho$ | Space charge density | C/ $cm^3$ |
| $\mu_n, \mu_p$ | Electron and hole mobilities | $Cm^2/V \cdot s$ |
| $T$ | Temperature | K |
| $q$ | Elementary charge | C |
| $k_B$ | Boltzmann constant | eV/K |
| $J_n, J_p$ | Electron and hole current densities | $A/cm^2$ |
| $Q_{th,n}, Q_{th,p}$ | Thermoelectric transport coefficients | W/cm·K |
| $E_t$ | Trap energy level | eV |
| $E_i$ | Intrinsic Fermi level | eV |
| $\gamma_n, \gamma_p$ | Degeneracy factors for electrons and holes | - |
| $\tau_n, \tau_p$ | Electron and hole lifetimes (SRH model) | s |
| $n_i$ | Intrinsic carrier concentration | $cm^{-3}$ |
| $\sigma_n, \sigma_p$ | Electron and hole capture cross sections | $cm^2$ |
| $v_{th,n}, v_{th,p}$ | Thermal velocity of electrons and holes | cm/s |
| $C_n, C_p$ | Capture constants $C = \sigma v_{th}$ | $cm^3/s$ |
| $g_t(E)$ | Trap density per unit energy | $cm^{-3} \cdot eV^{-1}$ |
| $f_t(E)$ | Trap occupancy function | - |
| $r_n, r_p$ | Electron and hole recombination rates (differential) | $cm^{-3} \cdot s^{-1}$ |
| $R_n, R_p$ | Total recombination rates (integrated over energy) | $cm^{-3} \cdot s^{-1}$ |
| $C_{ox}$ | Gate oxide capacitance | F |
| $C_s$ | Semiconductor capacitance | F |
| $G$ | Conductance due to traps | S ($\Omega^{-1}$) |
| $Y$ | Admittance (complex) | S |
| $Z$ | Impedance (complex) | Ω |
| $\omega$ | Angular frequency | rad/s |
| $j$ | Imaginary unit ($\sqrt{-1}$) | - |

Table 3: Materials Properties

| **Symbol** | **Description** | **Value** |
|---|---|---|
| $E_g$ | Bandgap (Ge) | 0.664 eV |
| $\epsilon$ | Relative Permittivity (Ge) | 16.2 |
| $\chi$ | Electron Affinity (Ge) | 4 eV |
| $N_c$ | Effective density of states, conduction band (Ge) | $1.04 \times 10^{19}$ $cm^{-2}$ $eV^{-1}$ |
| $N_v$ | Effective density of states, valence band (Ge) | $6 \times 10^{18}$ $cm^{-2}$ $eV^{-1}$ |
| $\mu_n$ | Electron Mobility (Ge) | 3900 $cm^2$ $V^{-1}$ $s^{-1}$ |
| $\mu_p$ | Hole Mobility (Ge) | 1900 $cm^2$ $V^{-1}$ $s^{-1}$ |
| $E_g$ | Bandgap ($Si_{0.2}Ge_{0.8}$) | 0.797 eV |

| $\epsilon$ | Relative Permittivity ($Si_{0.2}Ge_{0.8}$) | 15.56 |
|---|---|---|
| $\chi$ | Electron Affinity ($Si_{0.2}Ge_{0.8}$) | 4.01 eV |
| $N_c$ | Effective density of states, conduction band ($Si_{0.2}Ge_{0.8}$) | $2.75\times10^{19}$ cm$^{-2}$ eV$^{-1}$ |
| $N_v$ | Effective density of states, valence band ($Si_{0.2}Ge_{0.8}$) | $7.16\times10^{18}$ cm$^{-2}$ eV$^{-1}$ |
| $\mu_n$ | Electron Mobility ($Si_{0.2}Ge_{0.8}$) | 419 cm$^2$ V$^{-1}$ s$^{-1}$ |
| $\mu_p$ | Hole Mobility ($Si_{0.2}Ge_{0.8}$) | 420 cm$^2$ V$^{-1}$ s$^{-1}$ |
| $m^*_{n,DOS}$ | Electron density of states effective mass ($Si_{0.2}Ge_{0.8}$) | 1.06 $m_0$ |
| $m^*_{p,DOS}$ | Hole density of states effective mass ($Si_{0.2}Ge_{0.8}$) | 0.434 $m_0$ |
| $\epsilon$ | Relative Permittivity ($Al_2O_3$) | 8.5 |

## IV. Results and Discussion

This section presents the simulated electrical response of Ge/SiGe MIS and quantum well devices under both frequency-domain and time-domain excitation, with a focus on isolating the signatures of oxide interface, QW interface, and bulk traps. The results are organized to first establish baseline behavior in MIS capacitors, followed by analysis of how trap parameters influence conductance spectra and capacitance. We then examine the effect of device architecture and dielectric material, culminating in a study of time-domain current transients for DLTS-like analysis. The implications for qubit-relevant charge noise and device optimization are addressed in the following section.

### IV.A. Frequency Response of SiGe MIS Capacitors

Figure 2 and Figure 3 provide key insights into the frequency-dependent conductance behavior of the Au-$Al_2O_3$-SiGe MIS structure across different gate bias regimes. The normalized conductance $G/\omega$, which serves as a proxy for the density and dynamic response of interface and bulk traps, reveals distinct signatures under inversion and accumulation conditions.

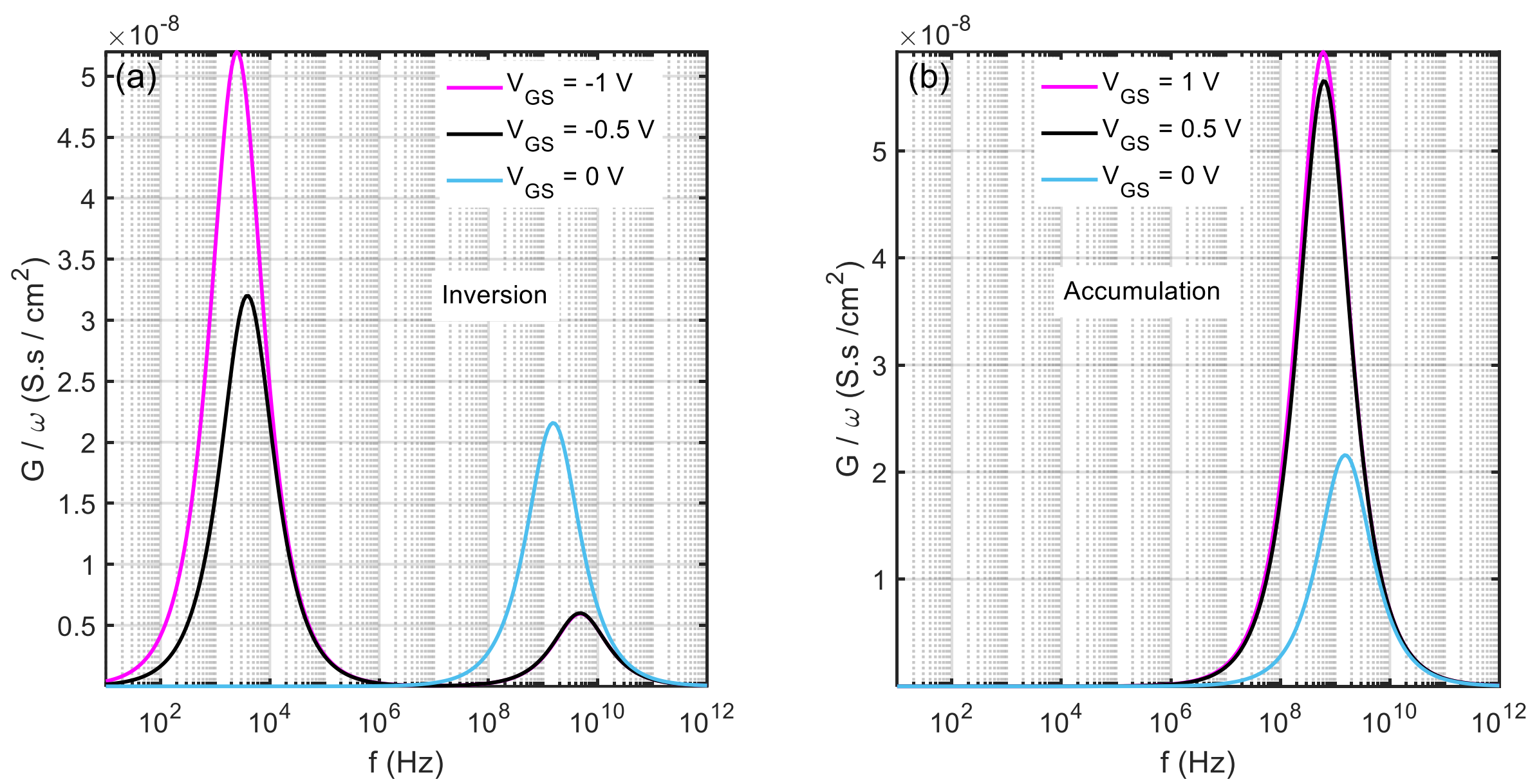

Figure 2: Simulated $G/\omega$ versus frequency for an Au-$Al_2O_3$-SiGe MIS device under (a) inversion and (b) accumulation conditions, assuming an n-type background doping of $10^{15}$ cm$^{-3}$ and a minority carrier lifetime of 10 ns based on the SRH recombination model.† In inversion, the peak in G/ω appears in the kHz range and is governed primarily by the minority carrier generation-recombination time. In contrast, under accumulation, the peak shifts toward the GHz range, dominated by the majority carrier response and strongly dependent on the doping level.

Under negative gate bias (inversion for an n-type substrate), the depletion region extends deep into the SiGe, and the AC response is governed by the availability and dynamics of minority carriers. In this regime, the dominant process is generation-recombination via trap-assisted transitions, which are inherently slow due to the limited concentration of thermally generated holes. As a result, the peak frequency of the G/ω spectrum is located in the low-kHz range (Figure 2a), consistent with the emission time constants associated with SRH dynamics. The amplitude of the G/ω peak remains relatively constant in strong inversion because the recombination rate saturates as the depletion width reaches its maximum extent and the minority carrier generation rate stabilizes.

In contrast, under positive gate bias (accumulation), the surface is flooded with majority electrons. Here, the dominant dynamic process is dielectric relaxation, where mobile carriers respond to the AC perturbation by screening the electric field. This is a much faster process than minority carrier generation and leads to a peak frequency in the GHz range, as shown in Figure 3b. The relaxation time $\tau_D$ in this regime is inversely related to the doping density and mobility, in agreement with the trends discussed in Section III and illustrated in Figure 5. As the frequency increases beyond the relaxation capability of the system, the carriers can no longer respond quickly enough, and the G/ω signal rolls off.

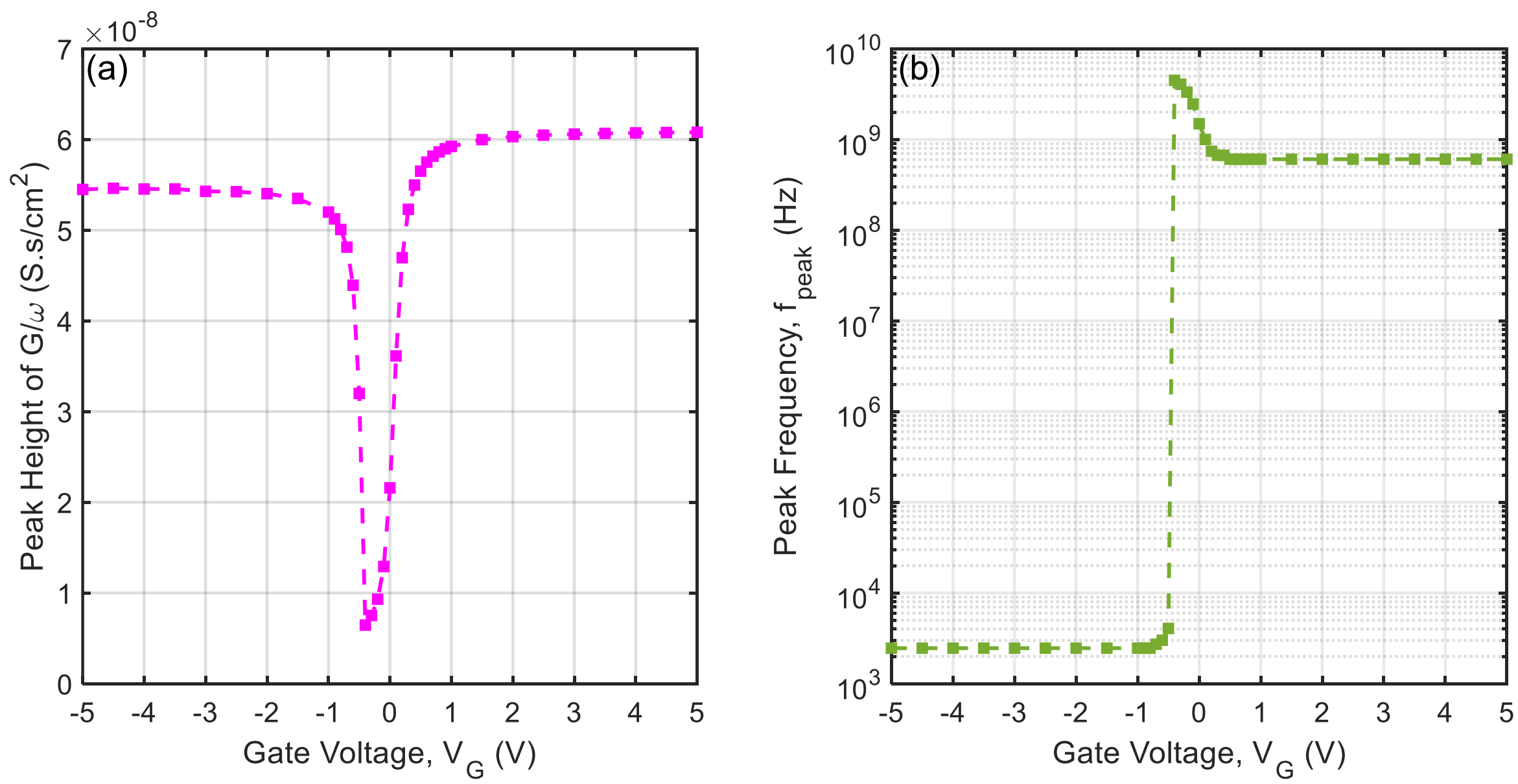

Figure 3: (a) Extracted G/ω peak height and (b) corresponding peak frequency $f_{peak}$ as functions of gate voltage $V_G$ for an n-type doping level of $10^{15}$ cm$^{-3}$ and a minority carrier lifetime of 10 ns. In the inversion regime, the peak height is lower and the response is slower due to limited minority carrier availability. In accumulation, the higher majority

† For a capture cross-section of $10^{-16}$ cm$^2$, trap density corresponding to carrier lifetime of 10 ns is $5\times10^{16}$ cm$^{-3}$

carrier density leads to increased peak height and faster response. Near the depletion region, the G/ω peak height drops significantly, while the peak frequency rises sharply, reflecting changes in carrier dynamics, charge distribution, and the influence of the internal electric field.

The behavior near zero bias reveals a transition zone where both majority and minority carriers have reduced influence. In this depletion region, the charge density is minimal, and the device exhibits an intermediate dielectric and recombination response. This is evidenced by the significant drop in G/ω peak height (Figure 3a) and a sharp increase in peak frequency (Figure 3b), as the system moves from slow minority-dominated recombination to fast majority-carrier relaxation. This sharp frequency transition arises due to the rapidly changing charge distribution and electric field profile in the channel, which affect both trap occupation and response time.

Figure 4 provides further support by illustrating how capacitance and conductance vary with frequency and gate voltage. It presents both the C-V and G/ω-V characteristics of the SiGe MIS structure, revealing important frequency-dependent behavior as a function of gate bias. These curves reflect the transition between inversion, depletion, and accumulation regimes, each dominated by distinct carrier dynamics.

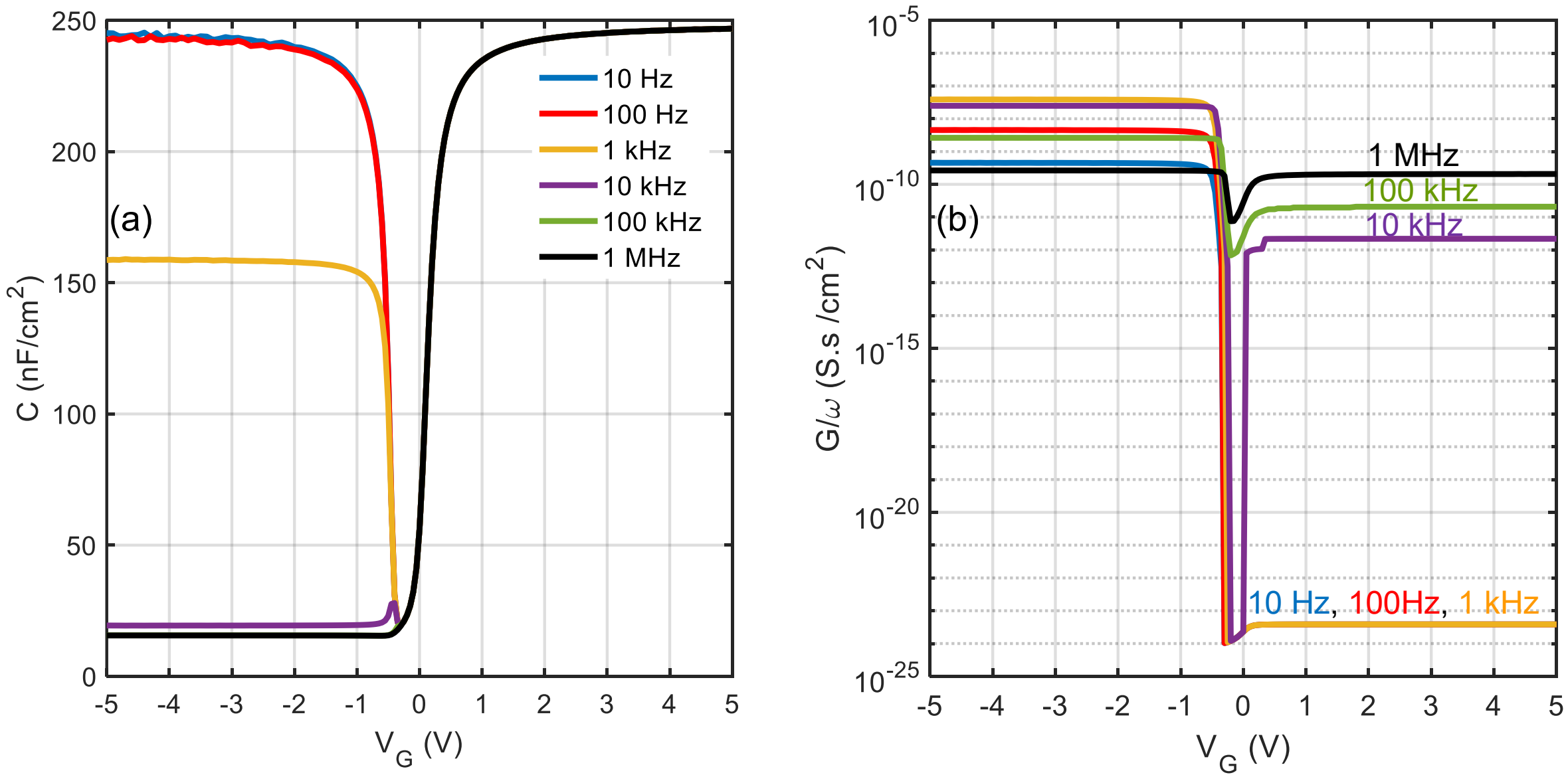

Figure 4: (a) Capacitance-voltage (C-V) and (b) conductance-voltage (G/ω-V) characteristics for an Au-$Al_2O_3$-SiGe MIS device with n-type doping of $10^{15}$ $cm^{-3}$ and a minority carrier lifetime of 10 ns. In the inversion regime, the capacitance decreases significantly as the frequency enters the kHz range due to the limited response of minority carriers. Within this frequency window, G/ω exhibits a peak corresponding to the trap-assisted generation-recombination dynamics. At low frequencies (10 Hz, 100 Hz, and 1 kHz), the simulated G/ω values reach a minimum that is constrained by the lower numerical computation limit of the solver.

In Figure 4a, the C-V curves demonstrate a pronounced frequency-dependent roll-off in the inversion region, particularly in the kHz range. This transition occurs when the AC signal becomes too fast for the minority carriers (holes, in the case of an n-type substrate) to follow. At low frequencies (e.g., 10-100 Hz), the minority carrier generation-recombination processes are able to track the applied signal, enabling the capacitance to remain relatively high. However, as the frequency increases, the minority carriers can no longer respond within a signal cycle, and the inversion-layer capacitance drops toward the depletion capacitance limit.

The corresponding G/ω plot in Figure 4b reflects this same transition. The conductance reaches a peak at the frequency where trap-assisted generation-recombination processes are most active. This peak frequency marks the characteristic emission time of minority carriers from trap states. The amplitude of the G/ω peak reflects both the density of traps and the strength of the generation-recombination pathway, making this an important diagnostic feature for trap characterization.

In contrast, in the accumulation regime, the AC response is dominated by the dielectric relaxation of the majority carrier population (electrons). The relevant time constant is given by:

$$\tau_D = \frac{\epsilon_r \epsilon_0}{q \mu_n N_d}$$

where $\varepsilon_r$ is the relative permittivity of the semiconductor, $\varepsilon_0$ is the vacuum permittivity, q is the elementary charge, $\mu_n$ is the electron mobility, and $N_D$ is the doping concentration. This dielectric relaxation time governs how quickly the mobile charge carriers can screen out an applied electric field. Higher doping concentrations result in shorter $\tau_D$, thereby shifting the system's frequency response to higher frequencies, as observed in Figure 5.

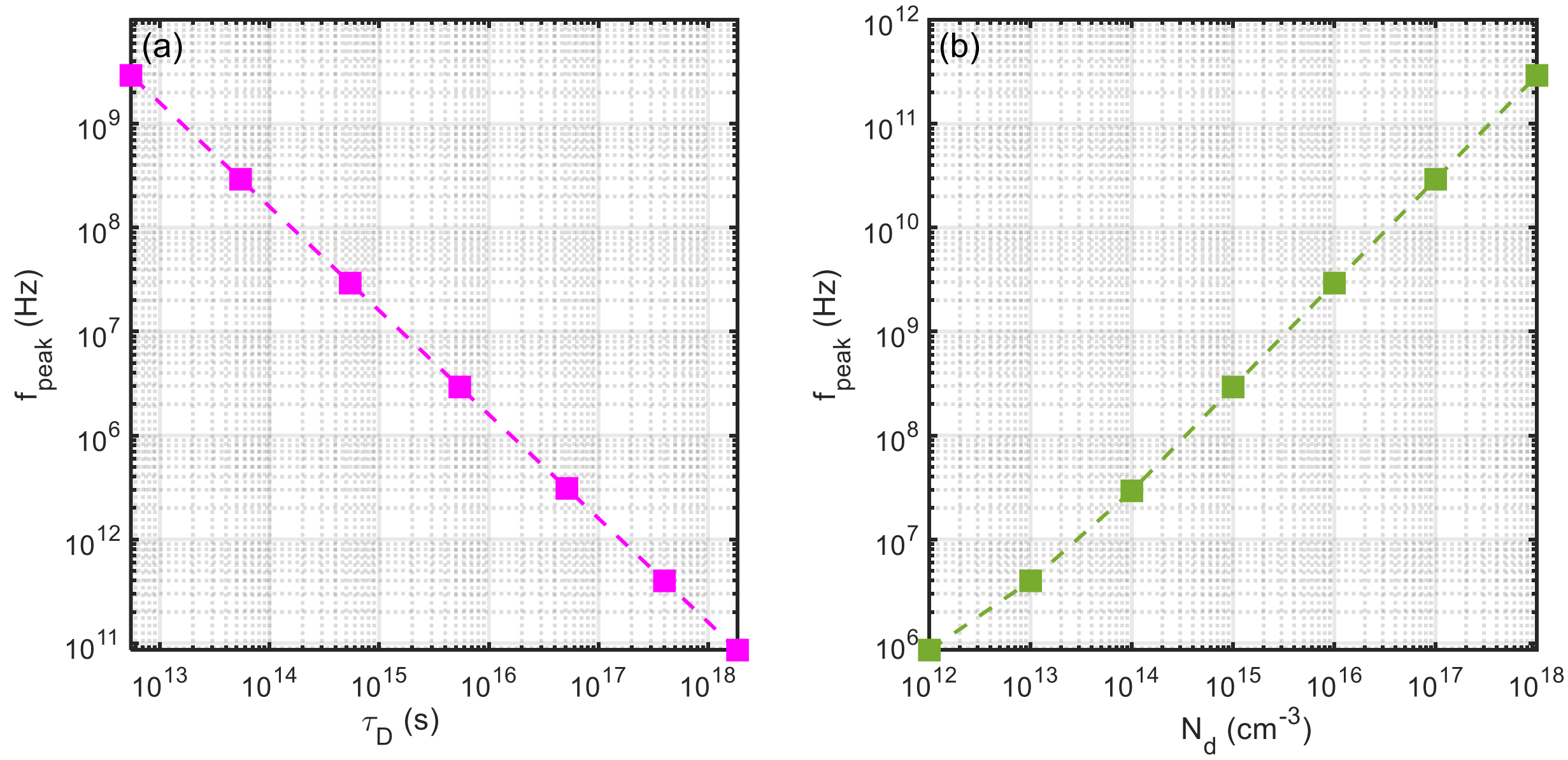

Figure 5: (a) Peak frequency, $f_{peak}$ versus carrier relaxation time $\tau_D$, and (b) $f_{peak}$ versus doping concentration $N_D$, for a gate bias $V_G$=3 (strong accumulation) in the absence of traps. The peak frequency in accumulation is strongly dependent on the doping level: as $N_D$ increases, the carrier relaxation time $\tau_D$ decreases due to enhanced conductivity, resulting in an increase in $f_{peak}$. The two subplots illustrate the inverse relationship between $f_{peak}$ and $\tau_D$, and the direct dependence of $f_{peak}$ on $N_D$, which together explain the opposing trends observed in (a) and (b).

The peak in G/ω therefore shifts to higher frequencies with increasing doping, consistent with the expected behavior from dielectric relaxation theory. Importantly, this shift occurs even in the absence of trap-related processes, demonstrating that both intrinsic carrier dynamics and extrinsic traps must be disentangled when analyzing experimental data.

These results reinforce the dual nature of frequency response in MIS structures: low-frequency behavior is trap- and recombination-limited, while high-frequency behavior is dictated by majority-carrier relaxation. The ability to resolve these regimes in simulation enables extraction of both minority carrier lifetimes and dielectric relaxation properties, making frequency-domain analysis a powerful tool for semiconductor device characterization.

Figure 6 illustrates how the frequency-dependent capacitance and G/ω response of the SiGe MIS device are modulated by the minority carrier lifetime $\tau_p$, a key parameter governing SRH recombination dynamics. In the inversion regime (e.g., at $V_G$=−5 V), the device's capacitive response is largely controlled by thermally generated minority carriers, since the channel is depleted of majority carriers. These minority carriers are generated and recombine via deep-level traps distributed within the depletion region. The ability of these carriers to respond to an AC signal depends critically on their emission and capture time constants, which are dictated by the SRH lifetime $\tau_p$.

As shown in Figure 6a, the frequency at which the capacitance begins to roll off, i.e., the transition frequency, shifts as a function of $\tau_p$. A longer minority carrier lifetime corresponds to a slower emission rate from traps, which delays the trap response to AC perturbations. Consequently, the capacitive response begins to decline at a lower frequency, since traps cannot follow the AC signal above that cutoff. In contrast, for shorter lifetimes, traps can rapidly capture and emit carriers, allowing them to contribute to the total capacitance even at higher frequencies.

This behavior is mirrored in Figure 6b, where the peak frequency of G/ω shifts accordingly. The conductance peak arises when the trap occupancy lags the AC signal by approximately 90 degrees, a condition that maximizes energy dissipation due to carrier exchange. The position of this peak provides a direct measure of the effective emission rate, which scales inversely with $\tau_p$:

$$f_{peak} \sim \frac{1}{2\pi\tau_p}$$

Thus, as the minority carrier lifetime increases, the peak in G/ω shifts to lower frequencies, reflecting the slower trap response. This is a key signature of trap-mediated recombination and is often used in both experimental and simulation-based DLTS analyses to extract lifetime values.

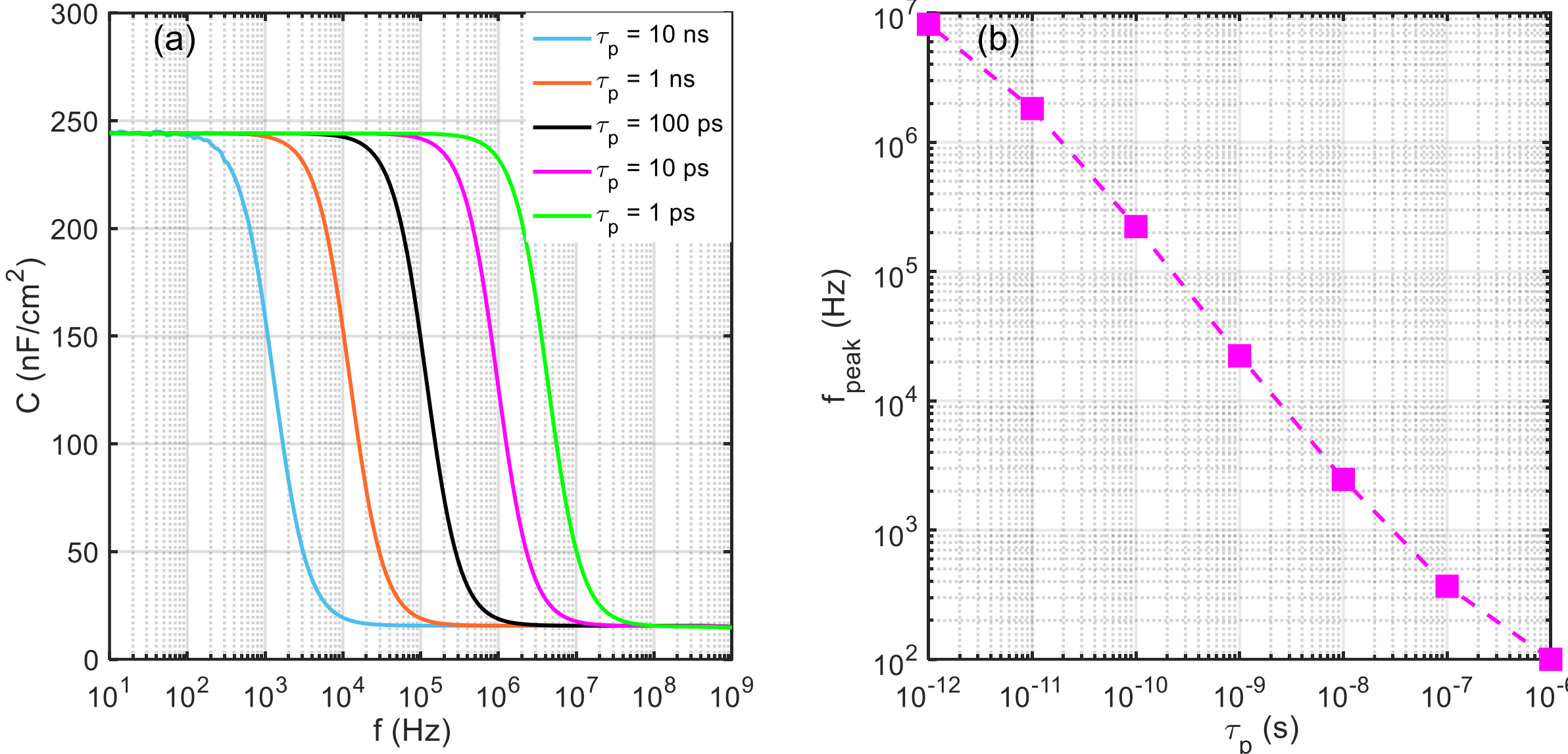

Figure 6: (a) Frequency dependence of capacitance under inversion bias ($V_G$=−5) showing the impact of bulk minority carrier lifetime $\tau_p$ on the transition frequency. Longer carrier lifetimes correspond to slower recombination rates, resulting in a delayed capacitive response and a shift of the roll-off frequency to lower values (as shown by the blue curve). (b) Peak frequency $f_m$ of the G/ω spectrum as a function of $\tau_p$, illustrating the inverse relationship between the minority carrier lifetime and the speed of the device's response.

These results demonstrate that minority carrier lifetime plays a dual role: it sets the transition frequency in the capacitance spectrum and determines the peak frequency of the trap-assisted conductance. By analyzing the position and magnitude of these features, one can extract both the dynamics of carrier trapping and the density of active traps, offering a comprehensive picture of defect behavior in the device.

### IV.B. Effect of Trap Parameters on AC Conductance Spectra

Figure 7 and Figure 8 examine how trap-related parameters, trap density $N_t$ , capture cross-section $\sigma_C$, trap energy level $E_t$, and trap distribution width $\sigma_E$, influence the frequency-dependent conductance spectrum G/ω, based on the SRH recombination framework with Gaussian energy-distributed traps.

As shown in Figure 7a, increasing the trap density $N_t$ results in a higher probability of carrier capture and emission events, thereby enhancing the overall recombination rate. This leads to a stronger conductance response, i.e., a higher G/ω peak. However, since a broader population of traps is introduced, each with its own time constant, the spectral response becomes more dispersed. The conductance peak not only becomes broader but also shifts to lower frequencies, reflecting the contribution of longer relaxation times associated with traps deeper in the bandgap or more weakly coupled to the channel.

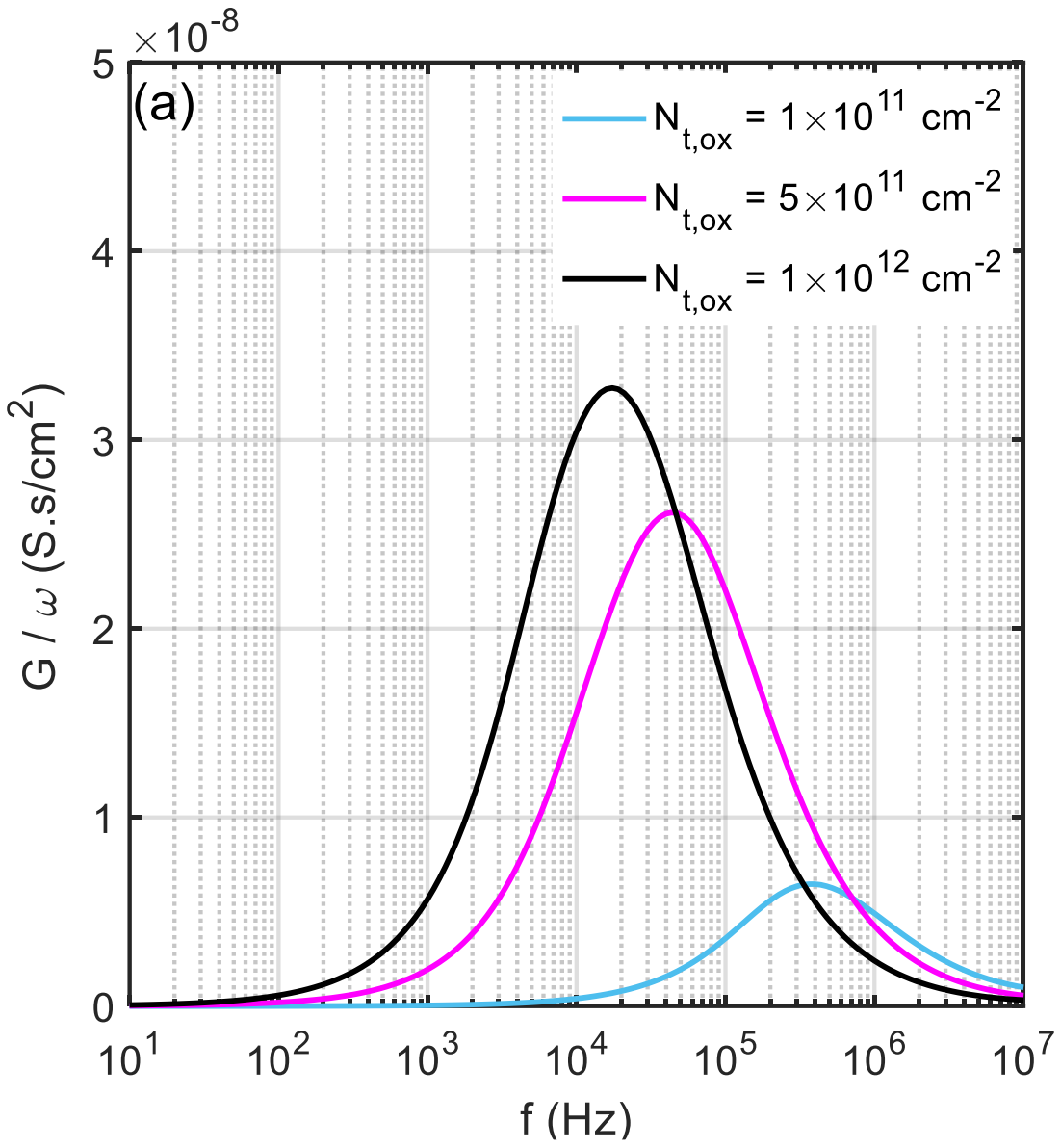

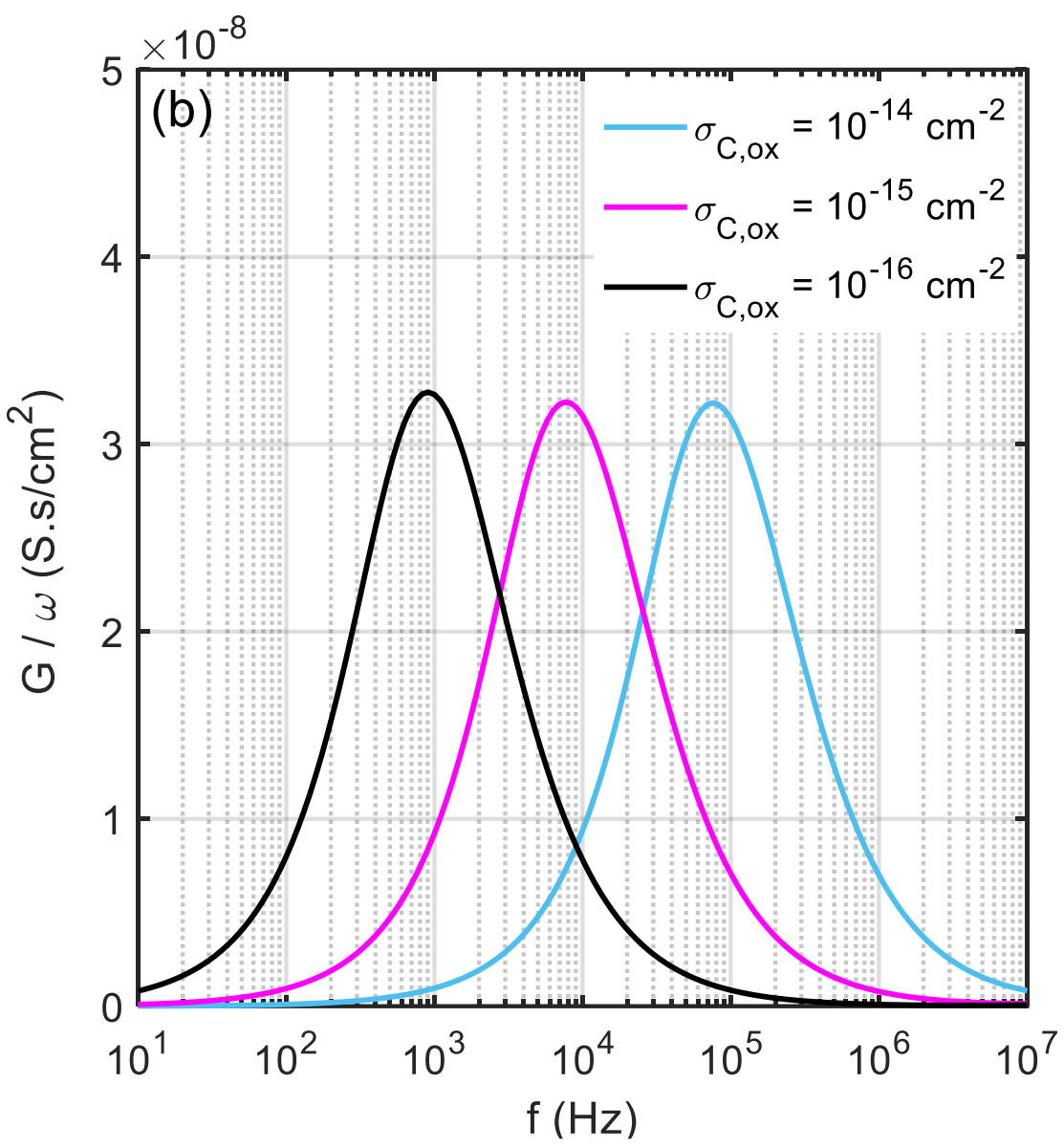

Figure 7: (a) Simulated G/ω versus frequency plots showing the effect of oxide trap density $N_{t,ox}$. An increase in trap density introduces a greater number of trap states with a distribution of relaxation times, including slower states, which shifts the peak frequency $f_m$ to lower values. (b) G/ω versus frequency at $V_G$=0, illustrating the effect of the capture cross-section $\sigma_{C,ox}$ . A larger capture cross-section enhances the trap-carrier interaction rate, allowing traps to respond more quickly and shifting the peak frequency to higher values.

The peak frequency, $f_{peak}$ in the G/ω spectrum corresponds to the inverse of the effective emission time constant of the trap population. For a Gaussian energy distribution of traps, this effective time constant reflects an energy-averaged relaxation process. As $N_t$ increases, traps with longer $\tau$ (slower response) begin to dominate, thereby lowering $f_{peak}$ and broadening the response.

In contrast, Figure 7b shows the effect of increasing the capture cross-section $\sigma_C$ . A larger cross-section implies that carriers are more likely to be captured by a given trap in a unit time, which translates to a shorter emission/capture time constant:

$$\tau \propto \frac{1}{\sigma_C v_{th}}$$

where $v_{th}$ is the thermal velocity. As a result, traps with larger $\sigma_C$ can respond more quickly to high-frequency AC signals, and the conductance peak shifts to higher frequencies. This allows for differentiation between fast and slow traps based not just on their density but also on their cross-sectional kinetics.

Figure 8a and Figure 8b further explore how the trap energy level $E_t$ and the energy width $\sigma_E$ of the Gaussian trap distribution impact the conductance behavior. In Figure 8a, moving the trap energy further from the valence band edge (toward midgap or closer to the conduction band) increases the recombination efficiency for n-type material. These midgap-like traps are more likely to be occupied by electrons and available to capture holes, leading to higher recombination rates and a more pronounced response. Because these traps are energetically positioned to emit carriers more easily, they also exhibit faster response times, shifting the conductance peak to higher frequencies.

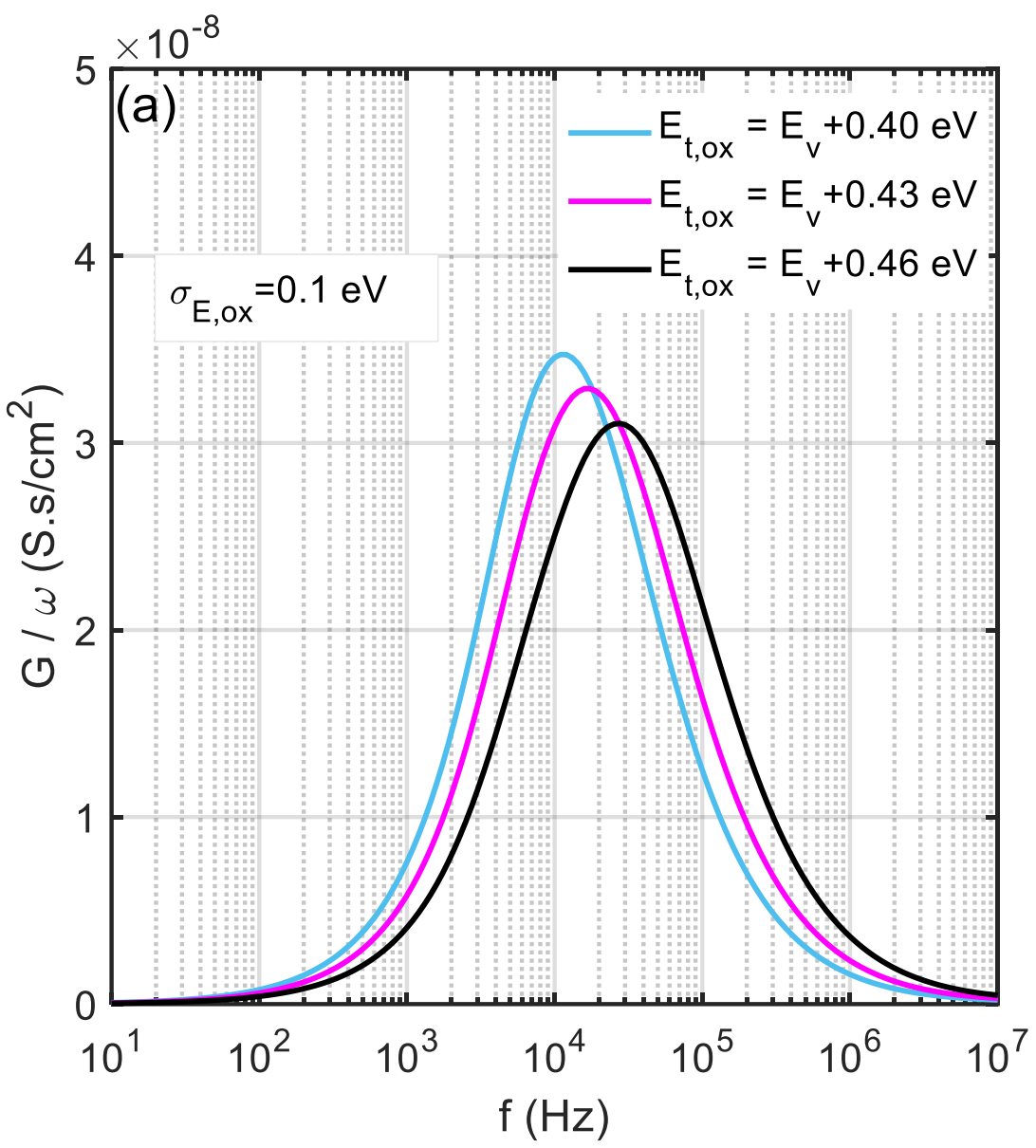

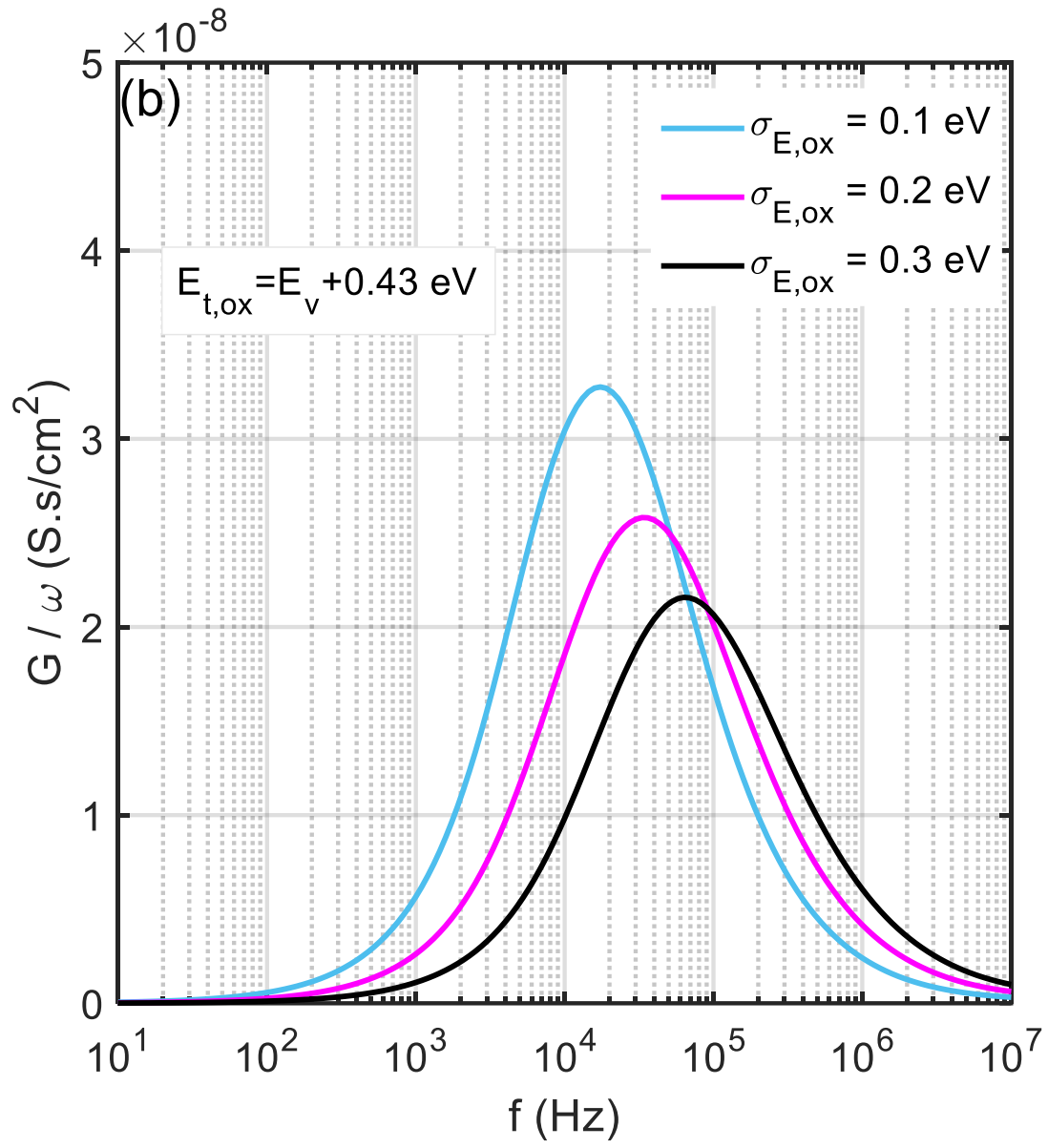

Figure 8: (a) Simulated G/ω versus frequency plots showing the effect of oxide trap energy level $E_{t,ox}$ . Traps located farther from the valence band edge (i.e., closer to the conduction band for an n-type substrate) have increased availability of carriers for recombination, enabling faster emission rates and shifting the peak frequency to higher values. (b) G/ω versus frequency at $V_G$=0, showing the influence of the Gaussian trap energy distribution width $\sigma_{E,ox}$ . A broader energy distribution increases the spread of trap states away from midgap, reducing the overall recombination rate through midgap traps and resulting in a net shift of the conductance peak toward higher frequencies.

In Figure 8b, increasing the width $\sigma_E$ of the Gaussian energy distribution spreads the trap states over a broader energy range. While this might seem to increase the number of active traps, it actually leads to a reduction in overall recombination efficiency, because fewer traps are centered near the optimal energy (typically midgap for maximum recombination). As a result, the peak amplitude of G/ω decreases. Furthermore, traps that are energetically farther from midgap tend to have shorter emission times, contributing more strongly at higher frequencies, which explains the rightward shift in the conductance peak.

Importantly, the shape of the G/ω spectrum, its width, peak height, and location, contains embedded information about the trap kinetics and distribution. Slower traps (i.e., those with longer $\tau$) exhibit stronger interaction with the AC signal near their resonance frequency, maximizing energy dissipation and giving rise to larger peaks in G/ω. This property is particularly useful in experimental analysis, where the frequency location and curvature of the conductance peak can be used to extract both the density and kinetic profile of traps.

Together, these results highlight the power of frequency-domain simulation in distinguishing between trap populations based on their physical and chemical characteristics. The interplay between density, energy level, capture cross-section, and energy dispersion governs not only the magnitude of trap-induced conductance but also the timescale over which different traps interact with the applied signal. These insights form the foundation for targeted trap engineering in quantum devices, where minimizing specific types of traps is essential to enhancing coherence and stability.

### IV.C. Impedance Response of Ge/SiGe Quantum Well Devices

To understand how different trap locations affect the AC electrical behavior of a Ge-based QW device, we simulate and analyze the frequency-dependent conductance of a gate-stacked structure consisting of $Si_{0.2}Ge_{0.8}$ barriers enclosing a Ge QW. This structure is representative of architectures used for gate-defined hole spin qubits and is sensitive to electrically active defects at multiple spatial locations.

Figure 9 presents a comparison of the normalized conductance G/ω versus frequency for devices with traps located either at the oxide/SiGe interface or at the SiGe/Ge QW interfaces. As expected from SRH recombination theory, a higher trap density in either case results in increased recombination and thus a larger conductance response. However, the detectability of QW interface traps is more limited in practice: the simulation shows that a minimum trap density of approximately $10^8$ cm$^{-2}$ is needed for a discernible peak to emerge above the computational background noise. Below this threshold, the trap contribution is too small to significantly modulate the impedance, especially under room-temperature conditions.

In contrast, oxide interface traps generally occur at much higher densities in fabricated devices, typically in the $10^9$ to $10^{11}$ cm$^{-2}$ range, and are located directly under the gate dielectric. Their proximity to the gate electrode enhances their capacitive coupling and makes their contribution to the conductance both stronger and faster. As a result, oxide traps produce well-resolved, high-amplitude peaks in the G/ω spectrum at moderate frequencies.

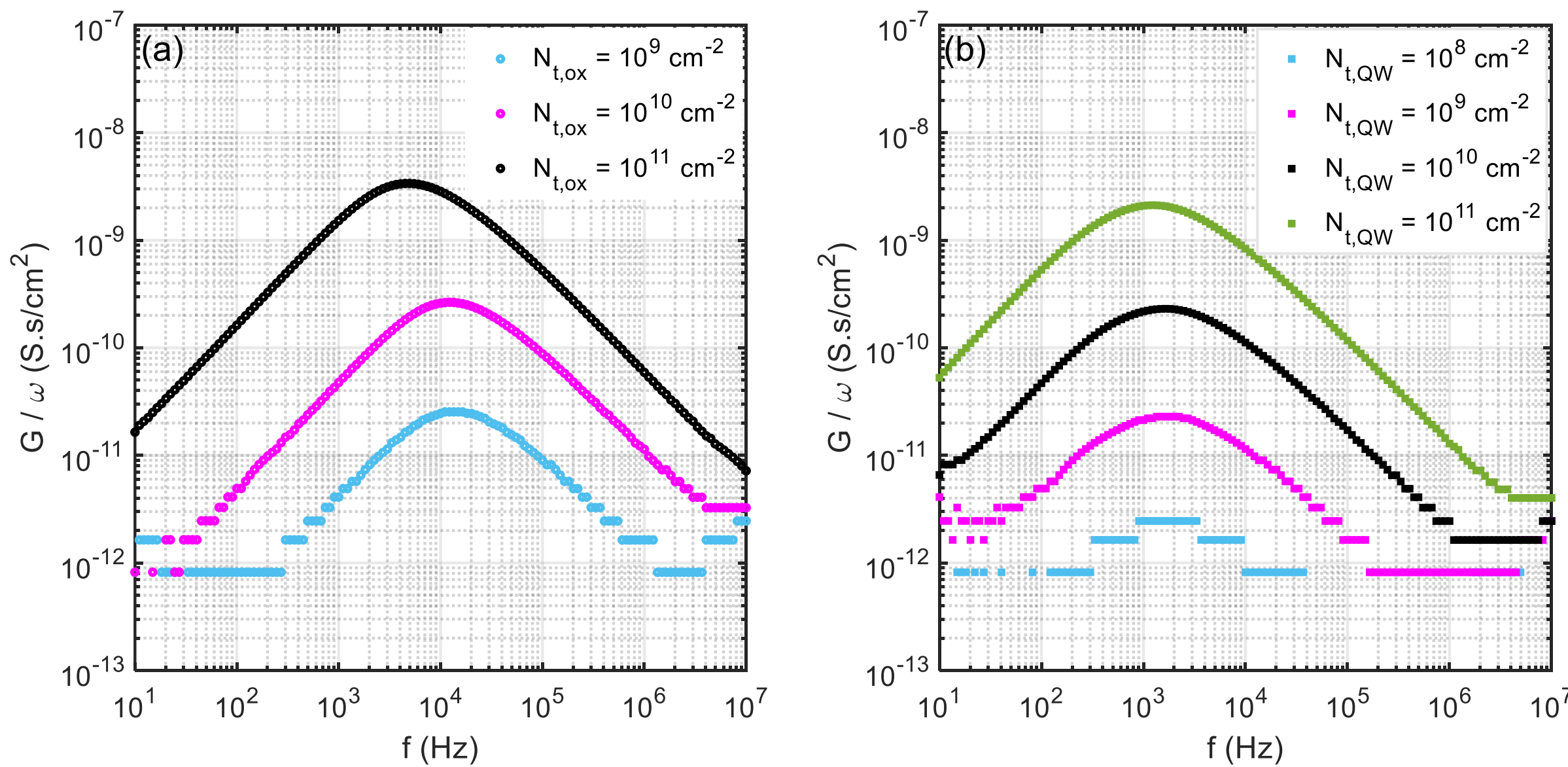

Figure 9 *:* (a) Simulated G/ω versus frequency plots illustrating the effect of oxide interface trap density $N_{t,ox}$. As the oxide trap density increases, the associated trap-assisted conductance becomes more pronounced, resulting in a stronger and more prominent peak in the frequency response. (b) G/ω versus frequency plots at $V_G$=0 showing the effect of QW interface trap density

$N_{t,QW}$. A higher-quality QW interface, with lower trap density, exhibits a weaker or negligible peak in the G/ω spectrum, indicating minimal trap-induced response.

Figure 10b provides a direct comparison of the oxide interface, QW interface, and bulk trap contributions under identical simulation conditions. The bulk traps, modeled with a volumetric density of $10^{15}$ $cm^{-3}$, contribute at higher frequencies due to their typically faster emission rates and spatial distribution deeper into the semiconductor. This creates a shoulder or secondary peak in the conductance spectrum, distinguishing them from interface-related features.

Importantly, the quality of the epitaxial SiGe/Ge interface affects both the density and spectral signature of the QW interface traps. High-quality interfaces, as expected in carefully grown heterostructures, result in lower trap densities, often below the detectability threshold of conductance-based methods. In these cases, the QW interface traps produce a broad, low-amplitude tail in the low-frequency range of the G/ω spectrum. In contrast, degraded or non-ideal interfaces, such as those resulting from lattice mismatch, misfit dislocations, or contamination, can yield trap densities sufficient to introduce detectable features.

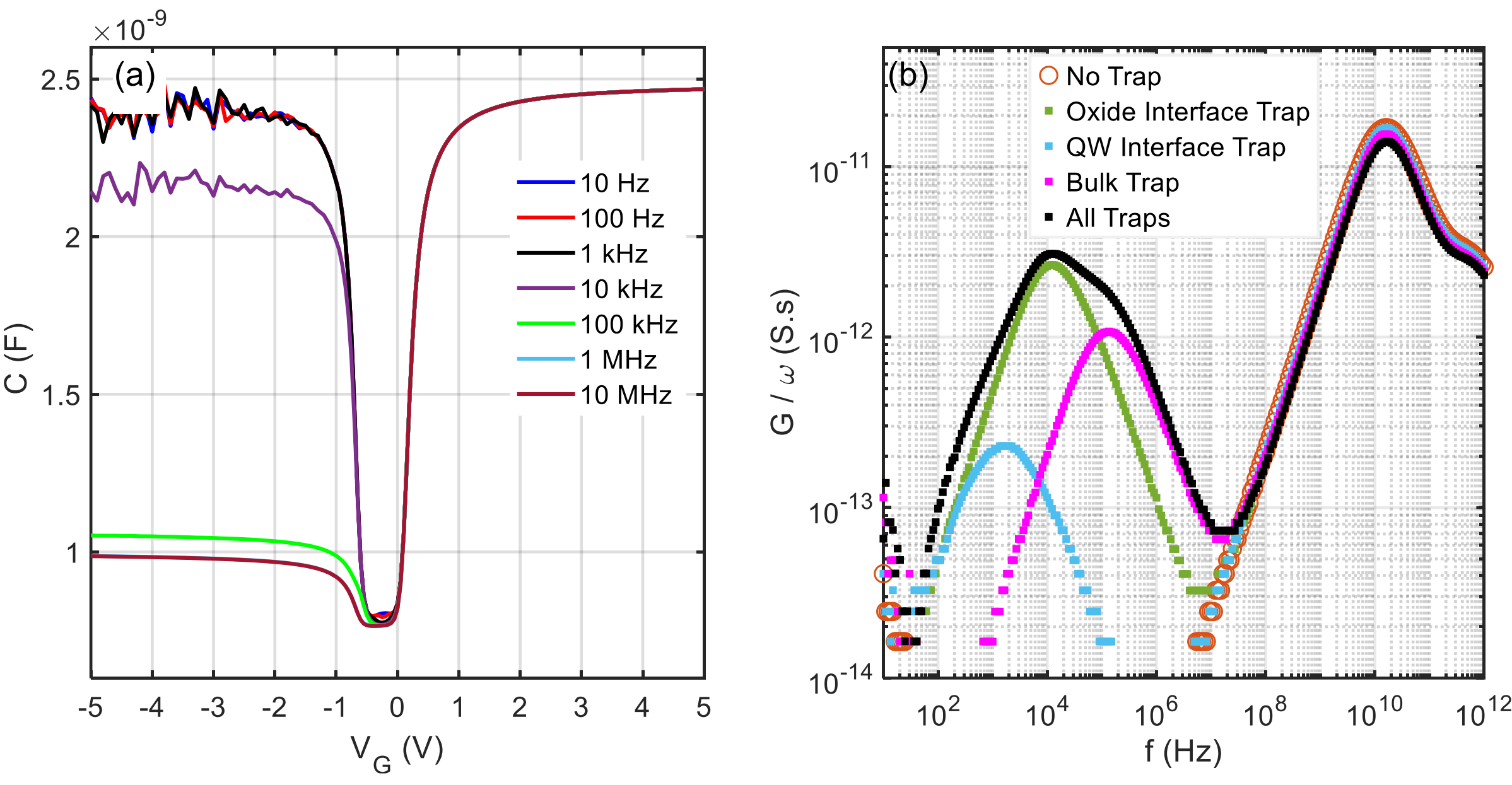

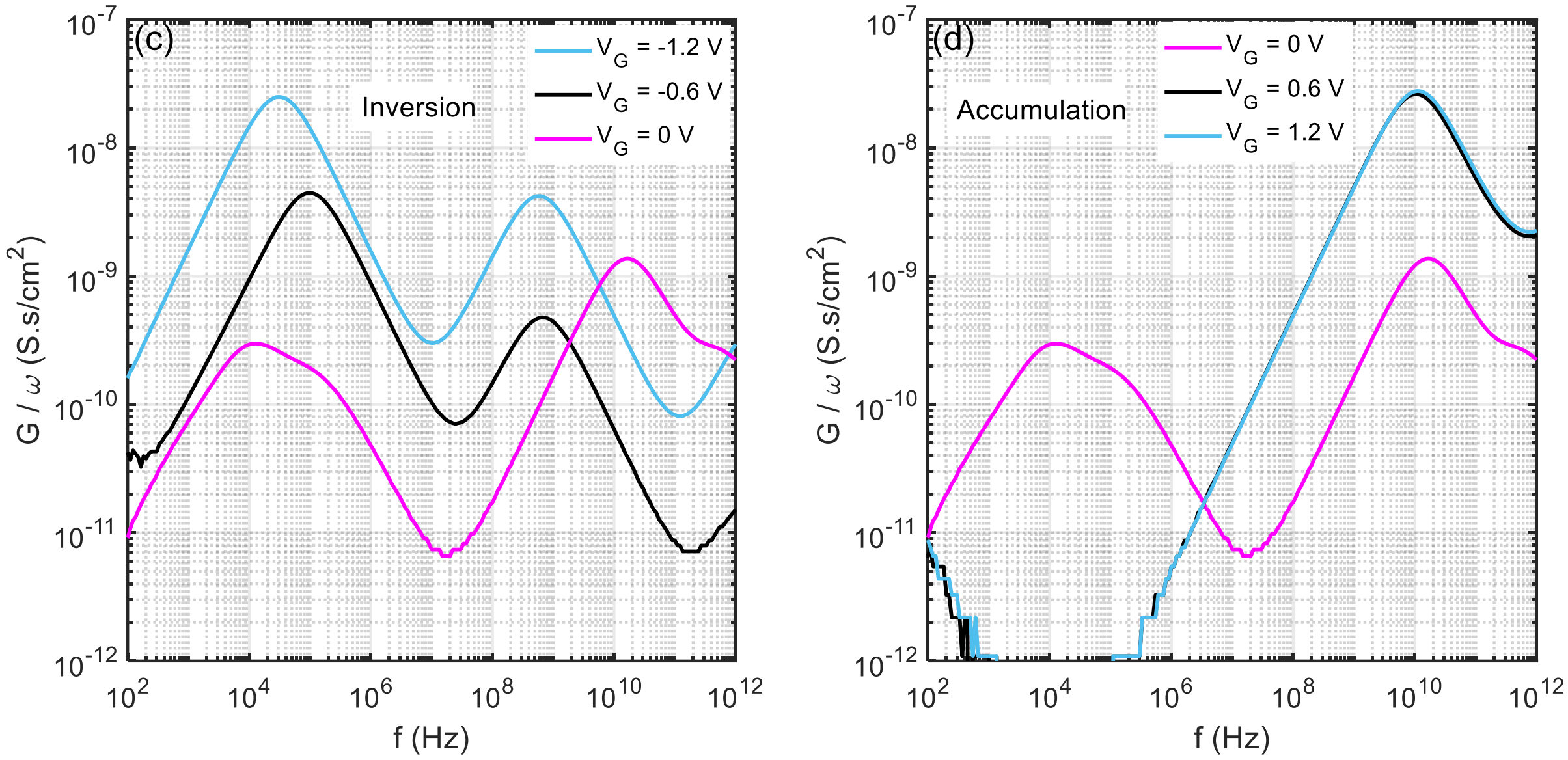

Figure 10: (a) Simulated capacitance-voltage (C-V) characteristics of a Ge/SiGe QW device including contributions from oxide interface, QW interface, and bulk traps. (b) G/ω versus frequency plots at $V_G$=0, showing the individual and combined effects of traps located at the oxide interface ($N_{t,ox}$=$10^{10}$ cm$^{-2}$, QW interface ($N_{t,QW}$=$10^9$ cm$^{-2}$), and in the bulk ($N_{t,bulk}$=$10^{15}$ cm$^{-3}$). Among the three, the oxide interface traps dominate the conductance response, producing a strong peak. QW interface traps contribute by broadening the low-frequency tail of the spectrum, while bulk traps introduce a high-frequency shoulder, reflecting their faster emission dynamics and distributed nature. Simulation without traps results in G/ω values below the computational limit. (c) and (d) show the peaks at various gate voltage, $V_G$ in inversion and accumulation respectively.

Figure 10a shows the C-V behavior of the QW device under low-frequency conditions (up to a few kHz). In the inversion regime, the capacitance is limited by the minority carrier response time, which is governed by the thermal generation and recombination kinetics of holes in the Ge quantum well. These dynamics are particularly slow when the depletion region is large and the hole population is sparse, leading to the suppression of inversion capacitance at higher frequencies.

Among the three types of traps considered, QW interface traps exhibit the slowest response, due to both their lower density and their location at a buried interface, which reduces their capacitive coupling to the gate. In contrast, oxide interface traps are highly responsive due to their surface proximity and high density, while bulk traps, though numerous, are less strongly coupled and contribute mainly at higher frequencies where emission rates are faster.

These results collectively demonstrate that impedance spectroscopy is a powerful tool for isolating and identifying the spatial origin and kinetic behavior of different trap types. By examining the shape, location, and amplitude of conductance peaks across bias and frequency ranges, one can deconvolve the contributions from oxide interfaces, buried QW interfaces, and the bulk, which is essential for improving the reliability and coherence properties of qubit devices based on Ge/SiGe platforms.

### IV.D. Impact of Gate Dielectric Choice on AC and DC Characteristics

To evaluate the impact of the gate dielectric material on the AC response of Ge/SiGe QW devices, we compare two dielectric stacks: $Al_2O_3$ and $SiO_2$. While $Al_2O_3$ is a high-κ dielectric with $\kappa \approx 8.5$, $SiO_2$ has a significantly lower dielectric constant ($\kappa \approx 3.9$). This difference directly influences the gate capacitance, electrostatic control, and ultimately, the device's ability to detect and resolve trap-related conductance signals.

Figure 11a shows the C-V curves for both gate dielectrics. As expected, the saturation capacitance in accumulation is lower for $SiO_2$, reflecting its smaller dielectric constant and reduced physical gate capacitance per unit area. This leads to weaker electrostatic coupling between the gate and the channel, which diminishes the modulation of channel potential and the gate's ability to sense subtle changes in trap occupancy. The reduced coupling also limits the maximum achievable capacitance and suppresses the frequency response of the channel, particularly for deeply buried traps such as those at the QW interfaces or in the bulk.

Figure 11b presents the corresponding $G/\omega$ versus frequency characteristics. The overall conductance response is also lower for the $SiO_2$ device, particularly in the depletion and inversion regimes. Since $G/\omega$ reflects the energy dissipated due to trap-assisted carrier exchange, its magnitude is strongly influenced by how effectively the gate potential modulates the trap occupancy. With $SiO_2$, this modulation is weaker, resulting in lower conductance peaks and more subtle spectral features, especially for lower-density traps.

To quantify these effects further, Figure 12 shows the extracted peak conductance height, peak frequency $f_m$, and the phase angle at the conductance peak, as functions of gate voltage for various trap configurations. In both the strong inversion (large negative gate bias) and strong accumulation (large positive gate bias) regimes, the extracted parameters are largely similar for $Al_2O_3$ and $SiO_2$. This occurs because, in these regions, the channel is either strongly depleted (inversion) or heavily populated (accumulation), and the trap-induced conductance is dominated by carrier availability and intrinsic kinetics rather than gate modulation strength.

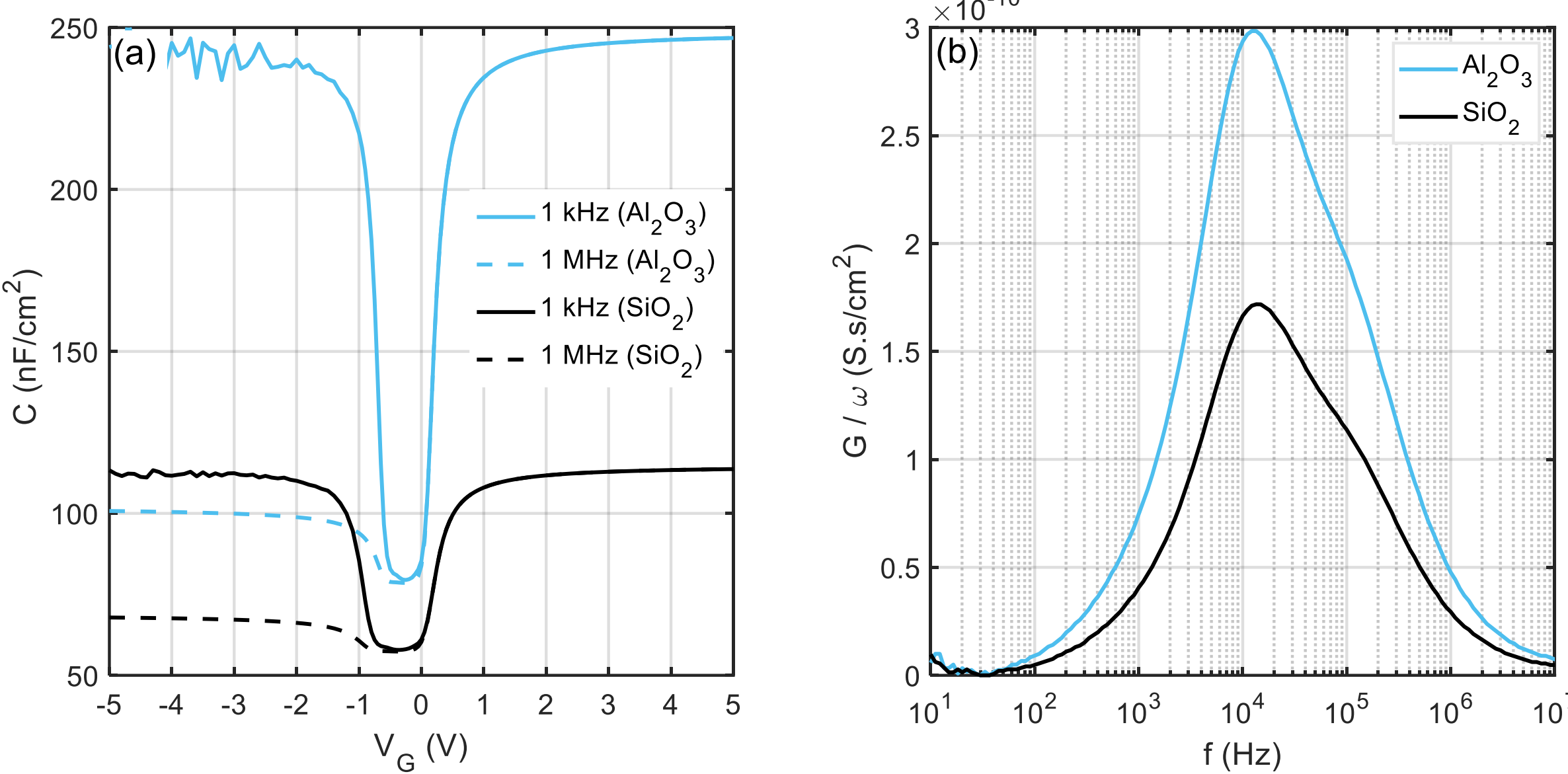

Figure 11: (a) C-V and (b) G/ω versus frequency characteristics comparing $Al_2O_3$ and $SiO_2$ as gate dielectrics in a Ge/SiGe QW device. Due to its lower dielectric constant (κ=3.9), $SiO_2$ provides weaker capacitive coupling between the gate and the channel compared to $Al_2O_3$ (κ=8.5). As a result, both the maximum capacitance and the overall conductance response are reduced for the $SiO_2$ gate stack. These differences in dielectric properties affect the gate control and signal amplitude, and must be considered when optimizing oxide materials for quantum device integration.

However, significant differences appear in the depletion region, where the channel is most sensitive to gate-induced changes and traps have the strongest capacitive coupling to the gate. In Figure 12b, the peak frequency $f_m$ diverges noticeably between the two dielectric materials for all trap types. This indicates that the frequency response is more sensitive to the gate dielectric in the depletion regime, where the field penetration depth and charge balance are finely modulated by the dielectric constant. The lower κ of $SiO_2$ results in reduced modulation efficiency, thereby altering the effective trap charging dynamics and the rate at which the AC signal perturbs the trap occupancy.

In contrast, the peak height and phase (Figure 12a and Figure 12c, respectively) show only minor variations between the two dielectrics in the depletion region. This suggests that, while the timing (i.e., frequency) of the trap response is affected by the gate coupling, the overall recombination or emission activity, as reflected in the integrated conductance magnitude, is less sensitive to the dielectric once a trap is active.

Overall, this comparison underscores a key design consideration: the choice of gate dielectric not only affects electrostatics and threshold voltage, but also has a measurable impact on the sensitivity and resolution of trap detection in frequency-domain analysis. While $Al_2O_3$ provides stronger gate control and better detectability of interface and bulk traps, $SiO_2$ may lead to underestimation or loss of spectral resolution, particularly for traps in buried interfaces or low-density regimes. This has direct implications for the selection of gate dielectrics in qubit devices, where accurate characterization of trap states is essential for understanding decoherence and charge noise.

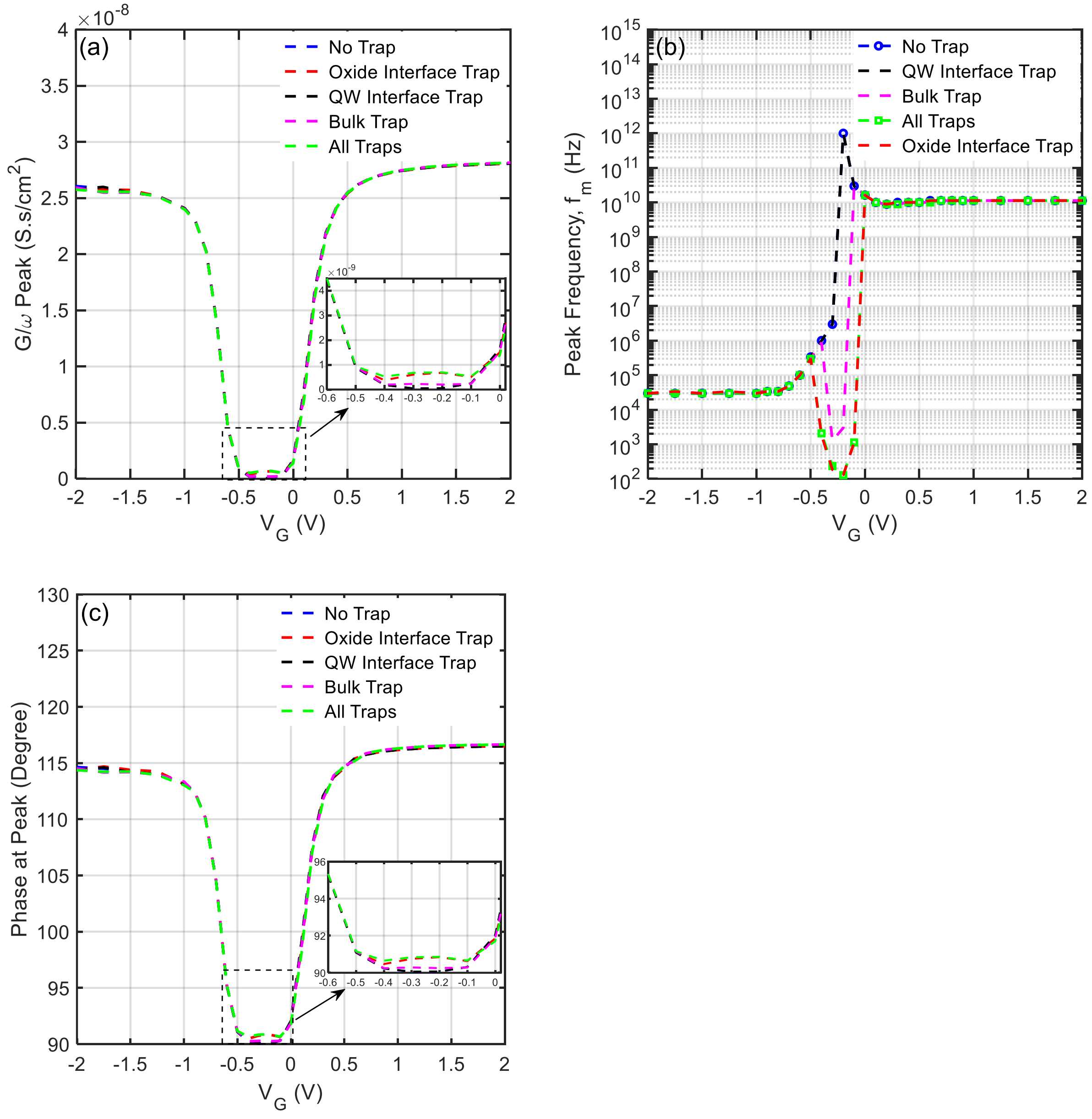

Figure 12: (a) Extracted G/ω peak height, (b) peak frequency $f_m$ , and (c) phase at the peak frequency as functions of gate voltage $V_G$ an Au-$Al_2O_3$-SiGe-Ge-SiGe quantum well device. Simulations compare five cases: no traps, oxide interface traps ($N_{t,ox}$=$10^{10}$ $cm^{-2}$), QW interface traps ($N_{t,QW}$=$10^{8}$ $cm^{-2}$), bulk traps ($N_{t,bulk}$=$10^{15}$ $cm^{-3}$), and all three traps combined. In the inversion regime, the device exhibits a slower response, characterized by a lower peak frequency, reduced conductance peak height, and smaller phase angle, due to the limited availability of minority carriers. Near the depletion region, both the peak height and phase drop sharply, while the peak frequency increases significantly, reflecting a transition in carrier dynamics and charge storage mechanisms.

### IV.E. Nyquist Analysis of Trap Contributions Across Gate Bias

In addition to frequency- and voltage-domain conductance plots, Nyquist plots offer a powerful complementary view for visualizing and interpreting the complex impedance behavior of semiconductor devices. Unlike G/ω spectra, which decompose the frequency-dependent real part of the admittance, Nyquist plots display the full complex impedance in the complex plane, plotting the imaginary component $\Im(Z)$ versus the real component $\Re(Z)$ as frequency varies. This approach emphasizes the dynamic interactions between capacitive storage and resistive dissipation mechanisms and provides an intuitive representation of trap-induced relaxation processes.

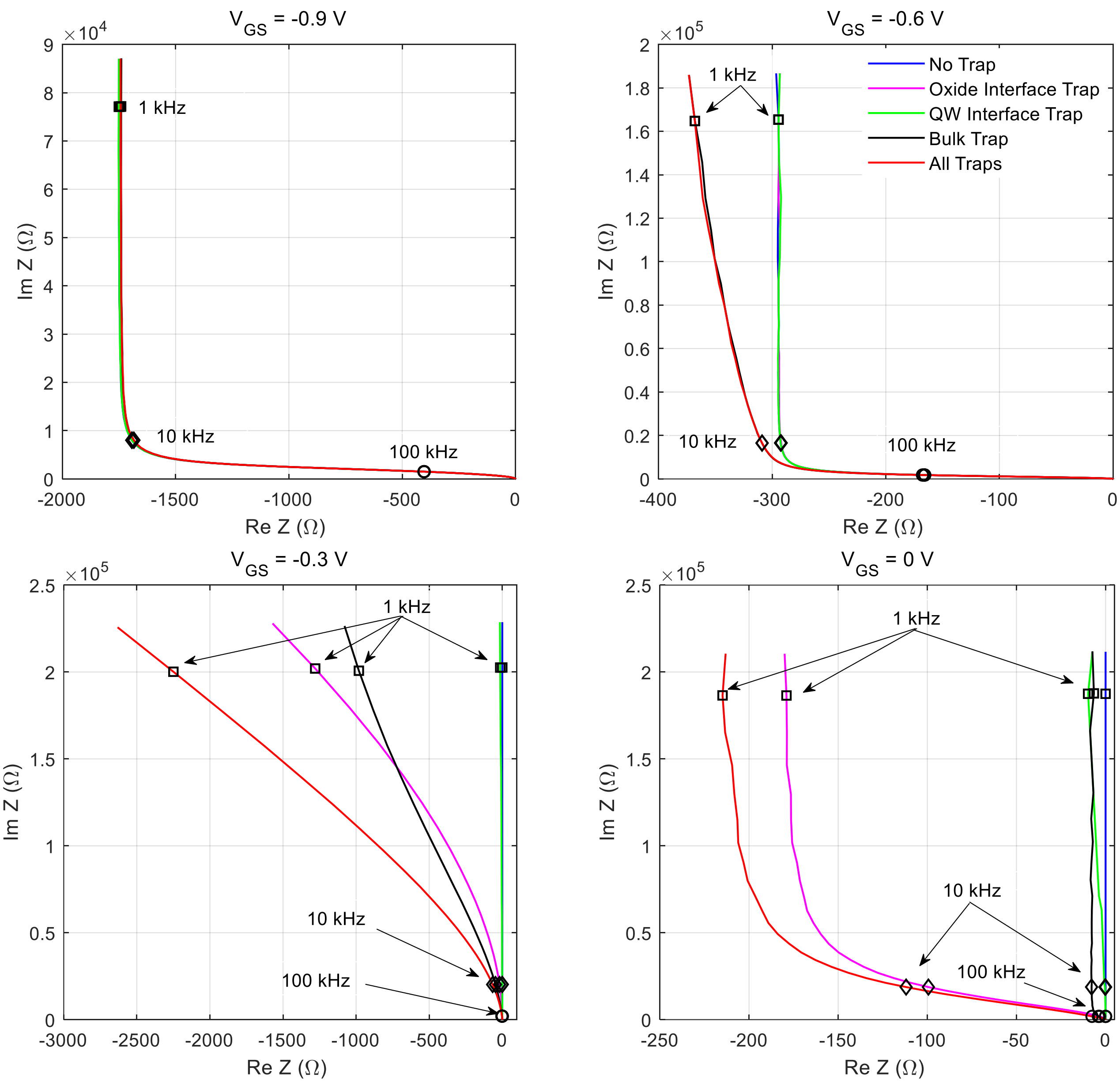

Figure 13: Simulated Nyquist plots (imaginary vs. real part of impedance) for the Ge/SiGe QW device under five trap configurations: no traps, oxide interface traps, QW interface traps, bulk traps, and all traps combined, evaluated at various gate voltages. Each plot traces the complex impedance response over a frequency sweep. Markers denote specific frequencies: circles for 1 kHz, diamonds for 10 kHz, and squares for 100 kHz. The shape and size of the Nyquist arc reflect the trap-induced relaxation dynamics, larger semicircles at lower frequencies indicate slower processes such as minority carrier recombination or deep trap response.

Figure 13 presents simulated Nyquist plots for the Ge/SiGe QW device, showing the influence of different trap types, oxide interface traps, QW interface traps, and bulk traps, as the gate bias is swept from depletion into strong inversion. Simulations include five cases: no traps, each trap type individually, and all traps combined. Markers at 1 kHz (circle), 10 kHz (diamond), and 100 kHz (square) help identify the frequency position along each curve.

Each semicircular arc in the Nyquist plot corresponds to a distinct RC time constant, typically arising from trap-related processes. The diameter of the semicircle is proportional to the magnitude of energy dissipation due to carrier trapping and emission, while the frequency position (from low frequency at the rightmost tip to high frequency at the leftmost tip) indicates the characteristic emission or relaxation time. A larger arc at lower frequencies is associated with slow trap dynamics, typically minority-carrier-related or arising from deeper traps; smaller or higher-frequency arcs indicate faster responses, such as dielectric relaxation or bulk trap activity.

At moderate negative gate voltages (depletion regime), the Nyquist plot reveals that oxide interface traps dominate the impedance behavior. Their high density and strong capacitive coupling to the gate lead to a well-defined semicircle with significant radius at low to mid frequencies. These traps are slow enough to strongly interact with the AC signal in the kHz range, producing pronounced arcs.

As the gate bias becomes more negative and the device enters strong inversion, the trap occupancy dynamics change. The depletion region widens, reducing the coupling of the oxide traps to the channel. Simultaneously, bulk traps, located deeper in the semiconductor and distributed throughout the Ge/SiGe heterostructure, begin to contribute more significantly. Their emission rates are generally faster and their spatial location makes them more sensitive under high field conditions. This leads to an additional semicircle or shoulder in the Nyquist plot at higher frequencies, indicating the presence of overlapping trap processes.

Interestingly, in the strong inversion regime, the Nyquist curves corresponding to different trap types begin to converge and overlap. This convergence is attributed to two effects: (1) the oxide interface traps become less active as the field across the oxide weakens and the surface is inverted, and (2) the inversion-layer minority carrier concentration saturates, reducing the sensitivity of the AC response to the exact location of trap states. As a result, all trap types contribute similarly to the overall impedance, and the system becomes kinetically limited by minority carrier generation, rather than spatial trap distribution.

Moreover, the shape and symmetry of the arcs provide qualitative information about the distribution of trap time constants. For example, a distorted or asymmetric arc indicates a non-uniform distribution of trap energies or spatial locations (as is often the case for bulk traps or QW interface traps with varying energy widths). In contrast, a symmetrical semicircle implies a relatively narrow and well-defined relaxation process, as seen with high-density oxide traps.

These results highlight several important advantages of Nyquist analysis:

- It allows visual differentiation of overlapping trap contributions across frequency ranges.
- It can reveal the dominant kinetic process in a given bias regime.
- It is highly sensitive to the presence of multiple trap species, each with distinct time constants and coupling strengths.

In experimental settings, fitting Nyquist plots with equivalent circuit models (e.g., parallel RC elements or distributed constant phase elements) can help extract quantitative estimates of trap densities, emission times, and capacitance contributions. In the simulation context, these plots serve as a powerful visual diagnostic of the spatial and energetic landscape of traps and complement frequency-domain and time-domain analyses.

In summary, the Nyquist analysis shown in Figure 13 confirms and reinforces the key conclusions from prior sections: oxide interface traps dominate at low to moderate gate biases, bulk traps emerge as the dominant mechanism in strong inversion, and QW interface traps contribute more subtly, primarily in intermediate regimes or when trap densities are elevated. The convergence of all cases in strong inversion emphasizes the unifying role of minority carrier dynamics in limiting device response at large reverse bias.

### IV.F. Time-Domain Signatures of Individual Traps in QW Devices

While frequency-domain techniques such as G/ω spectroscopy are effective for detecting oxide and bulk traps at moderate to high densities, they often lack the sensitivity needed to detect low-density trap states, particularly those located at buried quantum well interfaces, such as the epitaxial Ge/SiGe interface. In practice, these interfaces can have trap densities as low as $10^6$ $cm^{-2}$, which lie well below the detection threshold of AC conductance analysis due to weak capacitive coupling and insufficient recombination amplitude. However, time-domain techniques, especially those analogous to DLTS, can offer enhanced sensitivity by resolving small current transients induced by gate pulses.

Figure 14 demonstrates the time-resolved current response of the Ge/SiGe QW device subjected to a microsecond-scale gate voltage pulse of -0.5 V, placing the device into the depletion regime. Upon removal of the gate pulse and return to the equilibrium bias (e.g., $V_G$ = 0 V), the device exhibits a transient current decay as carriers are emitted from previously filled traps. This relaxation process reflects the emission time constants of different trap species. The shape and rate of the decay are strongly dependent on the spatial location, density, and emission kinetics of the traps, making the transient response a sensitive probe of trap properties.

The trailing (positive-going) edge of the gate pulse, corresponding to the return from VG = –0.5 V back to 0 V, induces a sudden change in the electrostatic potential across the device, initiating a transient period dominated by the emission of carriers from previously filled traps. As shown in Figure 14b, this results in a composite exponential decay that reflects the superposition of multiple relaxation processes associated with different trap types and spatial locations.

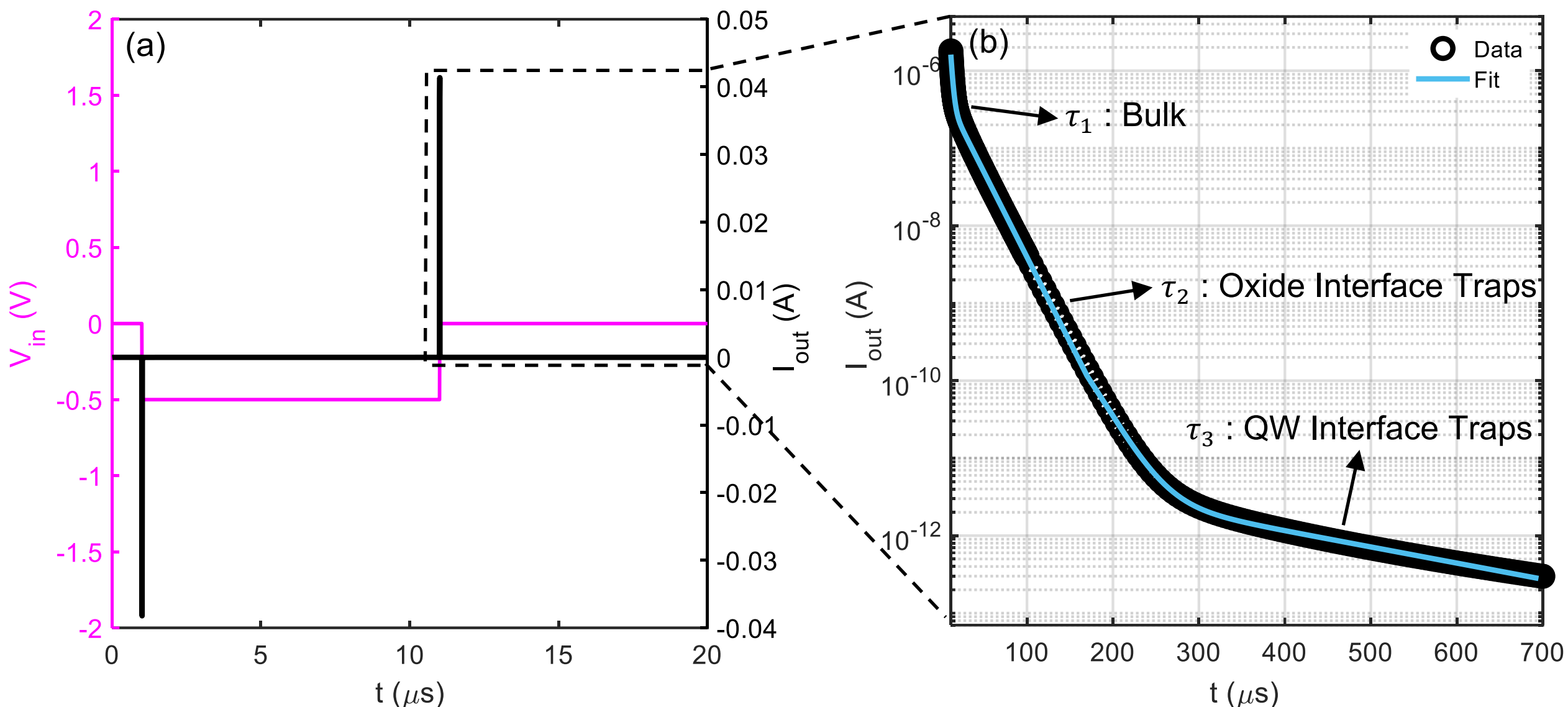

Figure 14: (a) Time-domain current response (black) of the Ge/SiGe QW device to a gate voltage pulse of -0.5 V (magenta), showing a transient followed by exponential decay. (b) Zoom-in of the current decay following the positive edge of the pulse, overlaid with multi-exponential fitting. The fit reveals three distinct decay time constants: $\tau_1$, $\tau_2$ and $\tau_3$, associated with bulk traps, oxide interface traps, and QW interface traps, respectively. $\tau_1$ corresponds to the fastest initial decay, while $\tau_3$ represents the slowest component that defines the long-term settling behavior of the transient current. Area of the QW device for all DLTS simulations is 1 mm$^2$.

### IV.G. Sensitivity of Time-Resolved Response to Trap Density and Gate Bias

To further assess the sensitivity of time-domain techniques in differentiating trap types in Ge/SiGe QW devices, we study how the gate voltage and the density of various traps influence the transient current response. As shown in Figure 14, the analysis focuses on time-resolved drain current decay following a gate pulse of -0.5 V with a pulse width of 10 μs, under different trap conditions.

To gain deeper insight into the sensitivity of the time-resolved response to trap density and gate bias, we extract and analyze three distinct decay components, labeled $\tau_1$, $\tau_2$ and $\tau_3$, corresponding to successive temporal segments of the transient current decay following the gate pulse (Figure 14b). The measured current response is modeled using a multi-exponential function of the form:

$$I(t) = C_{ox}\frac{dV}{dt} + C_1 e^{-\frac{t-t_1}{\tau_1}} + C_2 e^{-\frac{t-t_2}{\tau_2}} + C_3 e^{-\frac{t-t_3}{\tau_3}} + C_4$$

Here, $\tau_1$, $\tau_2$ and $\tau_3$ represent characteristic decay time constants associated with fast, intermediate, and slow trap emission pathways, respectively. The coefficients $C_1$, $C_2$, and $C_3$ reflect the relative amplitudes of each decay mode, and $\tau_1$, $\tau_2$ and $\tau_3$ are empirically defined onset times that demarcate the dominant region of each decay process. The constant term $C_4$ represents the steady-state current after all transient contributions have subsided.

These extracted time constants show small dependence on gate voltage within the depletion regime, reflecting low bias-sensitive occupation and emission rates of different trap species. Their evolution with gate bias is illustrated in Figure 15b.

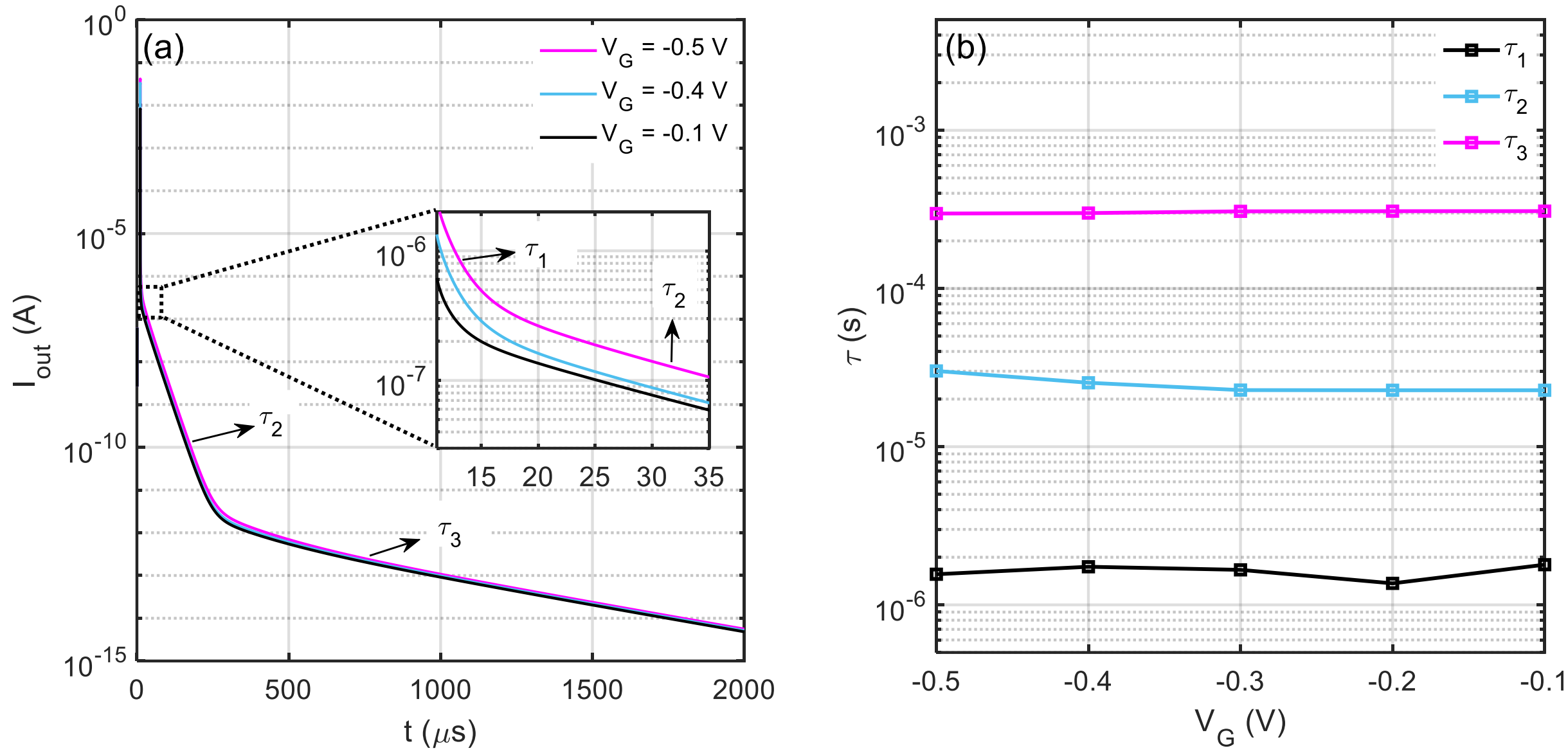

Figure 15: (a) Effect of varying gate voltage on the transient current response for oxide interface traps ($N_{t,ox}$=$10^{10}$ cm$^{-2}$), QW interface traps ($N_{t,QW}$=$10^7$ cm$^{-2}$), and bulk traps ($N_{t,bulk}$=$10^{15}$ cm$^{-3}$). Following the positive edge of the gate pulse from -0.5V to 0V, three distinct decay slopes are observed in the current transient, corresponding to different trap emission time constants. (b) Extracted fast decay time constant $\tau_1$, and slower decay time constant $\tau_2$ and $\tau_3$ as functions of gate voltage.

In Figure 15a, varying the gate voltage modulates the electric field across the QW and oxide layers, which in turn affects the occupation dynamics of traps during and after the pulse. Gate voltage thus serves as a tuning parameter that can enhance or suppress the coupling of different trap species to the channel, allowing selective probing of traps depending on their location and energy level. The decay slopes change in response to bias, highlighting the field-dependent kinetics of trap emission and capture processes.

The effect of varying trap densities across different trap types (Figure 16) reveals distinct influences on the characteristic decay time constants $\tau_1$, $\tau_2$ and $\tau_3$. The initial decay time constant, $\tau_1$, is most sensitive to bulk traps, which are distributed throughout the semiconductor and modulate carrier lifetimes and mobility over extended regions. Due to their relatively short emission times and higher volumetric densities ($\sim 10^{15}$ cm$^{-3}$), bulk traps predominantly determine the baseline recombination current immediately following the pulse.

In contrast, $\tau_2$ is strongly influenced by oxide interface traps, which contribute to the intermediate component of the decay. Their proximity to the gate and strong coupling to the vertical electric field result in moderate emission time constants that lie within the measurable transient window.

The final and slowest decay component, $\tau_3$, is dominated by QW interface traps, which primarily govern the long-time settling behavior of the transient current. Notably, this settling time remains observable even when QW interface trap densities are as low as $10^6$ $cm^{-2}$. This high sensitivity arises because these traps reside at the buried Ge/SiGe interface, where abrupt potential changes during the gate pulse induce significant occupation and emission activity. Additionally, their characteristic emission times lie in the microsecond-to-millisecond range, making them especially detectable via time-domain techniques such as DLTS.

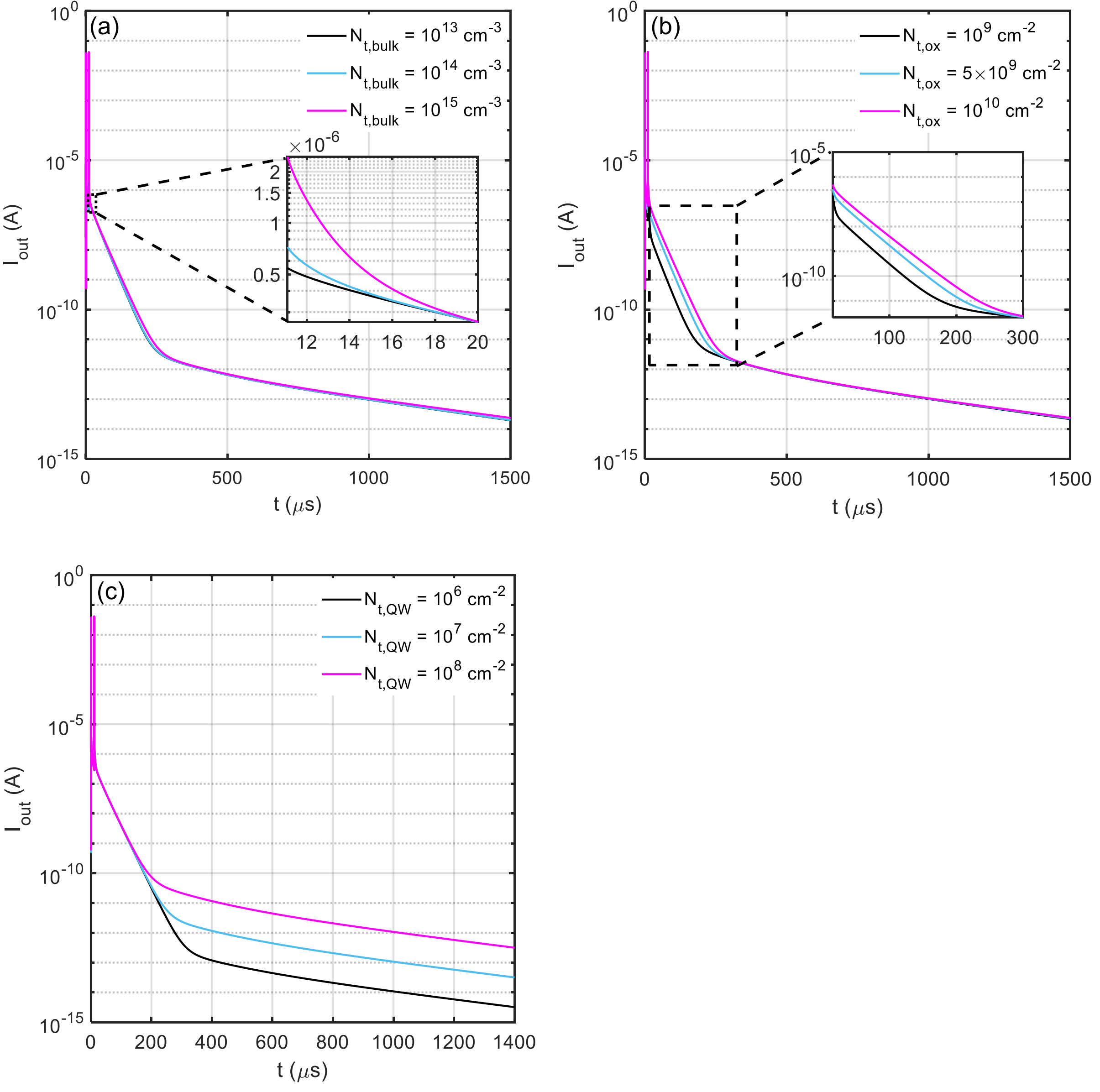

Figure 16: Time-domain current response to a gate pulse of $V_G$ = - 0.5 V for varying (a) bulk trap density ($N_{t,bulk}$), (b) oxide interface trap density ($N_{t,ox}$), and (c) QW interface trap density ($N_{t,QW}$). The distinct influence of each trap type on the current decay is evident: $\tau_1$ is primarily affected by bulk traps, $\tau_2$ by oxide interface traps, and $\tau_3$ by QW interface traps.

Figure 16a illustrates the impact of bulk trap density $N_{t,bulk}$. These traps are spatially distributed throughout the semiconductor volume and tend to affect initial transient decay of current. As $N_{t,bulk}$ increases, the initial current level decays faster due to enhanced recombination and leakage paths. This behavior is expected because bulk traps can continuously deplete or replenish carriers in the channel, especially under sustained electric fields, modifying the baseline current long after the fast trap processes have concluded.

Figure 16b shows the effect of oxide interface trap density $N_{t,ox}$ on the decay current. These traps, located at the gate dielectric/semiconductor interface, are directly influenced by the gate field and therefore respond prominently at both the positive and negative edges of the gate pulse. At higher densities, the current decay becomes slower, reflecting longer effective time constants due to increased recharging dynamics. Conversely, devices with lower $N_{t,ox}$ exhibit faster decay, as fewer interface traps are available to capture or emit carriers.

In Figure 16c, the influence of QW interface trap density $N_{t,QW}$ is highlighted. These traps are located at the buried SiGe-Ge heterojunction, where they are typically of high structural quality and hence low density, often around $10^6$ - $10^8$ $cm^{-2}$. Notably, the DLTS-like time-domain current analysis is able to resolve even these low-density QW traps, as they introduce distinct changes in the relaxation slope following the gate pulse. This is in stark contrast to impedance spectroscopy, where such low-density QW traps fail to manifest in G/ω plots due to weak capacitive coupling and limited recombination amplitude. This comparison underscores the superior sensitivity of DLTS methods in detecting traps at buried interfaces critical to qubit operation.

To further elucidate the dominant contributions of different trap types to the observed transient current decay, we systematically varied the density of each trap class (bulk traps, oxide interface traps, and QW interface traps) and analyzed the corresponding changes in the fitted decay time constants ($\tau_1$, $\tau_2$, $\tau_3$) and their amplitudes ($C_1$, $C_2$, $C_3$). The results are summarized in Figure 17 through Figure 19.

Figure 17 presents the effect of varying bulk trap density $N_{t,\text{bulk}}$ on the extracted decay parameters. As shown in Figure 17 (a)-(c), increasing $N_{t,\text{bulk}}$ has a pronounced effect on the fastest decay component, $\tau_1$, which decreases with higher trap density due to faster recombination and emission events occurring within the bulk. The amplitude $C_1$, shown in Figure 17(d), also increases significantly with bulk trap density, indicating a stronger contribution of the bulk-mediated relaxation pathway to the overall current decay. In contrast, $\tau_2$ and $\tau_3$, and their corresponding amplitudes $C_2$ and $C_3$ (Figure 17 (b), (c), (e), and (f)), remain relatively unaffected by changes in bulk trap density, consistent with the spatial and temporal decoupling of these mechanisms from the bulk region.

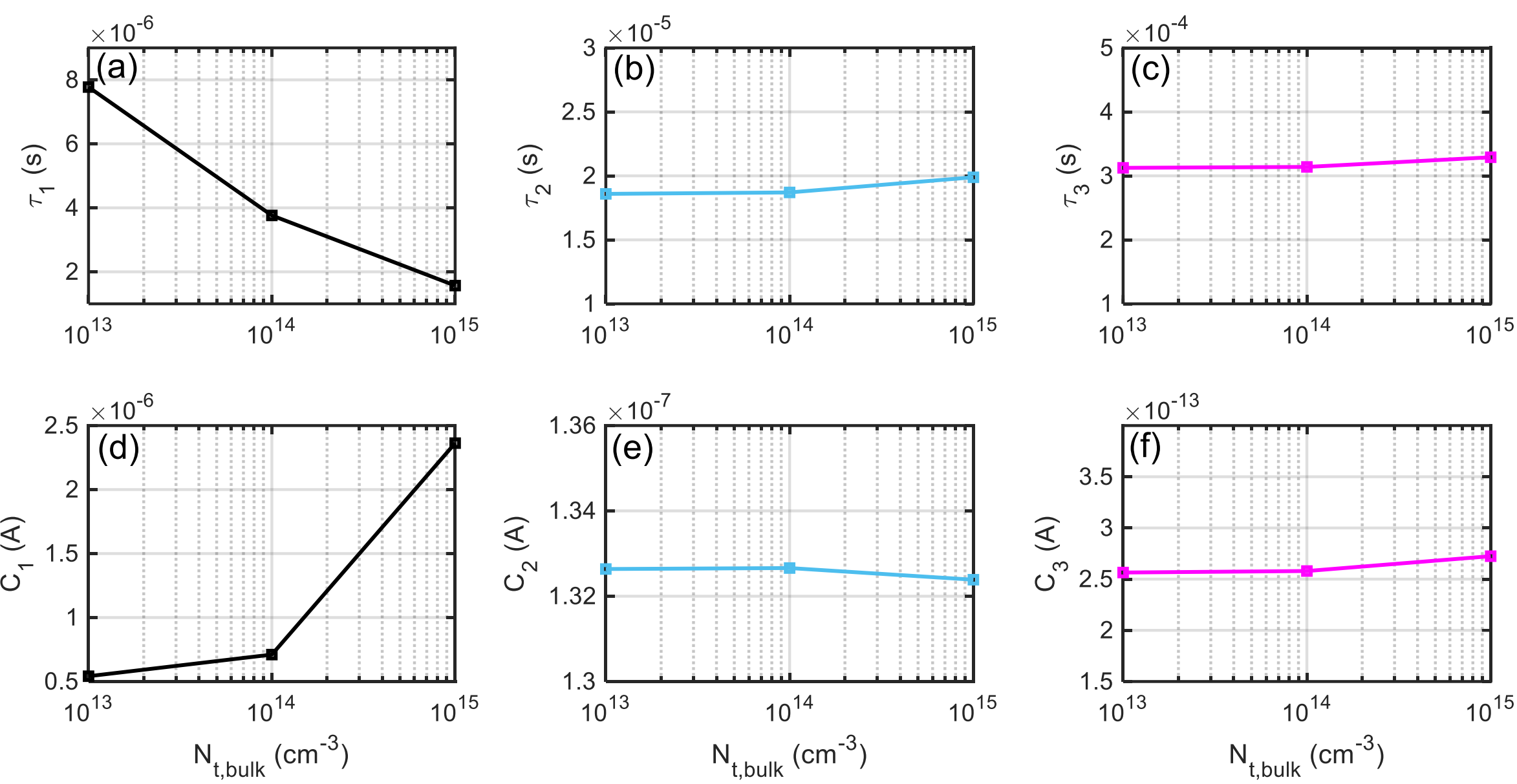

Figure 17: Extracted decay time constants $\tau_1$, $\tau_2$ and $\tau_3$ as a function of bulk trap density ($N_{t,bulk}$) are shown in (a), (b), and (c), respectively. The corresponding exponential amplitudes $C_1$, $C_2$, $C_3$ are shown in (d), (e), and (f). The results indicate that bulk trap density strongly influences the initial decay constant $\tau_1$ and its amplitude $C_1$, while having minimal effect on the slower components.

Figure 18 explores the effect of oxide interface trap density $N_{\text{t,ox}}$.

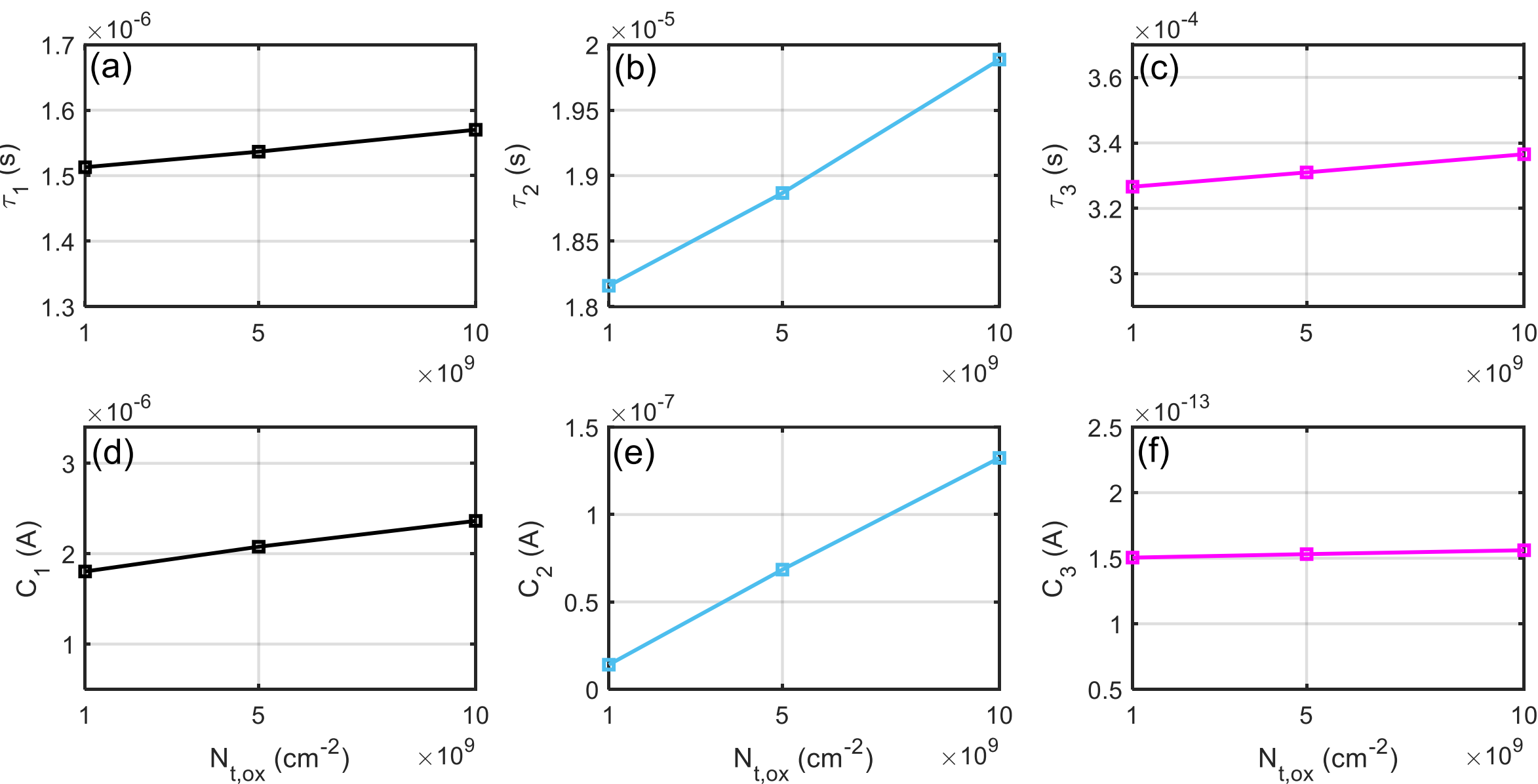

Figure 18: Extracted decay time constants $\tau_1$, $\tau_2$ and $\tau_3$ as a function of oxide interface trap density ($N_{t,ox}$) are shown in (a), (b), and (c), respectively. The corresponding exponential amplitudes $C_1$, $C_2$, $C_3$ are shown in (d), (e), and (f). The results demonstrate that oxide interface trap density strongly influences the intermediate decay constant $\tau_2$ and its amplitude $C_2$, with minimal effect on the other components.

As seen in Figure 18(b) and (e), the intermediate decay component $\tau_2$ and its amplitude $C_2$ are strongly modulated by $N_{t,ox}$. This behavior reflects the medium-range trap response times and their spatial positioning near the gate oxide, where electric field modulation is most significant during pulsed operation. In contrast, $\tau_1$ and $\tau_3$ (Figure 18 (a) and (c)) exhibit minimal changes with increasing oxide trap density, further reinforcing the association of $\tau_2$ with the oxide interface. The amplitude $C_1$ (Figure 18(d)) remains nearly constant, while $C_3$ (Figure 18(f)) shows only slight variation, indicating limited cross-coupling between oxide and QW interface traps.

Figure 19 shows the effect of varying QW interface trap density $N_{t,QW}$. As expected, the slowest decay time constant $\tau_3$, shown in Figure 19(c), is most sensitive to $N_{t,QW}$, increasing in magnitude with trap density. This slow component reflects the deep-level or spatially confined traps at the Ge/SiGe heterointerface that dominate long-term settling dynamics. Correspondingly, the amplitude $C_3$ (Figure 19 (f)) increases with $N_{t,QW}$, confirming enhanced charge exchange activity at the QW interface. The faster components $\tau_1$ and $\tau_2$, and amplitudes $C_1$ and $C_2$ (Figure 19 (a), (b), (d), and (e)), show only minor dependence on QW trap density, indicating that these dynamics are largely decoupled from QW interface processes.

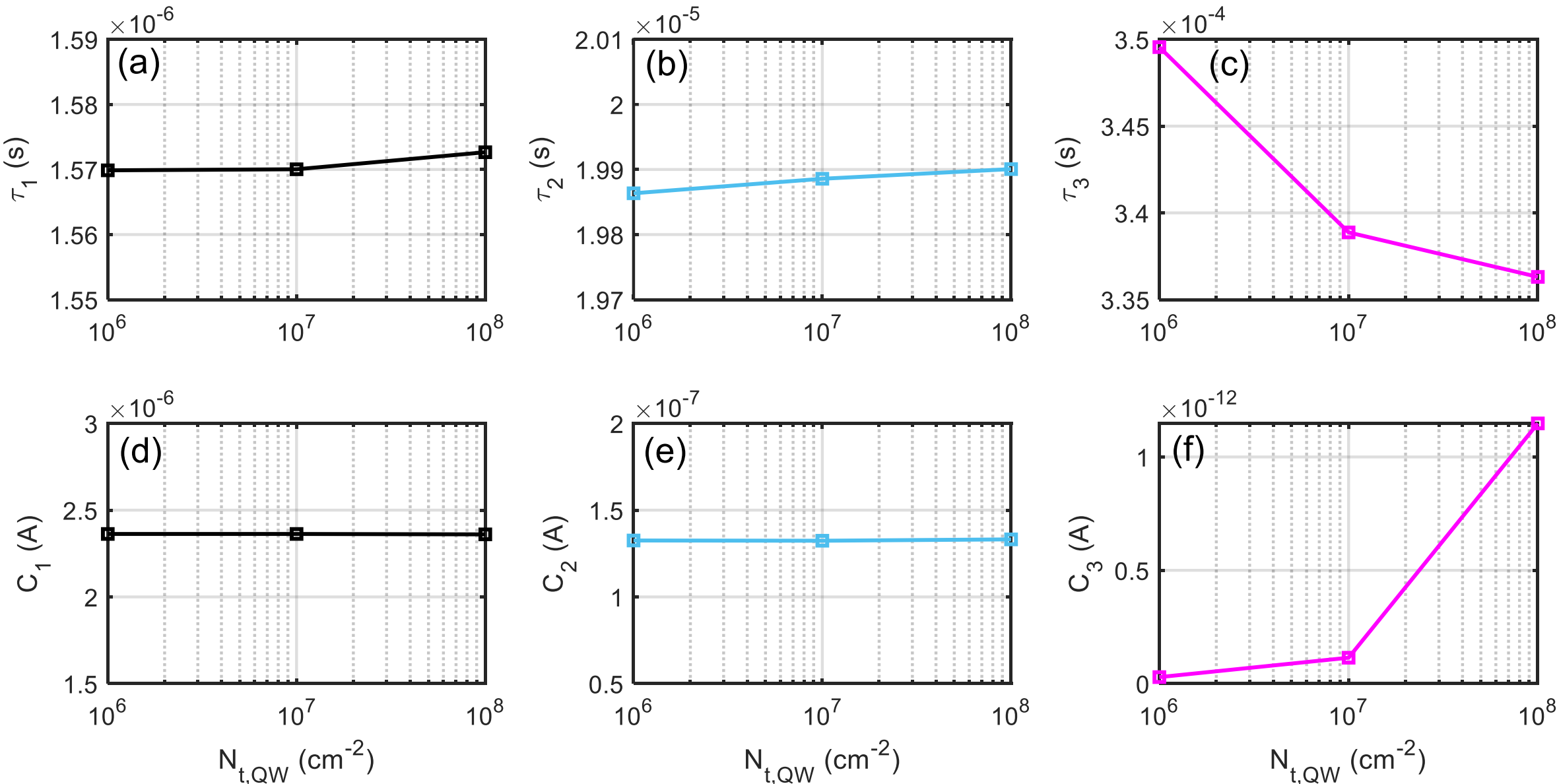

Figure 19: Extracted decay time constants $\tau_1$, $\tau_2$ and $\tau_3$ as a function of QW interface trap density ($N_{t,QW}$) are shown in (a), (b), and (c), respectively. The corresponding exponential amplitudes $C_1$, $C_2$, $C_3$ are shown in (d), (e), and (f). The results indicate that QW interface trap density primarily affects the slowest decay component $\tau_3$ and its amplitude $C_3$, with minimal influence on the faster components.

Thus, the time-domain current decay curve can be decomposed into multiple segments, each dominated by a distinct trap species:

- Early decay ($\tau_1$): Characterized by fast response and strong gate coupling, governed by bulk traps.

- Mid decay ($\tau_2$): Dominated by oxide interface traps, which exhibit intermediate emission dynamics near the oxide-semiconductor interface.
- Late decay ($\tau_3$ and settling time): Influenced by QW interface traps, which lie deeper in the structure and exhibit slower carrier exchange due to both spatial separation and intrinsic kinetics.

This decomposition of the transient behavior into trap-specific regimes provides a powerful diagnostic capability: by fitting the decay curve with multiple exponentials and analyzing their voltage dependence, one can infer the spatial distribution, energetic depth, and emission kinetics of trap species that are otherwise difficult to resolve in standard frequency-domain measurements.

From the perspective of quantum applications, these insights are particularly valuable. QW interface traps, despite being relatively silent in AC impedance spectroscopy, may still couple to gate-defined spin qubits via local charge fluctuations. Such interactions can degrade coherence and gate fidelity. The ability to detect and characterize these buried traps through time-domain DLTS-like methods thus offers a crucial tool for validating heterostructure quality and guiding material and process optimization for scalable spin qubit architectures.

Taken together, these observations reinforce a key insight: each trap type leaves a distinguishable signature in the time-domain current response, and different segments of the transient curve are each dominated by different trap mechanisms. The slope of decay after the positive pulse edge correlates with bulk and oxide trap densities, and the relaxation time reflects QW interface kinetics. These results have direct implications for quantum device design. Since even small populations of QW interface traps, undetectable in AC conductance measurements, can introduce fluctuating local fields that degrade spin coherence, the ability to resolve them using time-domain DLTS techniques becomes essential. This method provides a non-invasive, highly sensitive diagnostic for trap assessment in heterostructure-based qubit platforms, enabling process optimization and improved interface engineering.

### IV.H. AC Impedance Spectroscopy as an Advanced Alternative to Conventional C–V Measurements

Capacitance-voltage (C-V) measurements are a widely used technique for characterizing semiconductor junctions and dielectric interfaces. By sweeping the gate voltage and measuring the resulting capacitance, one can infer parameters such as doping profiles, oxide thickness, depletion width, and the presence of interface states. However, C-V measurements are typically performed at a fixed frequency (often ~1 MHz), limiting their sensitivity to dynamic processes that lie outside the measurement bandwidth. Furthermore, C-V accuracy can degrade in the presence of gate leakage, series resistance, or when overlapping mechanisms contribute to the measured capacitance.

AC impedance spectroscopy (also called admittance spectroscopy) generalizes the concept of C-V by measuring the full complex impedance over a broad frequency range. This technique allows both resistive and capacitive components to be extracted, often modeled with equivalent circuits. More importantly, impedance spectroscopy provides frequency-resolved information about charge storage, transport, and relaxation processes, enabling the identification of characteristic time constants associated with traps, dielectric relaxation, and carrier exchange.

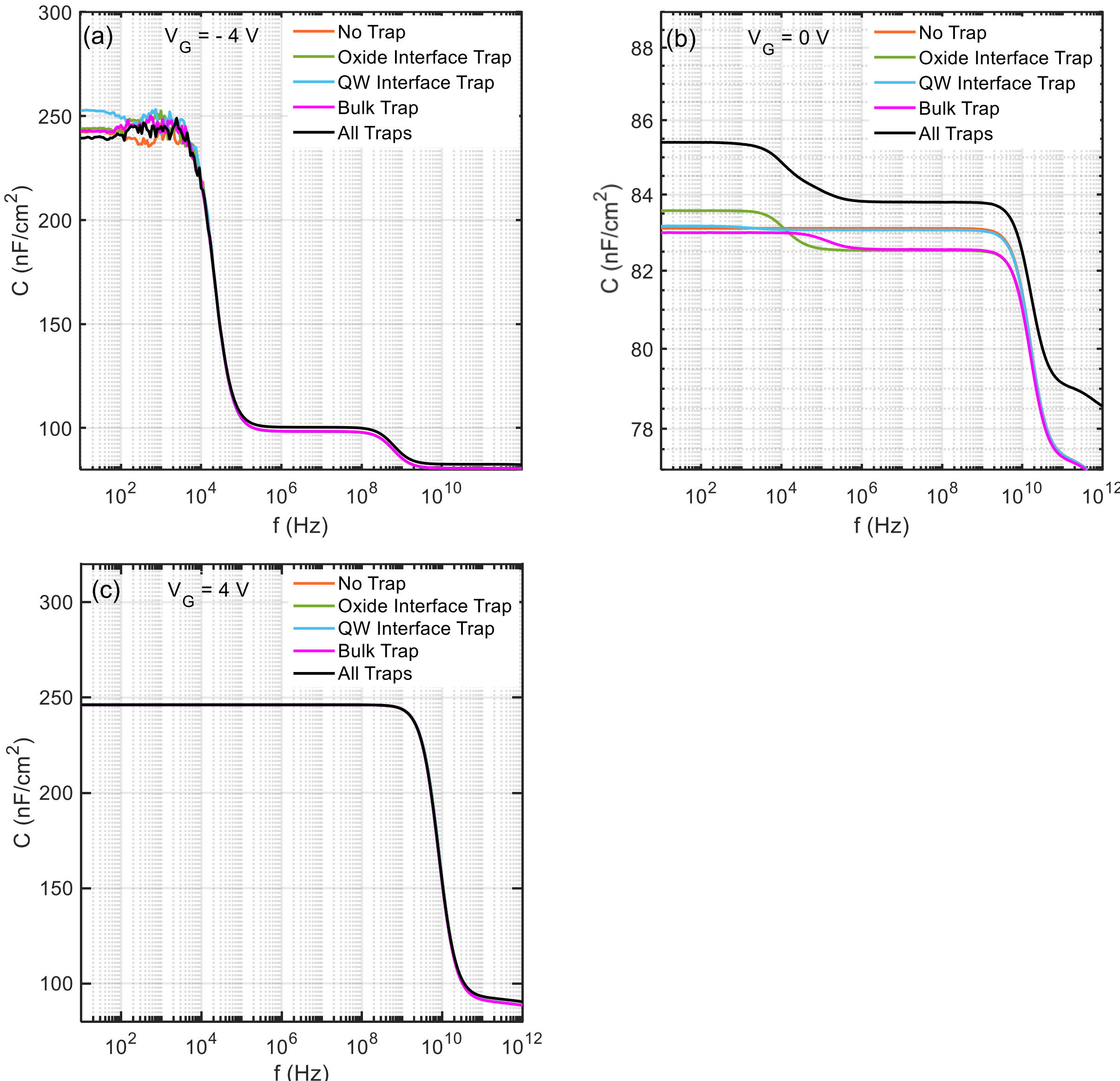

Figure 20: Capacitance vs frequency at three gate voltages: (a) inversion (VG = –4 V), (b) depletion (VG = 0 V), and (c) accumulation (VG = 4 V). Frequency-dependent changes are most pronounced in depletion, where oxide interface and bulk traps contribute significantly. Inversion and accumulation regimes show minimal sensitivity to traps.

A key advantage of impedance spectroscopy is its ability to distinguish between overlapping processes through the frequency dependence of the admittance. For example, trap-assisted carrier exchange manifests as frequency-dependent peaks in the conductance response, which are absent or indistinct in standard C–V data. Additionally, impedance spectroscopy is less sensitive to series resistance and can tolerate small leakage currents, making it better suited for emerging material systems like Ge/SiGe heterostructures.

The quantity G/ω (conductance normalized by angular frequency) is particularly useful for identifying trap states. It removes the purely capacitive contribution to the measured signal, yielding Lorentzian-like peaks at frequencies corresponding to the trap emission times ($\tau \approx 1/\omega_{\mathrm{p}}$). Each trap type (oxide interface, QW interface, or bulk) produces a distinct peak at a characteristic frequency, depending on its emission dynamics and coupling strength to the channel.

Figure 20 shows simulated capacitance vs frequency at three bias regimes: (a) inversion ($V_G$ = –4 V), (b) depletion ($V_G$ = 0 V), and (c) accumulation ($V_G$ = 4 V). While accumulation and inversion regimes show little frequency dependence, the depletion regime reveals strong variation in capacitance with frequency, particularly due to oxide and bulk traps. QW interface traps, however, do not significantly affect the capacitance spectrum, likely due to their lower density and weaker coupling at high frequencies.

Figure 21 plots G/ω vs frequency for the same bias points. Unlike capacitance, G/ω reveals clear trap-induced peaks in the depletion regime for all three trap classes—oxide, QW interface, and bulk. These peaks correspond to characteristic time constants of carrier exchange processes and are absent under inversion and accumulation. The distinct frequency signatures of each trap type highlight the superior selectivity of G/ω analysis for characterizing buried and low-density traps.

To extract quantitative values of trap parameters, we apply two models to fit the G/ω spectra: (1) a discrete-state model assuming a few well-defined trap levels, and (2) a continuum-state model assuming a broad distribution of trap states. The two expressions for G/ω are:

Discrete-state model: $G/\omega = \frac{\omega\tau_1 C_{it1}}{1+\omega^2\tau_1^2} + \frac{\omega\tau_2 C_{it2}}{1+\omega^2\tau_2^2} + \frac{\omega\tau_3 C_{it3}}{1+\omega^2\tau_3^2}$

Continuum-state model: $G/\omega = \frac{C_{it1}}{2\omega\tau_1}\ln(1+\omega^2\tau_1^2) + \frac{C_{it2}}{2\omega\tau_2}\ln(1+\omega^2\tau_2^2) + \frac{C_{it3}}{2\omega\tau_3}\ln(1+\omega^2\tau_3^2)$

where $\tau_1, \tau_2, \tau_3$ are the emission time constants and $C_{it1}, C_{it2}, C_{it3}$ are the trap-associated capacitances for bulk, oxide interface, and QW interface traps, respectively.

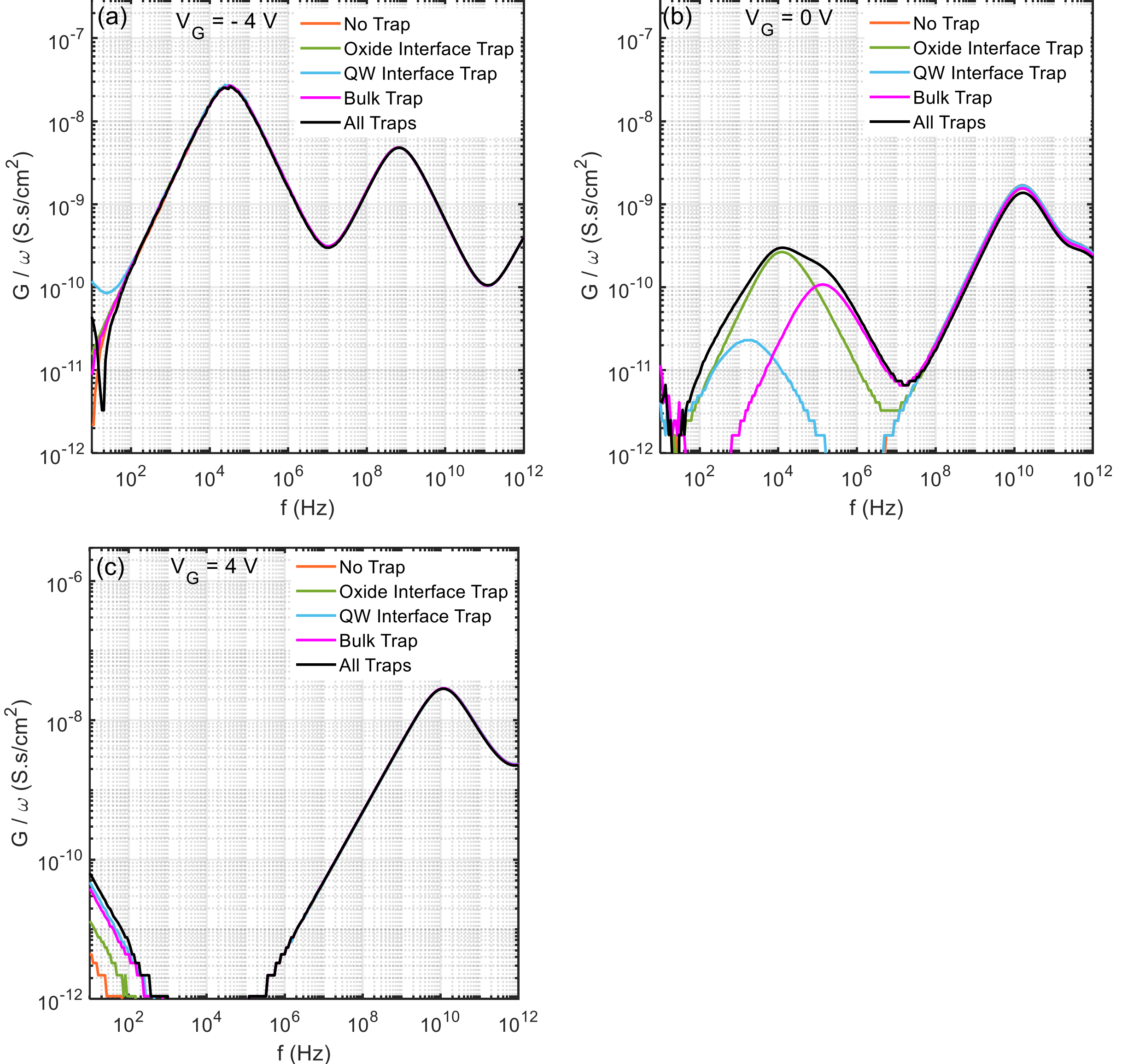

Figure 21: G/ω vs frequency under (a) inversion, (b) depletion, and (c) accumulation. Distinct peaks in the depletion regime correspond to oxide interface, QW interface, and bulk trap states. These features are absent in inversion and accumulation, emphasizing the sensitivity of G/ω analysis to trap dynamics in the depletion regime.

Figure 22 compares fits obtained using both models. The discrete-state model accurately reproduces the peak locations and amplitudes, with peaks centered at $\omega = 1/\tau$. However, it fails to capture the broader spectral features observed in the simulated data. The continuum-state model better matches the peak broadening due to its logarithmic dependence, though it systematically yields $\tau$ values nearly twice those of the discrete model due to its shifted peak condition ($\omega \approx 1.98/\tau$).

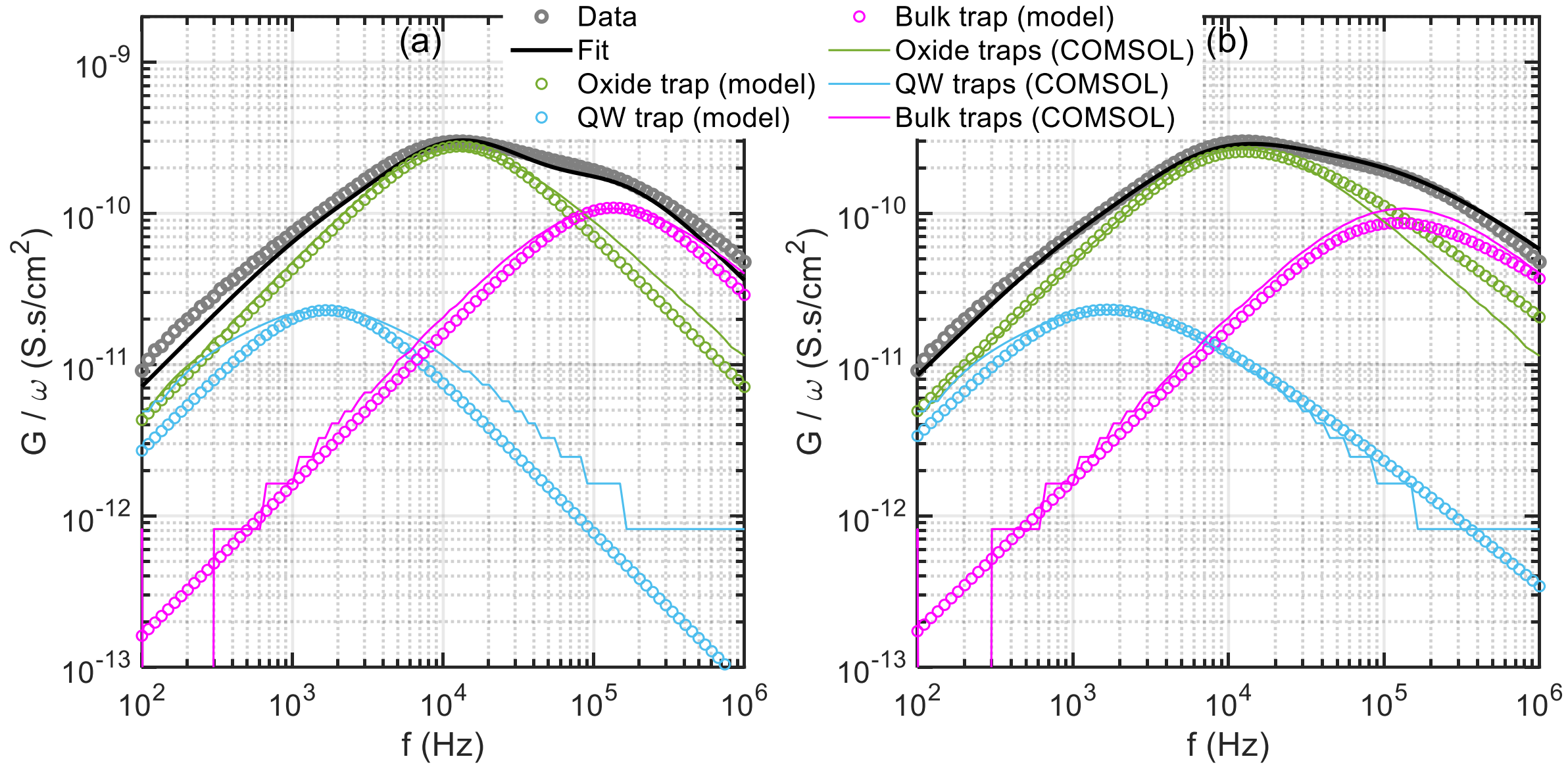

Figure 22: G/ω vs frequency fitted using (a) discrete-state and (b) continuum-state models. The discrete model matches peak position and height but underestimates spectral width. The continuum model captures both peak position and broadening, offering higher fidelity. Fitted $\tau$ values from the discrete model align better with simulated trap parameters, but the continuum model provides a more physically realistic representation of distributed trap states.

Table 4 compares the $\tau$ values extracted from both models with the actual parameters used in the simulations. While discrete models provide closer estimates to the input $\tau$ values, the continuum model better reflects the physical broadening expected from energy-distributed trap ensembles, particularly in real devices where traps are rarely monodisperse.

Table 4: Comparison of extracted decay time constants $\tau_1$, $\tau_2$, and $\tau_3$ using two analytical models, Discrete States Model and Continuum States Model, against the actual values from simulation. The values correspond to oxide interface traps, QW interface traps, and bulk traps, respectively.

| **Trap Type** | **Associated $\tau$** | **Discrete States** | **Continuum States** | **Actual** |
|---|---|---|---|---|
| Oxide Interface Traps | $\tau_1$ | $1.24\times10^{-5}$ s | $2.48\times10^{-5}$ s | $0.82\times10^{-5}$ s |
| QW Interface Traps | $\tau_2$ | $9.43\times10^{-5}$ s | $18.86\times10^{-5}$ s | $9.02\times10^{-5}$ s |
| Bulk Traps | $\tau_3$ | $1.18\times10^{-6}$ s | $2.56\times10^{-6}$ s | $0.88\times10^{-6}$ s |

In summary, AC impedance spectroscopy, and in particular, G/ω analysis, provides enhanced capability to resolve trap dynamics in semiconductor heterostructures. Compared to conventional C-V methods, it offers multiple technical advantages:

- Frequency resolution of dynamic processes, allowing detection of trap emission and recharging rates via their dielectric response.
- Improved tolerance to leakage currents, as the method separates resistive and capacitive components.
- Extraction of multiple trap parameters (density, cross-section, time constant) via modeling of the frequency-dependent conductance or capacitance.

- Detection of weakly coupled or deeply buried traps, which may be invisible in C-V but still measurably modulate impedance.
- Identification of series resistance effects, which can distort C-V but are explicitly modeled in impedance analysis.

These attributes make impedance spectroscopy an indispensable tool for evaluating trap-related defects in quantum device platforms, where even subtle relaxation processes can impact coherence and control fidelity.

### IV.I. Comparison of DLTS and AC Impedance Techniques for Trap Detection in Semiconductor Structures

Both DLTS and AC impedance spectroscopy techniques probe the dynamic response of trap states but differ fundamentally in their operating principles, sensitivity, and the type of information they provide. DLTS is a time-domain method that detects traps via their transient current or capacitance response following a voltage or current perturbation. A gate or junction pulse is used to fill traps with carriers; once the pulse is removed, the emission of these carriers produces a transient signal that decays exponentially as traps empty. The emission rate depends on the trap energy level and temperature as

$$r = \sigma v_{th} N_c \, exp[-(E_c - E_t)/kT],$$

where $\sigma$ is the capture cross-section, $v_{th}$ is the carrier thermal velocity, and $(E_c - E_t)$ is the trap activation energy. By measuring the transient over a temperature range, DLTS allows extraction of the trap activation energy, capture cross-section, and trap concentration, making it particularly powerful for deep-level defects and traps with emission times between microseconds and seconds. Because the transient signal integrates small changes over time, DLTS can resolve trap densities down to ~$10^6$–$10^7$ $cm^{-2}$, well below the detection limit of steady-state techniques.

In contrast, AC impedance spectroscopy is a frequency-domain technique that probes the small-signal response of a device under steady bias. It measures the full complex admittance Y(ω) = G(ω) + jωC(ω), where traps that can exchange charge with the semiconductor at the AC frequency contribute to both the conductance (G) and capacitance (C). By normalizing G(ω) by ω, trap contributions appear as Lorentzian peaks in G/ω vs $f$, centered at $f_p \approx 1/(2\pi\tau)$, where $\tau$ is the emission time constant. This allows direct extraction of $\tau$ and the interface-state capacitance $C_{it}$, and hence the trap density and characteristic time constant. AC impedance spectroscopy is therefore most sensitive to shallow and interface-related traps, including oxide interface traps, QW interface traps, and near-surface bulk traps, which typically have emission times in the kHz–MHz range.

A key advantage of DLTS lies in its superior sensitivity to low trap concentrations. Because it analyzes transient relaxation directly in the time domain, small variations in charge or capacitance ($\Delta C/C < 10^{-4}$) can be resolved above the background noise. This makes DLTS particularly effective for detecting buried or low-density QW interface traps (as seen in Figure 16c), which may be electrically silent in AC impedance spectra. Furthermore, the temperature dependence of DLTS provides direct access to activation energies and trap depth information, quantities that AC techniques cannot extract without complementary modeling.

Conversely, AC impedance spectroscopy offers distinct advantages in spectral resolution and selectivity. By spanning multiple frequency decades, it can simultaneously identify multiple trap species based on their characteristic frequency response (Figure 20 - Figure 22). This enables deconvolution of overlapping processes such as oxide interface, QW interface, and bulk traps, each contributing peaks at different frequencies in the G/ω spectrum. Impedance spectroscopy also operates under steady-state bias, avoiding thermal ramping, and can tolerate small gate leakage or series resistance that would otherwise obscure DLTS transients. In addition, equivalent circuit fitting allows separation of transport-related resistances and dielectric relaxation, giving broader insight into charge transport mechanisms beyond traps alone.

In practice, the two techniques are complementary. DLTS excels at identifying deep and low-density traps, quantifying their energetic position and capture cross-section, while AC impedance spectroscopy excels at frequency-resolved identification of shallow and interface traps, and at mapping distributed trap ensembles under steady operating conditions.

In QW systems, this complementarity is particularly valuable:

- AC impedance spectroscopy (G/ω–f) can efficiently identify oxide and bulk trap behavior influencing charge transport and electrostatic stability.
- DLTS provides the necessary time-domain and temperature-dependent resolution to detect QW interface traps at densities (as low as $10^{6}$ $cm^{-2}$ in SiGe/Ge QWs), which directly impact spin qubit coherence through local charge fluctuations.

When combined, the two approaches yield a comprehensive trap profile spanning both spatial location (oxide, QW, bulk) and energetic depth, establishing a unified framework for diagnosing charge noise and reliability in advanced quantum heterostructures.

Taken together, these three methods provide a hierarchical framework for defect characterization in semiconductor and quantum devices (Table 5). C-V serves as a baseline electrostatic probe, AC impedance spectroscopy enables high-resolution mapping of frequency-dependent trap behavior, and DLTS delivers the highest sensitivity to deep and low-density traps. In Ge/SiGe QW architectures, combining AC and DLTS approaches provides both spatial and energetic insight into the traps responsible for charge noise and decoherence in hole spin qubits, offering a powerful diagnostic toolkit for material and process optimization.

**Table 5**. Comparison of C-V, AC Impedance Spectroscopy, and DLTS Techniques

| Feature / Parameter | **C–V Measurement** | **AC Impedance Spectroscopy (G/ω–f)** | **Deep-Level Transient Spectroscopy (DLTS)** |
|---|---|---|---|
| **Measurement Domain** | Frequency fixed (typically 1 MHz) | Frequency swept (1 Hz – 1 MHz or more) | Time-domain transient (μs – s) |
| **Primary Quantity Measured** | Capacitance vs gate voltage | Complex admittance (G and C vs frequency) | Transient capacitance or current after pulse |
| **Operating Condition** | Steady-state bias | Steady-state small-signal AC bias | Transient following bias pulse |
| **Sensitivity Range** | Moderate (≥ $10^{9}$ $cm^{-2}$ traps) | Good for interface traps (≥ $10^{8}$ $cm^{-2}$) | Very high (≥ $10^{6}$ $cm^{-2}$) |

| **Trap Types Detected** | Interface and oxide traps (if active at test frequency) | Shallow, interface, and near-surface bulk traps | Deep and buried traps with slow emission |
|---|---|---|---|
| **Extracted Parameters** | Oxide thickness, flat-band voltage, interface density (approx.) | Trap density, time constant ($\tau$), relaxation frequency | Activation energy ($E_t$), $\sigma_t$, $N_t$, $\tau_t$ |
| **Frequency / Time Resolution** | Single point | Multi-decade frequency sweep | Time decay and temperature dependence |
| **Energy-Level Resolution** | Poor | Indirect (from $\tau$) | Excellent (from temperature dependence) |
| **Sensitivity to Leakage / Series R** | High (measurement degrades easily) | Low (can tolerate small leakage) | Moderate (leakage affects transient amplitude) |
| **Complexity** | Simple setup and interpretation | Moderate (requires circuit modeling) | High (requires pulse control and temperature sweep) |
| **Best Use Case** | Doping and oxide profiling | Dynamic trap and relaxation spectroscopy | Deep-level trap energetics and low-density detection |

## V. Relevance to Charge Noise in Ge Hole Spin Qubits

The utility of impedance spectroscopy and DLTS-like time-domain techniques, as developed in this work, becomes particularly salient when considered in the context of charge noise in QD hole spin qubits. In Ge/SiGe systems, hole-based spin qubits have gained significant attention due to their strong SOC [15,16,17], enabling fast all-electrical spin control [18,19]. However, this same SOC renders hole spins intrinsically more susceptible to charge fluctuations, especially when compared to electron spin qubits, where spin-charge hybridization is weaker [20]. In this section, we outline the origins of charge noise in gate-defined Ge QD systems and establish how the electrical characterization techniques developed in this paper provide a quantitative pathway to diagnose and suppress the underlying physical mechanisms of decoherence.

### V.A. Physical Origins of Charge Noise in Gate-Defined Ge QDs

Charge noise in semiconductor qubits arises primarily from temporal fluctuations in the local electrostatic environment. These fluctuations couple to the qubit through the gate-defined confinement potential and modulate the qubit's energy levels via the electric field dependence of the spin-orbit Hamiltonian. The primary sources of such charge noise in Ge/SiGe heterostructures include:

1. Oxide Interface Traps: At the dielectric/semiconductor boundary (e.g., $Al_2O_3$/SiGe), dangling bonds, fixed charges, and interface states can trap and de-trap carriers, creating low-frequency noise (1/f-type) that directly modulates the confinement potential of a QD located nearby. These states are thermally activated and their response time is governed by capture cross-section, trap energy, and spatial proximity.

2. Quantum Well Interface Traps: Though typically lower in density due to epitaxial growth, traps at the Ge/SiGe heterointerface can act as fluctuating dipoles near the QD confinement plane. While their contribution to static leakage is often negligible, their dynamic response within the qubit operation bandwidth (~kHz-GHz) can introduce significant decoherence, especially via spin-orbit-mediated EDSR channels.

3. Bulk Traps and Background Impurities: Deep-level states in the Ge or SiGe channel, as well as residual dopants, can modulate the local potential landscape. These effects are more pronounced in devices with imperfect growth or post-processing damage, and are typically associated with short-range noise in the tens to hundreds of MHz range.

4. Remote Charge Fluctuators in Gate Dielectrics and Metals: Charges in the gate oxide or within the gate metal can couple capacitively to the QD and cause both quasistatic offsets and dynamic fluctuations. While often overlooked, these have been implicated in observed low-frequency noise in planar silicon qubits [21]and likely contribute in Ge systems.

These noise sources have frequency-dependent spectral densities and may couple differently to gate voltage, spin splitting, or Rabi frequencies depending on the qubit geometry and control scheme.

### V.B. Using Spectroscopic Techniques to Isolate Noise Sources

The measurement techniques developed in this paper, frequency-dependent admittance spectroscopy and time-domain DLTS analysis, offer a direct route to isolating the microscopic origin and timescale of charge noise sources in Ge/SiGe quantum dot systems. Specifically:

1. Impedance Spectroscopy (G/ω Analysis)
   - Extracts trap density, response time, and energy distribution by fitting frequency-domain data to trap-assisted SRH models.
   - Peak position and height in the G/ω curves reflect the emission time constants and density of traps. These quantities correlate with the noise power spectral density expected in qubit operation bands (10 Hz to 1 GHz).
   - As shown in Figure 2 to Figure 12, oxide interface traps dominate in depletion and moderate bias regimes, contributing strongly to slow fluctuators (Hz-kHz), while bulk traps introduce high-frequency shoulders relevant to GHz-scale operations.
2. DLTS-Based Time-Domain Analysis
   - Decomposes the total current response into distinct exponential components ($\tau_1$, $\tau_2$, relaxation time) that correlate with traps at different spatial locations.
   - As demonstrated in Figure 14 to Figure 19, time constants associated with QW traps and oxide interface traps can be separated and quantified based on the temporal signature of their carrier emission.
   - DLTS techniques are particularly sensitive to low-density QW interface traps (~$10^6$-$10^8$ $cm^{-2}$), which are electrically silent in frequency-domain methods, but can cause slow electric field fluctuations near the QD and drive decoherence via SOC mixing.

3. Nyquist Plot Interpretation

- o Allows visualization of distinct trap relaxation arcs, corresponding to overlapping or distributed trap responses (Section IV.E).
- o By correlating arc curvature and area with specific trap types, the method gives an estimate of loss mechanisms that would correspond to local heating or energy fluctuations in qubit operation.

### V.C. Translating Trap Signatures to Qubit Dephasing and Gate Errors

The extracted trap parameters, such as trap density $N_{t,i}$, emission time constant $\tau_i$, capture cross-section $\sigma_i$, and energy distribution, can be quantitatively linked to the qubit's charge noise spectral density $S_\epsilon(f)$, which impacts both the dephasing time $T_2^*$ and coherence time $T_2$ of spin qubits under electrical control.

A general expression for the energy fluctuation spectrum due to charge traps is [22]:

$$S_\epsilon(f) \propto \sum_i \left(\frac{\partial \epsilon}{\partial q_i}\right)^2 \frac{N_{t,i}}{1 + (2\pi f \tau_i)^2}$$

Here, $\partial\epsilon/\partial q_i$ quantifies the sensitivity of the qubit energy to the charge state of trap species i; $N_{t,i}$ is the trap density for that class; $\tau_i$ is the associated emission time constant; and $f$ is the frequency of the noise component. This Lorentzian summation reflects that each trap class contributes to $S_\epsilon(f)$ near its characteristic switching frequency $f \sim 1/2\pi\tau_i$, modulated by how strongly its charge state couples to the qubit.

In practice, full quantum electrostatic modeling is required to accurately compute the coupling factor $\partial\epsilon/\partial q_i$, which quantifies how strongly a localized trapped charge $q_i$ at location $i$ perturbs the qubit energy splitting $\epsilon$. This coupling generally decays with increasing distance $d_i$ between the trap and the qubit. In layered semiconductor structures, this decay can be approximated either as exponential attenuation, $\partial\epsilon/\partial q_i \propto exp\,(-d_i/\lambda)$ [23], where $\lambda$ is a characteristic electrostatic screening length, or as a power-law form $\partial\epsilon/\partial q_i \propto 1/d_i^\alpha$ [22], where $\alpha$ is an attenuation exponent typically in the range $1 \leq \alpha \leq 2$. The exponential attenuation $exp\,(-d_i/\lambda)$ is more appropriate when electrostatic screening dominates the decay (e.g., in high-dielectric or metallic environments or very short screening lengths). Since Ge/SiGe systems have modest dielectric constants (~16) and no strong metallic screening near the QD, electrostatic field lines are not exponentially suppressed over short distances and the power law form is a better approximation. The choice of exponent depends on geometry and screening: $\alpha = 1$ is appropriate for planar capacitive systems, while $\alpha = 2$ is used for 3D point-charge interactions. In this study, we adopt the power-law approximation with $\alpha = 1$, which is well suited for gate-defined quantum dot architectures with planar capacitive coupling [24]. This formulation captures essential spatial dependencies while preserving analytical tractability. Let the relevant layer thicknesses be defined as:

- $t_{ox}$: thickness of the gate dielectric (e.g., $Al_2O_3$),
- $t_{bar}$: thickness of the top SiGe barrier layer above the QW,
- $t_{QW}$: total thickness of the Ge QW,

- $t_{stack}$=$t_{bar\ (top)}$+$t_{QW}$+ $t_{bar\ (bottom)}$

Then the trap-to-qubit distances are expressed as:

- Oxide interface traps:

$$d_{ox} = t_{ox} + t_{bar} + \frac{t_{QW}}{2}$$

- QW interface traps (assumed symmetric, at top and bottom interfaces of the Ge QW):

$$d_{QW} = \frac{t_{QW}}{2}$$

- Bulk traps (assumed uniformly distributed within the active QW region of total thickness $t_{stack}$:

$$d_{bulk} = t_{stack}/\sqrt{12} \text{ (RMS distance for uniform slab)}$$

The relative weighting factors become:

$$w_i \propto \frac{1}{d_i^2}$$

and the normalized noise spectral density becomes:

$$S_{\epsilon}(f) \propto \sum_i w_i \frac{N_{t,i}}{1+(2\pi f \tau_i)^2} \quad \text{with} \quad w_i = \frac{1/d_i^2}{1/d_{QW}^2}$$

so that $w_{QW} = 1$, and all other weights are relative to the QW interface coupling.

This formulation captures the intuition that while oxide interface traps may have high densities and long time constants, their large separation from the qubit results in weak capacitive coupling. In contrast, QW interface traps, even at moderate densities, can dominate $S_{\epsilon}(f)$ in the relevant kHz–MHz frequency range due to their proximity to the qubit and typical time constants in the microsecond-to-millisecond range.

To combine the bulk and interface contributions consistently in the noise spectral density expression, the volumetric trap density of bulk traps $N_{t,bulk}$ [$cm^{-3}$] must be converted to an areal density [$cm^{-2}$] by multiplying with the effective thickness over which the traps contribute to noise:

$$N_{t,bulk}(2D) = N_{t,bulk}(3D) \times t_{stack}$$

where $t_{stack}$= $t_{bar\ (top)}$+$t_{QW}$+ $t_{bar\ (bottom)}$ is the total thickness of the quantum well and surrounding barrier layers. This effective projection assumes that traps throughout the stack contribute to charge fluctuations seen by the qubit, with electrostatic weighting accounted for separately in the coupling factor $w_{bulk}$. This conversion allows the bulk contribution to be directly added to those from the oxide and QW interface traps in the location-weighted spectral summation.

For the device geometry considered in this study (with a 30 nm $Al_2O_3$ gate dielectric, 55 nm $Si_{0.2}Ge_{0.8}$ barriers, and a 16 nm Ge quantum well), we estimate trap-to-qubit distances and corresponding weights as $w_{QW} = 1$, $w_{bulk} \approx 0.0484$, and $w_{ox} \approx 0.0074$. Thus, for this structure:

- The oxide interface traps are ~135X weaker in electrostatic coupling compared to QW interface traps,
- Bulk traps couple ~20X weaker than QW interface traps.

These normalized weights are directly applied in Figure 23 to represent the relative impact of different trap classes on the noise spectral density $S_\epsilon(f)$. Oxide traps appear as mid-frequency bumps (~10–100 kHz), while QW interface traps dominate in the low-frequency regime relevant for gate operations and dynamical decoupling bandwidths. Bulk traps contribute broadband background noise due to their intermediate coupling and relaxation characteristics.

This location-weighted model provides a useful approximation for incorporating trap spatial distribution into dephasing analysis and highlights the need for interface-level control during fabrication. In particular, maintaining QW interface trap densities below $10^7$ cm$^{-2}$ may be critical for achieving low-noise hole spin qubit operation.

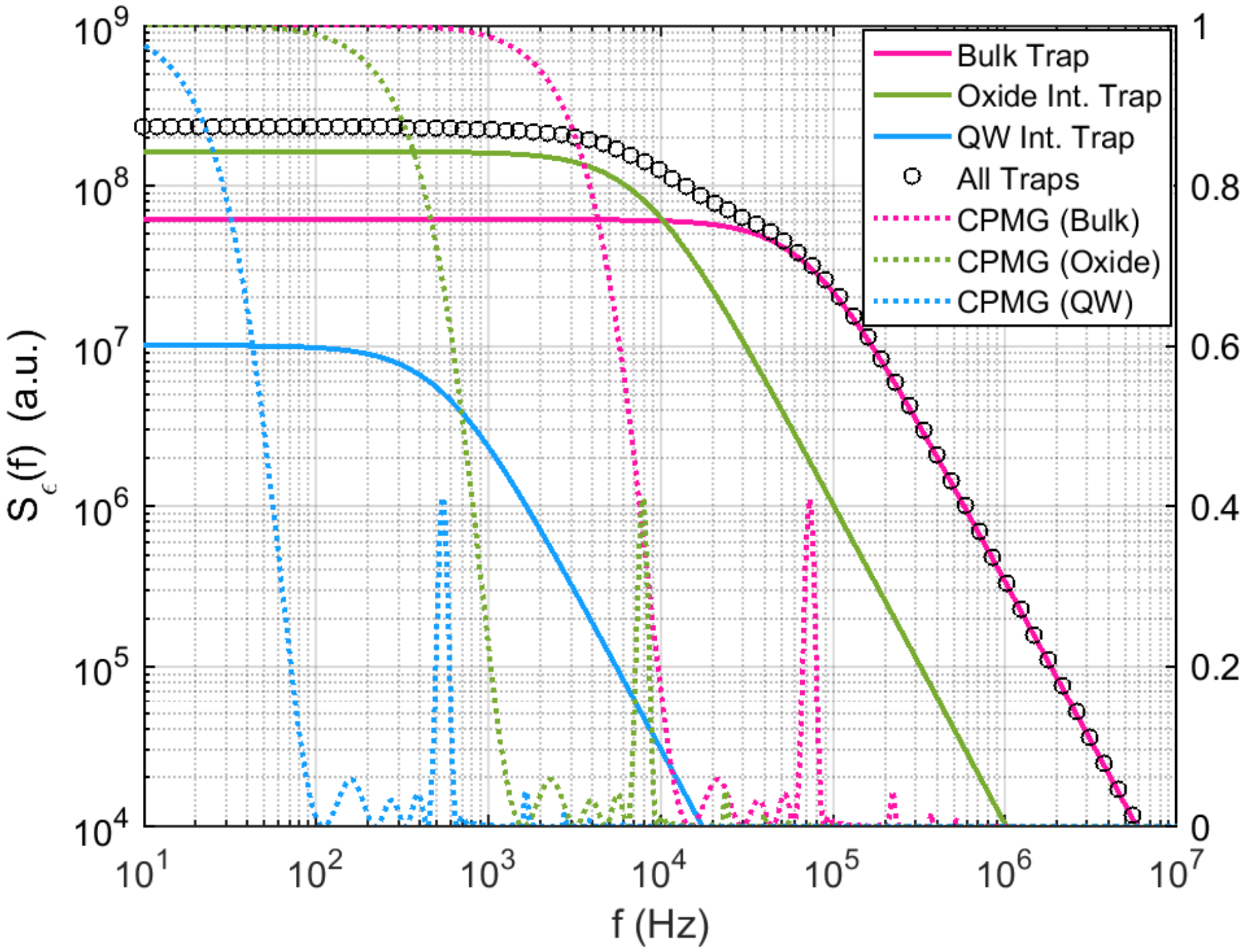

Figure 23: Spectral noise density $S_\epsilon(f)$due to different trap types in the Ge/SiGe QW device, overlaid with CPMG-10 filter functions targeting each dominant noise source. The contributions from bulk traps (magenta), oxide interface traps (green), and QW interface traps (blue) are shown along with their combined noise spectrum (black circles). The CPMG filter functions (dotted lines) are tuned to suppress noise near the characteristic knee frequency $f_c = 1/(2\pi\tau)$ of each trap class, effectively attenuating decoherence-inducing charge noise in their respective frequency ranges. The model assumes trap densities of $N_{t,\text{bulk}} = 10^{14}$ cm$^{-3}$, $N_{t,\text{ox}} = 10^{10}$ cm$^{-2}$, and $N_{t,\text{QW}} = 10^7$ cm$^{-2}$, with respective emission time constants of $\tau_{\text{bulk}} = 2.14\,\mu$s, $\tau_{\text{ox}} = 19.9\,\mu$s, and $\tau_{\text{QW}} = 287\,\mu$s. The CPMG filters are designed using $n = 10$π-pulses to target the spectral knee frequencies of each trap class.

To further illustrate how trap-induced charge noise couples to qubit coherence, we designed dynamical decoupling filters based on the Carr–Purcell–Meiboom–Gill (CPMG) pulse sequence and applied them to the modeled spectral noise density $S_{\epsilon}(f)$.[25] As shown in Figure 23, each CPMG-10 filter (constructed with ten equally spaced $\pi$-pulses) introduces frequency-domain nulls that selectively suppress noise at and around the characteristic knee frequency $f_c = 1/(2\pi\tau)$ of a given trap species. For instance, the oxide interface traps with $\tau \approx 20\,\mu s$ produce spectral peaks near 8 kHz, which can be attenuated using a CPMG filter with total duration $T \approx 125\,\mu s$. Similarly, QW interface traps, which exhibit $\tau \approx 0.3$ ms, generate noise near 500 Hz and require a longer pulse sequence ($T \approx 2$ ms) for optimal suppression. Bulk traps, having $\tau \approx 2\,\mu s$, contribute at higher frequencies (~60 kHz) and are more effectively mitigated by shorter pulse intervals. The overlap between $S_{\epsilon}(f)$ and the CPMG filter response demonstrates that multi-pulse dynamical decoupling can be tailored to reject noise arising from specific trap types.

These observations reinforce the importance of understanding the electrostatic environment of the qubit. In prior work, we quantified how charge fluctuations translate into g-factor modulation using self-consistent electrostatic and spin-orbit modeling, and identified electric-field sweet spots where dephasing is suppressed.[26]

It is important to note that another class of traps, border traps, reside physically within the oxide, typically 1-2 nm from the semiconductor interface. These traps can exchange charge with the semiconductor via tunneling and exhibit a broad distribution of time constants. While not explicitly modeled in this work, we acknowledge their potential contribution to low-frequency charge noise and bias drift. Modeling these traps with a finite spatial distribution rather than as a lumped 2D sheet would result in a broader and asymmetric $G/\omega$ response. Future studies incorporating distributed oxide trap profiles using tunneling-based charge exchange models are warranted to capture these effects more accurately.

### V.D. Temperature Dependence of Trap Response and Model Validity

While quantum dot spin qubits operate at millikelvin temperatures, the simulations in this work are performed at 300 K to maximize the visibility of trap states through thermally activated emission and to ensure model validity under the Shockley-Read-Hall (SRH) formalism. At cryogenic temperatures, many deep-level traps become inactive due to exponentially suppressed emission, rendering classical DLTS and admittance models ineffective. However, the trap densities and spatial locations extracted at room temperature remain fixed in the material stack and influence low-temperature charge noise through frozen-in potential fluctuations. Moreover, the emission time constants $\tau$ extracted at 300 K allow us to identify which traps will remain active in the kHz-MHz frequency bands relevant for spin coherence. Lowering the temperature shifts conductance peaks to lower frequencies and sharpens C-V transitions due to reduced carrier screening, but the qualitative conclusions regarding trap origin, coupling, and mitigation strategies remain unchanged.

### V.E. Guidelines for Reducing Charge Noise: Trap Mechanisms, Impacts, and Mitigation

The spectroscopic analyses presented in this work establish a quantitative pathway for identifying trap populations that contribute most strongly to qubit dephasing, thereby informing materials optimization and device design. In gate-defined spin qubit platforms, particularly in Ge/SiGe quantum wells, charge noise originates from multiple electrically active defects at various locations and on varying timescales. These defects modulate the local electrostatic environment, leading to spin decoherence via spin–orbit coupling and fluctuating qubit energy splittings. Understanding the physical origin, time constants, and coupling strengths of these traps is therefore critical to engineering robust, scalable quantum devices.

Below we summarize and contextualize six major categories of trap mechanisms that influence spin qubit fidelity in Ge/SiGe heterostructures. For each, we highlight typical densities, locations, spectral impact, and mitigation strategies, along with relevant literature references.

**1. Oxide Interface Traps ($N_{t,ox}$): Interface-Localized, Mid-Timescale**

These traps reside on the semiconductor side of the $Al_2O_3$/SiGe interface and are modeled as fixed-energy defects just inside the SiGe layer. They arise from interfacial dangling bonds, interdiffusion, or structural mismatch. With typical densities ranging from $10^{10} - 10^{12}\ cm^{-2}eV^{-1}$ and time constants in the μs–ms range, these traps dominate the low-frequency regime (~1–10 kHz) in impedance spectra and contribute to early or mid-stage transients in DLTS.

Mitigation strategies include atomic layer deposition (ALD) with in situ surface pretreatment,[1] post-deposition annealing in forming gas to passivate dangling bonds, and improved pre-ALD surface terminations in Ge systems.[27] Reducing Nt,ox below $10^{10}\ cm^{-2}$ is critical to suppressing low-frequency charge noise, especially given the strong capacitive coupling of these states to gate electrodes.

**2. QW Interface Traps ($N_{t,QW}$): Buried, Long-Timescale**

These traps originate from strain relaxation, roughness, or intermixing at the $Ge/Si_{0.2}Ge_{0.8}$ heterointerface. Their densities typically range from $< 10^6$ to $10^8\ cm^{-2}$, and their time constants often span tens to hundreds of microseconds. Despite their low densities, they lie close to the qubit and exhibit strong capacitive coupling, significantly contributing to noise in the kHz–MHz regime—relevant for Rabi oscillations and dynamical decoupling.

These traps often evade detection in frequency-domain measurements but are clearly revealed in time-resolved decay analyses. Strategies to suppress $N_{t,QW}$ include low-temperature Ge growth, strain-balanced heterostructure design, and improved growth interruption control.[28-30] Maintaining interface trap densities below $10^7\ cm^{-2}$ is likely essential for scalable, fault-tolerant qubit operation.

**3. Bulk Traps ($N_{t,bulk}$): Distributed, Fast to Mid-Timescale**

Modeled as uniformly distributed in the SiGe or Ge layers, these traps arise from residual implantation damage, processing-induced defects, and background dopants. Densities typically range from $10^{13} - 10^{16}\ cm^{-3}$, with time constants spanning nanoseconds to microseconds. They contribute to the high-frequency tail of $G/\omega$ spectra and to early decay slopes ($\tau_1$) in transient current analyses.

These traps are especially problematic when the quantum dot overlaps with deep regions of the heterostructure. Mitigation strategies include hydrogen passivation, low-damage etching, flash annealing for dopant activation, and controlled dry processing.[31-33]

**4. Border Traps: Near-Interface Oxide Traps**

Not explicitly modeled in this work, border traps lie within a few nanometers inside the oxide layer and can tunnel to or from the semiconductor. They exhibit a broad distribution of time constants due to their spatial variation and are known to cause hysteresis, slow transient drift, and telegraph noise.[34] Though less coupled than interface traps, their contribution may become significant under long bias pulses or slow gate operations.

Future modeling frameworks should incorporate energy- and distance-dependent tunneling probabilities to capture the full impact of border traps. Techniques such as high-temperature annealing and carefully controlled ALD stoichiometry may reduce their prevalence.

**5. Fast Interface States: GHz Regime Noise Sources**

These ultrafast traps, often overlooked in traditional characterization, reside directly at or very near the oxide–semiconductor interface. They respond at gigahertz frequencies, possibly impacting spin readout and ultrafast gate control. While not prominent in DC or slow AC characterization, they may significantly influence high-speed spin control fidelity and deserve further study in future GHz-scale circuit modeling.[35,36]

**6. Mobile Ionic Species and Charge Migration**

Mobile ions, such as $H^+$ or $Na^+$, and migrating oxygen vacancies within high-$\kappa$ dielectrics like $Al_2O_3$ can induce slow, long-timescale drift or $1/f$-like charge noise. These processes manifest over seconds or minutes and may introduce baseline fluctuations or slow decoherence mechanisms.[37] While less relevant to short coherence times, they can impact qubit tuning stability and device reproducibility. In addition to materials mitigation (e.g., gettered ALD precursors, annealing), environmental shielding and low-humidity processing are recommended to limit these effects.[38]

**Process Implications and Outlook**

Together, these trap categories span nearly eight decades in time constant and multiple spatial scales, from buried interfaces to distributed bulk volumes to remote ionic motion. Our combined use of impedance spectroscopy and DLTS-inspired time-domain modeling enables the extraction of distinct trap parameters, including density, location, and time constant, offering an empirical basis for targeted process improvement.

In particular, oxide interface traps can be identified through their prominent peaks in frequency-domain spectra and their moderate time constants ($\tau$), and they can be effectively suppressed using ALD and post-deposition annealing techniques. QW interface traps are best detected via the slowest decay component ($\tau_3$) in time-domain DLTS measurements, and their density can be minimized through careful epitaxial growth optimization, including interface engineering and strain management. Bulk traps manifest in the early-stage transient current decay ($\tau_1$) and are mitigated through low-damage fabrication processes such as gentle etching and hydrogen passivation. Border traps and mobile ionic species, although not directly addressed in this study, are important contributors to long-term drift and low-frequency noise, making them relevant for maintaining bias stability and long-term device endurance.

From a quantum design perspective, these insights enable quantitative metrics for fabrication success, such as maintaining $N_{t,QW} < 10^7\ cm^{-2}$ and $\tau_{QW}$ outside the 1–100 kHz range to reduce spin dephasing. Additionally, identifying material-induced decoherence channels supports better filtering strategies, bias regime selection, and device scaling.

As quantum dot platforms advance toward multi-qubit arrays and error correction thresholds, the inclusion of these broader trap mechanisms, particularly border traps and fast interface states, will become essential for predictive noise modeling and fidelity benchmarking.

## Conclusion

In this work, we presented a comprehensive investigation of electrically active defects in strained Ge/SiGe QW heterostructures, focusing on their impact on both frequency- and time-domain device responses relevant to gate-defined QD hole spin qubits. While Ge/SiGe serves as a case study, the characterization framework and diagnostic techniques we developed are broadly applicable to other material platforms.

Conventional C-V measurements, while widely used, often provide limited or ambiguous information in the presence of multiple trap species or spatially distributed defects. To overcome these limitations, we employed a two-pronged approach: (i) frequency-domain analysis using AC impedance spectroscopy (especially G/ω and phase characteristics), and (ii) time-domain transient techniques inspired by deep-level transient spectroscopy (DLTS). This combined methodology enables disentanglement of bulk, interface, and quantum well trap contributions with greater sensitivity and specificity.

Our frequency-domain results reveal that:

- Oxide interface traps dominate the low-frequency conductance response, particularly in depletion and weak inversion, with signatures that scale predictably with trap density and capture cross-section;
- Bulk traps primarily modulate the high-frequency shoulder in G/ω spectra, becoming most apparent under strong accumulation;
- Quantum well interface traps, especially when their densities fall below ~$10^8$ $cm^{-2}$, produce broad or weak features that are often invisible in standard impedance spectra due to poor capacitive coupling and long emission times.

To address this resolution gap, especially for buried or low-density traps, we implemented a time-domain analysis of gate current transients following voltage pulses. This DLTS-inspired method provides enhanced sensitivity to trap location and emission dynamics. Our analysis decomposes the transient response into three distinct components:

- The fast decay constant $\tau_1$ is sensitive to bulk traps and is most pronounced in deep depletion,

- The intermediate decay $\tau_2$ arises from oxide interface traps, active across a wide range of depletion biases,
- The slowest component $\tau_3$, which governs the settling time, is dominated by QW interface traps, despite their low densities and weak frequency-domain signatures.

These time-resolved features remain detectable even when the same traps leave negligible signatures in G/ω plots, confirming the superior sensitivity of time-domain methods for identifying deeply buried, fast-switching, or sparsely distributed traps.

Importantly, we contextualized these trap dynamics within the operational needs of gate-defined Ge hole spin qubits, which are particularly vulnerable to charge noise due to strong spin-orbit coupling. Our framework enables the identification of trap species and locations that are most detrimental to spin qubit coherence, provides a means to quantitatively estimate acceptable trap density thresholds, informs the selection of materials and dielectrics that minimize noise coupling, and supports the optimization of gate biasing schemes to avoid activating harmful trap states.

This work represents the first unified analytical framework that connects classical semiconductor trap spectroscopy with the quantum-coherent operation of spin-based qubits, particularly in group IV hole-based architectures. While focused on Ge/SiGe, the approach is readily extendable to Si/SiGe, GeSn, InAs, and hybrid superconductor-semiconductor systems where charge noise poses a significant challenge.

In summary, our study delivers both a diagnostic toolkit and a predictive design strategy for improving coherence and fidelity in scalable spin qubit platforms through detailed understanding and control of trap-induced charge noise.

## V. Acknowledgment

This study was supported by AFOSR and the Laboratory for Physical Sciences (LPS) under contract numbers FA9550-23-1-0302 and FA9550-23-1-0763.

### References

[1] Nicollian, Edward H., and John R. Brews. *MOS (metal oxide semiconductor) physics and technology*. John Wiley & Sons, 2002.

[2] Lang, D. V. "Deep-level transient spectroscopy: A new method to characterize traps in semiconductors." *Journal of applied physics* 45, no. 7 (1974): 3023-3032.

[3] Blood, Peter, and John Wilfred Orton. "The electrical characterization of semiconductors: majority carriers and electron states." *Techniques of physics* 14 (1992): i-xxiii.

[4] Moeen, Mahdi, Arash Salemi, M. Kolahdouz, Mikael Östling, and Henry H. Radamson. "Characterization of SiGe/Si multi-quantum wells for infrared sensing." *Applied Physics Letters* 103, no. 25 (2013).

[5] Arai, Yasutomo, Yoshifumi Katano, Koji Tsubaki, Shigeki Uchida, and Kyoichi Kinoshita. "Infrared properties of interstitial oxygen in homogeneous bulk Si1− XGeX crystals." *Journal of Crystal Growth* 565 (2021): 126128.

[6] Gatti, Eleonora. "Recombination processes and carrier dynamics in Ge/SiGe multiple quantum wells.", PhD Dissertation, Università degli Studi di Milano-Bicocca (2012).

[7] Hutchins-Delgado, Troy A., Andrew J. Miller, Robin Scott, Ping Lu, Dwight R. Luhman, and Tzu-Ming Lu. "Characterization of shallow, undoped Ge/SiGe quantum wells commercially grown on 8-in.(100) Si wafers." *ACS Applied Electronic Materials* 4, no. 9 (2022): 4482-4489.

[8] Dsouza, K., Del Vecchio, P., Rotaru, N., Moutanabbir, O., & Vashaee, D. (2024). Heavy Hole vs. Light Hole Spin Qubits: A Strain-Driven Study of SiGe/Ge and GeSn/Ge. *arXiv preprint* arXiv:2412.16734.
[9] Madia, Oreste, Jacek Kepa, V. V. Afanas' ev, Jacopo Franco, Ben Kaczer, Andrii Hikavyy, and Andre Stesmans. "Dangling bond defects in silicon-passivated strained-Si 1− x Ge x channel layers." *Journal of Materials Science: Materials in Electronics* 31 (2020): 75-79.
[10] Bashir, A., K. Gallacher, R. W. Millar, D. J. Paul, A. Ballabio, J. Frigerio, G. Isella et al. "Interfacial sharpness and intermixing in a Ge-SiGe multiple quantum well structure." *Journal of applied physics* 123, no. 3 (2018).
[11] Costa, Davide, Lucas EA Stehouwer, Yi Huang, Sara Martí-Sánchez, Davide Degli Esposti, Jordi Arbiol, and Giordano Scappucci. "Reducing disorder in Ge quantum wells by using thick SiGe barriers." *Applied Physics Letters* 125, no. 22 (2024).
[12] Kavrik, Mahmut S., Peter Ercius, Joanna Cheung, Kechao Tang, Qingxiao Wang, Bernd Fruhberger, Moon Kim, Yuan Taur, Paul C. McIntyre, and Andrew C. Kummel. "Engineering high-K/SiGe interface with ALD oxide for selective GeO x reduction." *ACS applied materials & interfaces* 11, no. 16 (2019): 15111-15121.
[13] Varley, Joel B., Keith G. Ray, and Vincenzo Lordi. "Dangling bonds as possible contributors to charge noise in silicon and silicon–germanium quantum dot qubits." *ACS Applied Materials & Interfaces* 15, no. 36 (2023): 43111-43123.
[14] Hutchins-Delgado, Troy A., Andrew J. Miller, Robin Scott, Ping Lu, Dwight R. Luhman, and Tzu-Ming Lu. "Characterization of shallow, undoped Ge/SiGe quantum wells commercially grown on 8-in.(100) Si wafers." *ACS Applied Electronic Materials* 4, no. 9 (2022): 4482-4489.
[15] Hendrickx, N. W., D. P. Franke, A. Sammak, G. Scappucci, and M. Veldhorst. "Fast two-qubit logic with holes in germanium." *Nature* 577, no. 7791 (2020): 487-491.
[16] Scappucci, Giordano, Christoph Kloeffel, Floris A. Zwanenburg, Daniel Loss, Maksym Myronov, Jian-Jun Zhang, Silvano De Franceschi, Georgios Katsaros, and Menno Veldhorst. "The germanium quantum information route." *Nature Reviews Materials* 6, no. 10 (2021): 926-943.
[17] Watzinger, Hannes, Josip Kukučka, Lada Vukušić, Fei Gao, Ting Wang, Friedrich Schäffler, Jian-Jun Zhang, and Georgios Katsaros. "A germanium hole spin qubit." *Nature communications* 9, no. 1 (2018): 3902.
[18] Maurand, R., X. Jehl, D. Kotekar-Patil, Andrea Corna, Heorhii Bohuslavskyi, R. Laviéville, L. Hutin et al. "A CMOS silicon spin qubit." *Nature communications* 7, no. 1 (2016): 13575.
[19] Veldhorst, Menno, C. H. Yang, JCCea Hwang, W. Huang, J. P. Dehollain, J. T. Muhonen, S. Simmons et al. "A two-qubit logic gate in silicon." *Nature* 526, no. 7573 (2015): 410-414.
[20] Yoneda, Jun, Kenta Takeda, Tomohiro Otsuka, Takashi Nakajima, Matthieu R. Delbecq, Giles Allison, Takumu Honda et al. "A quantum-dot spin qubit with coherence limited by charge noise and fidelity higher than 99.9%." *Nature nanotechnology* 13, no. 2 (2018): 102-106.
[21] Freeman, Blake M., Joshua S. Schoenfield, and HongWen Jiang. "Comparison of low frequency charge noise in identically patterned Si/SiO2 and Si/SiGe quantum dots." *Applied Physics Letters* 108, no. 25 (2016).
[22] Paladino, E., Y. M. Galperin, G. Falci, and B. L. Altshuler. "1/f noise: Implications for solid-state quantum information." *Reviews of Modern Physics* 86, no. 2 (2014): 361-418.
[23] Ithier, G., E. Collin, P. Joyez, P. J. Meeson, Denis Vion, Daniel Esteve, F. Chiarello et al. "Decoherence in a superconducting quantum bit circuit." *Physical Review B—Condensed Matter and Materials Physics* 72, no. 13 (2005): 134519.
[24] Dial, O_E, Michael Dean Shulman, Shannon Pasca Harvey, H. Bluhm, V. Umansky, and Amnon Yacoby. "Charge noise spectroscopy using coherent exchange oscillations in a singlet-triplet qubit." *Physical review letters* 110, no. 14 (2013): 146804.
[25] Connors, Elliot J., J. Nelson, Lisa F. Edge, and John M. Nichol. "Charge-noise spectroscopy of Si/SiGe quantum dots via dynamically-decoupled exchange oscillations." *Nature communications* 13, no. 1 (2022): 940.
[26] Duong, N. & Vashaee, D. (2025). Design and Optimization of Spin Dynamics in Ge Quantum Dots: g-Factor Modulation, Dephasing Sweet Spots, and Phonon-Induced Relaxation. *arXiv preprint* arXiv:2509.10731
[27] Xie, Qi, Shaoren Deng, Marc Schaekers, Dennis Lin, Matty Caymax, Annelies Delabie, Xin-Ping Qu, Yu-Long Jiang, Davy Deduytsche, and Christophe Detavernier. "Germanium surface passivation and atomic

layer deposition of high-k dielectrics—a tutorial review on Ge-based MOS capacitors." *Semiconductor Science and Technology* 27, no. 7 (2012): 074012.
[28] Sammak, Amir, Diego Sabbagh, Nico W. Hendrickx, Mario Lodari, Brian Paquelet Wuetz, Alberto Tosato, LaReine Yeoh et al. "Shallow and undoped germanium quantum wells: a playground for spin and hybrid quantum technology." *Advanced Functional Materials* 29, no. 14 (2019): 1807613.
[29] Lodari, M., O. Kong, M. Rendell, A. Tosato, A. Sammak, M. Veldhorst, A. R. Hamilton, and G. Scappucci. "Lightly strained germanium quantum wells with hole mobility exceeding one million." *Applied Physics Letters* 120, no. 12 (2022).
[30] Costa, Davide, Lucas EA Stehouwer, Yi Huang, Sara Martí-Sánchez, Davide Degli Esposti, Jordi Arbiol, and Giordano Scappucci. "Reducing disorder in Ge quantum wells by using thick SiGe barriers." *Applied Physics Letters* 125, no. 22 (2024).
[31] Matsuoka, Ryotaro, Eriko Shigesawa, Satoru Miyamoto, Kentarou Sawano, and Kohei M. Itoh. "Fabrication of Ge MOS with low interface trap density by ALD of Al2O3 on epitaxially grown Ge." *Semiconductor Science and Technology* 34, no. 1 (2018): 014004.
[32] Xu, Yang, Genquan Han, Huan Liu, Yibo Wang, Yan Liu, Jinping Ao, and Yue Hao. "Ge pMOSFETs with GeO x passivation formed by ozone and plasma post oxidation." *Nanoscale research letters* 14 (2019): 1-7.
[33] Zhang, R., J. C. Lin, X. Yu, M. Takenaka, and S. Takagi. "Impact of plasma post oxidation temperature on interface trap density and roughness at GeOx/Ge interfaces." *Microelectronic engineering* 109 (2013): 97-100.
[34] Fleetwood, Daniel M. "'Border traps' in MOS devices." *IEEE transactions on nuclear science* 39, no. 2 (1992): 269-271.
[35] Gu, Siyuan, Jie Min, Yuan Taur, and Peter M. Asbeck. "Characterization of interface defects in ALD Al2O3/p-GaSb MOS capacitors using admittance measurements in range from kHz to GHz." *Solid-State Electronics* 118 (2016): 18-25.
[36] Hellenbrand, Markus, Erik Lind, Olli-Pekka Kilpi, and Lars-Erik Wernersson. "Effects of traps in the gate stack on the small-signal RF response of III-V nanowire MOSFETs." *Solid-State Electronics* 171 (2020): 107840.
[37] Bersuker, Gennadi. "Reliability implications of fast and slow degradation processes in high-k gate stacks." In *High Permittivity Gate Dielectric Materials*, pp. 309-341. Berlin, Heidelberg: Springer Berlin Heidelberg, 2013.
[38] Zhou, Binze, Mengjia Liu, Yanwei Wen, Yun Li, and Rong Chen. "Atomic layer deposition for quantum dots based devices." *Opto-Electronic Advances* 3, no. 9 (2020): 190043-1.